\documentclass[a4paper,11pt]{article}
\usepackage{jheppub}
\usepackage{amsmath,amssymb}
\usepackage{dsfont}
\usepackage{braket}
\usepackage{slashed} 

\usepackage{graphicx}
\usepackage[ansinew]{inputenc}
\usepackage[usenames,dvipsnames]{pstricks}
\usepackage{graphicx}
\usepackage{xcolor,color}
\usepackage{bm}
\newcommand{\pt}{\ensuremath{p_{\mathrm{T}}}}

\renewcommand{\d}{\ensuremath{\mathrm{d}}}
\newcommand{\mchi}{\ensuremath{\mathrm{m}_{\chi}}}
\newcommand{\mz}{\ensuremath{\mathrm{m}_{\mathrm{Z}}}}
\newcommand{\sz}{\ensuremath{s_{\mathrm{Z}}}}
\newcommand{\sx}{\ensuremath{s_{\mathrm{X}}}}
\newcommand{\sy}{\ensuremath{s_{\mathrm{Y}}}}

\newcommand{\hs}{\ensuremath{\hat{s}}}

\newcommand{\qt}{\ensuremath{q_{\mathrm{T}}}}
\newcommand{\pz}{\ensuremath{p_{\mathrm{Z}}}}
\newcommand{\py}{\ensuremath{p_{\mathrm{Y}}}}
\newcommand{\xz}{\ensuremath{x_{\mathrm{T,Z}}}}

\newcommand{\xy}{\ensuremath{x_{\mathrm{T,Y}}}}

\newcommand{\yz}{\ensuremath{y_{\mathrm{Z}}}}

\newcommand{\yy}{\ensuremath{y_{\mathrm{Y}}}}
\newcommand{\ee}{\ensuremath{\mathrm{e}}}
\newcommand{\mys}{\ensuremath{\mathrm{m}_{\mathrm{Y_0}}}}
\newcommand{\myv}{\ensuremath{\mathrm{m}_{\mathrm{Y_1}}}}
\newcommand{\myt}{\ensuremath{\mathrm{m}_{\mathrm{Y_2}}}}

\newcommand{\ptvecmiss}{\ensuremath{\mathbf{p}^{\text{miss}}_\mathrm{T}}}
\newcommand{\etmiss}{\ensuremath{E^{\text{miss}}_\mathrm{T}}}
\newcommand{\ptll}{p^l_{\mathrm{T}}}
\newcommand{\fbinv}{\mathrm{fb}^{\mathrm{-1}}}

\title{Probing the Dark Sector through Mono-Z Boson Leptonic Decays}

\author{Daneng Yang,}
\author{Qiang Li}

\affiliation{Department of Physics and State Key Laboratory of Nuclear Physics and Technology, \\
Peking University, Beijing, 100871, China}

\emailAdd{yangdn@pku.edu.cn,qliphy0@pku.edu.cn}

\date{\today}

\abstract{Collider search for dark matter production has been performed over the years based on high $\pt$ standard model signatures balanced by large missing transverse energy.
The mono-Z boson production with leptonic decay has a clean signature with the advantage that the decaying electrons and muons can be precisely measured.
This signature not only enables reconstruction of the Z boson rest frame, but also
makes possible recovery of the underlying production dynamics through the decaying lepton angular distribution.
In this work, we exploit full information carried by the leptonic Z boson decays to set limits on coupling strength parameters of the dark sector.
We study simplified dark sector models with scalar, vector, and tensor mediators and
observe among them different signatures in the distribution of angular coefficients.
Specifically, we show that angular coefficients can be used to distinguish different scenarios of the spin-0 and spin-1 models, including the ones with parity-odd and charge conjugation parity-odd operators. 
To maximize the statistical power, we perform a matrix element method study with a dynamic construction of event likelihood function.
We parametrize the test statistic such that sensitivity from the matrix element is quantified through a term measuring the shape difference. 
Our results show that the shape differences provide significant improvements in the limits, especially for the scalar mediator models.           
We also present an example application of a matrix-element-kinematic-discriminator, an easier approach that is applicable for experimental data. 
}
\begin{document}
\maketitle

\flushbottom
%---------------------------------------------------------------------

\tableofcontents

\section{Introduction}

The existence of dark matter (DM) is now well established.
Current measurement gives a cold DM density of $25.8$\%, which is much significant than the $4.84$\% baryon density~\cite{Patrignani:2016xqp,Ade:2015xua}.
Despite being an essential constituent of the universe,
intrinsic properties of the DM, like mass, spin and nongravitational interaction between the standard model (SM) particles are still elusive at present.
Assuming that DM is weakly interacting with the SM particles, 
the DM annihilation cross section will be constraint by the precisely measured relic DM abundance
and a weak-scale DM candidate is usually expected for consistency~\cite{PhysRevLett.39.165}.
The WIMP DM candidate can be produced at the LHC, and its missing from detection typically leads to large missing transverse energy, 
 resulting in mono-X signatures, 
where X may denote a jet~\cite{Beltran:2010ww,Aaboud:2017buf,Sirunyan:2017hci}, especially t-/b-jet~\cite{Aaboud:2017rzf,Sirunyan:2017xgm}, a photon~\cite{Sirunyan:2017ewk}, a Z boson~\cite{Carpenter:2012rg,Sirunyan:2017onm,Aaboud:2017bja}, a W boson~\cite{Bai:2012xg,Aaboud:2017efa} or a Higgs boson~\cite{Aaboud:2017yqz,Sirunyan:2017hnk}. 
Numerous efforts have been performed at the LHC searching for the DM, many results from 13 TeV collisions are now available~\cite{Sirunyan:2017onm,Sirunyan:2016iap,Sirunyan:2017hci,Khachatryan:2016jww,Sirunyan:2017hnk,Sirunyan:2017xgm,Sirunyan:2017ewk,Sirunyan:2017nvi,Aaboud:2017rzf,Aaboud:2017bja,Aaboud:2017buf,Aaboud:2017buh,Aaboud:2017efa,Aaboud:2017yqz},
with strategies and benchmark models described in Ref.~\cite{Abercrombie:2015wmb}.

In this analysis, we explore the effectiveness of the Z boson leptonic decay with mono-Z signature in probing properties of the dark sector.
Compared with other search channels, this channel has a relatively lower cross section 
and may not be the most powerful one at the stage of searching.
However, precisely measured electrons and muons provide a clean signature and can be used to increase the signal feasibility. 
Phenomenology of this channel has been explored in Ref.~\cite{Neubert:2015fka,Petriello:2008pu,Alves:2015dya,Han:1999ne,Yu:2014ula}, 
including higher-order QCD predictions, multivariate analysis, a search for extra dimension and effects on electron-positron colliders. 
LHC measurements are also available, and limits have been set on several dark sector models~\cite{Sirunyan:2017onm,Aaboud:2017bja,Aad:2014wca}.
To better exploit the powerfulness of the lepton angular distribution, 
we study systematically information carried by the angular distribution and
how they are affected by the dark sector.

The modeling of the dark sector can be implemented in many models.
As there is no strong support for the correctness of a specific model, it is now popular to set limits on parameters of effective or simplified theories~\cite{Goodman:2010ku,Goodman:2010yf,Abercrombie:2015wmb,Cao:2009uw}.
Despite the simplicity, these models may not be realistic if we are not applying them in a suitable case.
Either oversimplification nor overdress of the theory can lead to ineffectual results.
For example, going to very high energy can result in the violation of unitarity in effective theories~\cite{Abercrombie:2015wmb,Cotta:2012nj}.
On the other hand, some features are general among models and can have less dependence on the variations of model parameters,
e.g., spin and mass of the dark mediator, parity or charge conjugation parity (CP) of the couplings. 
If applying carefully, those effective or simplified models can help us better understand the phenomenology of the dark sector.

Motivated by this, we look for specific variables that can
have discrimination power on general features of the dark sector.
We consider the associated production of a Z boson and a dark mediator,
where the Z boson decays to a pair of electrons or muons
and the dark mediator decays to a pair of dark matter.
As the dark matter is unmeasurable, the typical feature of the event is a
single leptonically decaying Z boson, with $\pt$ balanced by the missing transverse momentum vector.
With precisely measured electron or muon momenta, one can reconstruct the Z boson rest frame
and study in detail information carried by the Z boson spin density matrix.
We consider simplified models for spin-0, spin-1, and spin-2 mediators~\cite{Mattelaer:2015haa,Backovic:2015soa,Neubert:2015fka,Das:2016pbk,Kraml:2017atm}.
In each case, only a few benchmark scenarios are considered with representative parameter values.
For the spin-0 model, we assume the dark mediator can only weakly interact with bosons
through a set of dimension-5 operators as described in Ref.~\cite{Neubert:2015fka}.
In this case, the mono-Z boson channel is advantageous as a triple boson coupling is necessary for the production.
If introducing couplings to the SM fermions assuming minimal flavor violation, their effects are suppressed
due to proportionalities to the Yukawa couplings~\cite{Backovic:2015soa,Cheung:2010zf,Lin:2013sca}.
The spin-1 mediator model is chosen to be consistent with the one adopted in the LHC experiment~\cite{Abercrombie:2015wmb}.
A spin-2 mediator model described in Ref.\cite{Kraml:2017atm} is also tested.

To maximally exploit the statistical power of the data,
we present a framework to use the matrix element method (MEM) with a dynamical construction of event likelihood function and set unbinned limits on parameters of the dark sector~\cite{doi:10.1143/JPSJ.57.4126,doi:10.1143/JPSJ.60.836,Gao:2010qx,Chatrchyan:2012sn,DeRujula:2010ys}.
We parametrize the test statistic in a way such that the sensitivity of MEM can be quantified through a term proportional to the KL-divergence of two probability density functions~\cite{kullback1951}.
Limits on the coupling strengths of the dark sector models are set at 95\% confidence level (CL) based on the asymptotic approximation.
As the spin-2 scenarios are found to have similar angular coefficients to the one of a spin-independent spin-1 model,
they are not considered in the limit setting.
An example application of a matrix-element-kinematic-discriminator is also demonstrated with simulated events. 

This paper is organized as follows:
Section~\ref{sec:param} introduces the parametrization of lepton angular distribution.
Section~\ref{sec:angC} describes computational details and presents numerical results of angular coefficients in the Collins-Soper frame.
Section~\ref{sec:limits} explained the statistical method for setting limits and present results on the coupling strengths of dark sector models.
Section~\ref{sec:summary} summarizes our major findings and outlooks aspects of the study.

%\clearpage

\section{Parametrization of lepton angular distribution}
\label{sec:param}

A probability density function (pdf) for a single event can be defined through the matrix element as~\cite{Alwall:2010cq}

\begin{eqnarray}
\rho(\mathbf{p}^{\text{vis}}|\lambda) = \dfrac{1}{\sigma_{\lambda}} \sum_{a,b} \int \d x_1 \d x_2 f_a(x_1,\mu_{\mathrm{F}}) f_b(x_2,\mu_{\mathrm{F}}) 
\int \d \Phi \dfrac{\d\hat{\sigma}}{\d \Phi} \prod_{i\in\text{vis}} \delta(\mathbf{p}_i-\mathbf{p}_i^{vis}),
\label{eq:pdf}
\end{eqnarray}

where $\Phi$ represents the Lorentz invariant phase space, in our case, a four body version $\Phi_4(k_{l},k_{\bar{l}},k_{\chi},k_{\bar{\chi}})$
with $l=e,\mu$ and $\chi$ for the DM particle. 
$f_a(x,\mu_{\mathrm{F}})$ corresponds to the parton distribution function of parton $a$, 
with an energy fraction of $x$ and a factorization scale $\mu_{\mathrm{F}}$. 
$\lambda$ stands for a set of parameters of interest. 
The visible part of the phase space is determined through observables, while the invisible part is integrated over. 
The general cross section formula is written as:

\begin{eqnarray}
\sigma &=& \sum_{a,b} \int \d x_1 \d x_2 f_a(x_1,\mu_{\mathrm{F}}) f_b(x_2,\mu_{\mathrm{F}}) 
\int \d \Phi_4(k_{l},k_{\bar{l}},k_{\chi},k_{\bar{\chi}}) \dfrac{\d \hat{\sigma}}{\d \Phi_4(k_{l},k_{\bar{l}},k_{\chi},k_{\bar{\chi}})}. 
\end{eqnarray}

For the same process, it follows that the $\rho(\mathbf{p}^{\text{vis}}|\lambda)$ is indeed a probability density function for the visible kinematics:

\begin{eqnarray}
\left(\prod_{i\in\text{vis}}\int\d^3p_i\right) \rho(\mathbf{p}^{\text{vis}}|\lambda) = 1.
\end{eqnarray}

To calculate the production of a Z boson in association with a DM mediator,
we parametrize the four-momenta as follows:

\begin{eqnarray}
p_1^{\mu} =& x_1 \dfrac{\sqrt{s}}{2} (1,0,0,1)^T &= \dfrac{\sqrt{\hs}}{2} \sqrt{\dfrac{x_1}{x_2}} (1,0,0,1)^T , \\ \nonumber
p_2^{\mu} =& x_2 \dfrac{\sqrt{s}}{2} (1,0,0,-1)^T &= \dfrac{\sqrt{\hs}}{2} \sqrt{\dfrac{x_2}{x_1}} (1,0,0,-1)^T , \\ \nonumber
\py^{\mu} =& (\py^0, -\qt, 0, \py^3)^T &= \left( \dfrac{\sqrt{s}}{2} \xy \cosh \yy, -\qt, 0, \dfrac{\sqrt{s}}{2} \xy \sinh \yy \right)^T, \\ \nonumber
\pz^{\mu} =& (\pz^0, \qt, 0, \pz^3)^T &= \left( \dfrac{\sqrt{s}}{2} \xz \cosh \yz, \qt, 0, \dfrac{\sqrt{s}}{2} \xz \sinh \yz \right)^T,
\end{eqnarray}

where $\xz=\dfrac{2\sqrt{\sz+\qt^2}}{\sqrt{s}}$, $\xy=\dfrac{2\sqrt{\sy+\qt^2}}{\sqrt{s}}$.

\begin{figure}[htbp]
  \centering
  \includegraphics[width=9.0cm]{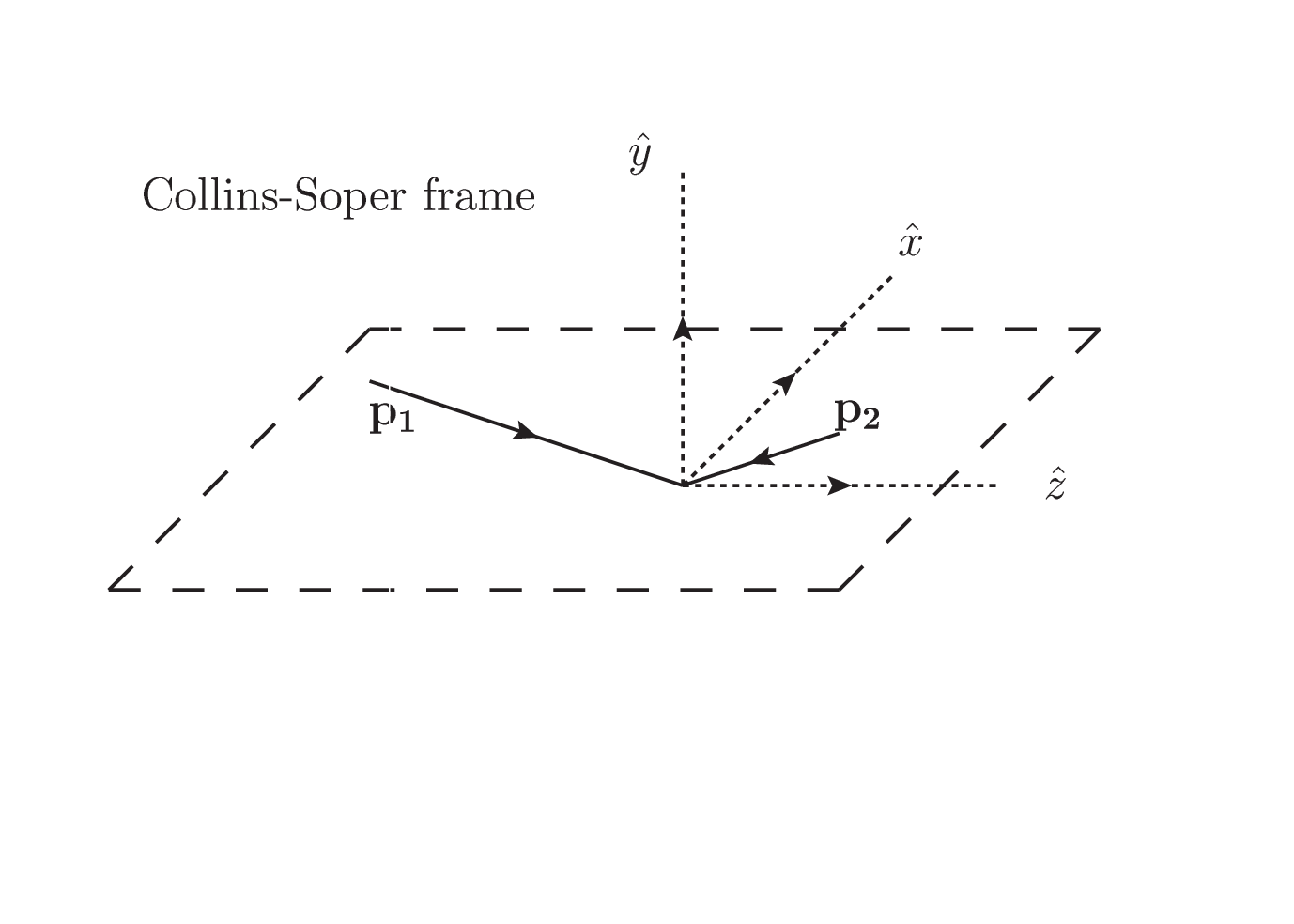}
  \caption{\label{fig:frames} Sketch of the Collins-Soper frame. $\mathbf{p}_1, \mathbf{p}_2$ correspond to the three momenta of the right- and left- flying protons.}
\end{figure}

It is common to study the decaying lepton angular distribution in the Collins-Soper (CS) frame~\cite{PhysRevD.16.2219}.
The Collins-Soper frame, as shown in Fig.\ref{fig:frames}, is a Z boson rest frame, with the z-axis lying in a way bisects the opening angle $\theta_{ab}$ between the beam and negative target momenta directions.
In this frame, momenta of the two incoming partons become:

\begin{eqnarray}
\label{eq:kinCS}
p^{CS}_1 &=& \dfrac{x_1}{2} \sqrt{\dfrac{s}{\sz}} \ee^{-\yz} ( \sqrt{\sz+\qt^2} , -\qt,0,\sqrt{\sz}), \\ \nonumber
p^{CS}_2 &=& \dfrac{x_2}{2} \sqrt{\dfrac{s}{\sz}} e^{\yz} ( \sqrt{\sz+\qt^2} , -\qt,0,-\sqrt{\sz}), 
\end{eqnarray}

where the $x_{1,2}$ and $\yz$ dependences have been factorized out.
Determined by these two momenta, the z-axis of this frame treats the in- and out-partons equally and $\tan\frac{\theta_{ab}}{2}=\frac{|\bf{\qt}|}{\sqrt{\sz}}$
is invariant under the longitudinal boost. 
This feature makes it suitable for the study of effects at finite $|\bf{\qt}|$.
To avoid possible dilutions by the initial states swapped processes, 
we performed a rotation of $\pi$ around the x-axis for events with $\yz<0$~\cite{Aad:2016izn,Khachatryan:2015paa}. 
This rotation makes all angular coefficients distribute symmetric in $\yz$.

In experiment, only the two decaying lepton pair are measurable, giving 
a set of visible variables $\yz,\qt,\sz,\cos\theta_{CS},\phi_{CS}$, where the latter two denote polar and azimuthal angles of the charged lepton in the CS frame.
We parametrize the Lorentz invariant phase space in a way such that the invisible part
$\sy,\yy,\cos\theta_{\chi},\phi_{\chi}$ can be integrated over:

\begin{eqnarray}
\int \d \Phi_4(k_{l},k_{\bar{l}},k_{\chi},k_{\bar{\chi}}) &=& 
\int \dfrac{\d \sz}{2\pi} \dfrac{\d \sx}{2\pi} \int \d\Phi'_2(\py,\pz) \d\Phi_2(k_{l},k_{\bar{l}}) \d\Phi_2(k_{\chi},k_{\bar{\chi}}), \\ 
\int \d\Phi'_2(\py,\pz) &=& \int \dfrac{\d^3 \pz}{(2\pi)^3 2 \pz^0} \dfrac{\d^3 \py}{(2\pi)^3 2 \py^0} (2\pi)^4 \delta^4(p_1+p_2-\pz-\py), \\ \nonumber
      &=& \dfrac{1}{4\pi s} \int \d \yz \d \yy \d \qt \cdot \qt  \\ \nonumber
       && \delta(x_1-\dfrac{\xz}{2}\ee^{\yz}-\dfrac{\xy}{2}\ee^{\yy}) \delta(x_2-\dfrac{\xz}{2}\ee^{-\yz}-\dfrac{\xy}{2}\ee^{-\yy}) \\ 
\int \d \Phi_2(k_1,k_2) &=& \dfrac{1}{8\pi} \bar{\beta}(\dfrac{\mathrm{m}_1^2}{s_{12}},\dfrac{\mathrm{m}_2^2}{s_{12}}) \dfrac{\d \cos\theta}{2} \dfrac{\d \phi}{2\pi}, \\ 
      \bar{\beta}(a,b) &=& \sqrt{\lambda(1,a,b)} = \sqrt{1+a^2+b^2-2 a-2 b-2 a b}. \nonumber
\end{eqnarray}

Then we factorize the decay angular distribution in terms of nine harmonic polynomials and eight angular coefficients $A_i,i=0,...,7$~\cite{Aad:2016izn,Khachatryan:2015paa}:

\begin{eqnarray}
\dfrac{\d\sigma}{\d\qt\d\yz\d\sz\d\cos\theta\d\phi} 
&&= \left( \int \d\cos\theta \d\phi \dfrac{\d\sigma}{\d\qt\d\yz\d\sz\d\cos\theta\d\phi} \right)
\dfrac{3}{16 \pi} \\ \nonumber 
&& \Bigg\{ (1+\cos^2\theta) + \dfrac{1}{2} A_0 (1-3\cos^2\theta) +
 A_1 \sin2\theta \cos\phi \\ \nonumber 
&& + \dfrac{1}{2} A_2 \sin^2\theta \cos2\phi 
 + A_3 \sin\theta \cos\phi + A_4 \cos\theta \\ \nonumber
&& + A_5 \sin^2\theta \sin2\phi + A_6 \sin2\theta \sin\phi
 + A_7 \sin\theta\sin\phi \Bigg\},
\end{eqnarray}

where the polar and azimuthal angles $\theta,\phi$ are measured in the CS frame. 
Coefficients $A_5-A_7$ are parity-odd and do not contribute at tree level and are found to be
very small for a Z boson production~\cite{Aad:2016izn,Khachatryan:2015paa}.  
Therefore in this analysis, we consider only $A_0-A_4$.

%\clearpage

\section{Numerical results of angular coefficients in the Collins-Soper frame}
\label{sec:angC}

As we are not directly searching for a resonance, the $\sz$ is expected to give no sensitivity and a narrow width approximation (NWA) is applied for convenience.
Apart from that, we have four observables from the Z boson decay: $\yz,\qt,\cos\theta_{CS},\phi_{CS}$.
To study the features of this four-dimensional data, we calculate angular coefficients in the $\yz-\qt$ plane for
both the major background process $\mathrm{Z Z}\to 2l2\nu$ production and different dark sector models.
The angular coefficients can be extracted using the method of moments~\cite{Beaujean:2015xea}.
In the experiment, it is more straightforward to extract from a likelihood fit~\cite{Aad:2016izn,Khachatryan:2015paa}.

Applying NWA for the Z boson, the cross section can be calculated through spin density matrices of the Z boson production ($\rho^{\rm{P}}$) and decay ($\rho^{\rm{D}}$): 

\begin{eqnarray}
\dfrac{\d\sigma}{\d\yz\d\qt\d\sy\d\Phi_2(k_{\chi},k_{\bar{\chi}})\d\cos\theta\d\phi} = 
\dfrac{\d\sigma_{P}}{\d\yz\d\qt\d\sy\d\Phi_2(k_{\chi},k_{\bar{\chi}})} \cdot 
\mathrm{Br}(\mathrm{Z}\to l^+ l^-) \cdot 3 \sum_{s,s'} \rho^{\mathrm{P}}_{s s'} \rho^{\mathrm{D}}_{s s'}. \nonumber 
\end{eqnarray}

The Z boson production density matrix is defined in a specific range (${\cal R}$) of $\yz-\qt$ as follows:

\begin{eqnarray}
\rm{Tr}\rho^{\rm{P}} &=& \int_{\cal R} \d\Phi'_2(\py,\pz) \d\Phi_2(k_{\chi},k_{\bar{\chi}}) 
\sum_{a,b} f_a(x_1,\mu_{\mathrm{F}}) f_b(x_2,\mu_{\mathrm{F}}) \dfrac{1}{2\hs} 
\overline{\sum_{\text{ext}}} \sum_{s} \left| {\cal M}_s \right|^2, \\ \nonumber
\rho^{\rm{P}}_{s s'} &=& \dfrac{1}{\rm{Tr}\rho^{\rm{P}}} \int_{\cal R} \d\Phi'_2(\py,\pz) \d\Phi_2(k_{\chi},k_{\bar{\chi}})
\sum_{a,b} f_a(x_1,\mu_{\mathrm{F}}) f_b(x_2,\mu_{\mathrm{F}}) \dfrac{1}{2\hs} 
\overline{\sum_{\text{ext}}} {\cal M}_s {\cal M}^*_{s'} 
\end{eqnarray}

where $\overline{\sum}_{\text{ext}}$ means sum over spins and colors of all external particles other than the Z boson and averaged for the initial state ones. 
The decay density matrix is obtained using the Z boson decay amplitudes and parametrized similar as in Ref.\cite{Dutta:2008bh}.
The production and decay density matrices are both normalized such that the trace is one. 

To obtain the amplitudes, we start from the {\sc FeynRules} models implimented by authors of Ref.\cite{Mattelaer:2015haa,Backovic:2015soa,Neubert:2015fka,Das:2016pbk,Kraml:2017atm}
and use {\sc ALOHA} in the {\sc MadGraph} framework to generate {\sc HELAS} subroutines for the helicity amplitudes~\cite{Alloul:2013bka,deAquino:2011ub,Alwall:2014hca,HAGIWARA19861,DREINER20101}. 
In the CS frame, we choose the z-axis as spin quantization axis,
hence a rotation is necessary to bring the helicity frame results to the CS frame ones. 
We choose the y-axis to be common for the two frames and find the opening angle between 
the two frames $\omega$ can be obtained through
\begin{eqnarray}
\cos\omega &=& \dfrac{2\sqrt{\tau_{\mathrm{Z}}}\sinh\yz}{\sqrt{\xz^2\cosh^2\yz-4\tau_{\mathrm{Z}}}},
\end{eqnarray}

where $\tau_{\mathrm{Z}} \equiv \dfrac{\sz}{s}$ and $\omega \in [0,\pi)$. 
The density matrices are then rotated according to Wigner's d-functions:

\begin{eqnarray}
\rho_{s s'}^{\mathrm{P},HEL} &=& \sum_{\alpha, \beta} d^{J=1}_{\alpha s} (\omega)   d^{J=1}_{\beta s'} (\omega) \rho^{\mathrm{P},CS}_{\alpha \beta}, \\ \nonumber
\rho_{s s'}^{\mathrm{P},CS}  &=& \sum_{\alpha, \beta} d^{J=1}_{\alpha s} (-\omega)  d^{J=1}_{\beta s'} (-\omega) \rho^{\mathrm{P},HEL}_{\alpha \beta}, \\ \nonumber
\end{eqnarray}

where we have used the following notations:

\begin{eqnarray}
g_{\alpha\beta} &=& - \sum_{s} \epsilon^*_{\alpha}(p,s) \epsilon_{\beta}(p,s) \\ \nonumber 
\epsilon^{\mu}(p,s) \epsilon_{\mu}(p,s') &=& - d^{J=1}_{s s'}(\theta_{s,s'}), \\ \nonumber
d^{J=1}_{s=+,-,0;s'=+,-,0} (\theta) &=& 
\left(
\begin{array}{ccc}
\dfrac{1+\cos\theta}{2} & \dfrac{1-\cos\theta}{2} & -\dfrac{\sin\theta}{\sqrt{2}} \\
\dfrac{1-\cos\theta}{2} & \dfrac{1+\cos\theta}{2} & \dfrac{\sin\theta}{\sqrt{2}} \\
\dfrac{\sin\theta}{\sqrt{2}} & -\dfrac{\sin\theta}{\sqrt{2}} & \cos\theta \\
\end{array}
\right).
\end{eqnarray}

The phase space is prepared analytically, and integration is performed using {\sc BASES}~\cite{Kawabata:1995th} and {\sc GNU Scientific Library}.  
We mapped the phase space variables to increase integration efficiencies. 
Specifically, for a massive propagator with mass m and width $\Gamma$, 
the invariant mass is generated with

\begin{eqnarray}
s &=& \mathrm{m}^2 + \mathrm{m} \Gamma \tan( x (y_{max}-y_{min}) + y_{min} ), \text{ where } \\ 
y_{min/max} &=& \arctan(\dfrac{s_{min/max}-\mathrm{m}^2}{\mathrm{m}\Gamma}),  \\ \nonumber
\mathrm{Jacobian} &=& \dfrac{y_{max}-y_{min}}{\mathrm{m}\Gamma} \left( (s-\mathrm{m}^2)^2 + (\mathrm{m}\Gamma)^2 \right),
\end{eqnarray}

and $x$ is a uniformly generated random number. 

The simulation considers $\sin\theta_W=0.23129$, $\mz=91.1876$~GeV, 
$\Gamma_{\mathrm{Z}}=2.4952$~GeV and $\alpha(\mz)^{-1}=127.95$~\cite{Patrignani:2016xqp}.
%$G_F=1.16637876\times 10^{-5}$~\cite{Patrignani:2016xqp}.
The W boson mass is obtained through $\mz \cos\theta_W$, assuming $\rho$ parameter equals to one.
The $\alpha_S$ is chosen to be consistent with the one in the parton distribution functions (PDF).
We use PDF set {\sc NNPDF23}~\cite{Ball:2013hta} with $\alpha_S(\mz)=0.130$ at leading order.
The factorization scale is set to be equal to the Z boson transverse energy $E_{\mathrm{T}}=\sqrt{\qt^2+\sz}$.
Cross sections in this section consider the visible Z boson decays to electrons and muons 
with NWA and $\mathrm{Br}(\mathrm{Z}\to l^+ l^-)=6.73$\%~\cite{Patrignani:2016xqp}.  
The advantage of our program is that high statistical accuracy can be achieved through a direct integration. 
To validate our program, we checked our angular coefficients through toy measurements based on {\sc MadGraph5\_aMC@NLO (MG5)} generated events.

\subsection{SM $\mathrm{Z Z}\to 2l2\nu$ background}

\begin{figure}[htb]
  \centering
  \includegraphics[width=10.0cm]{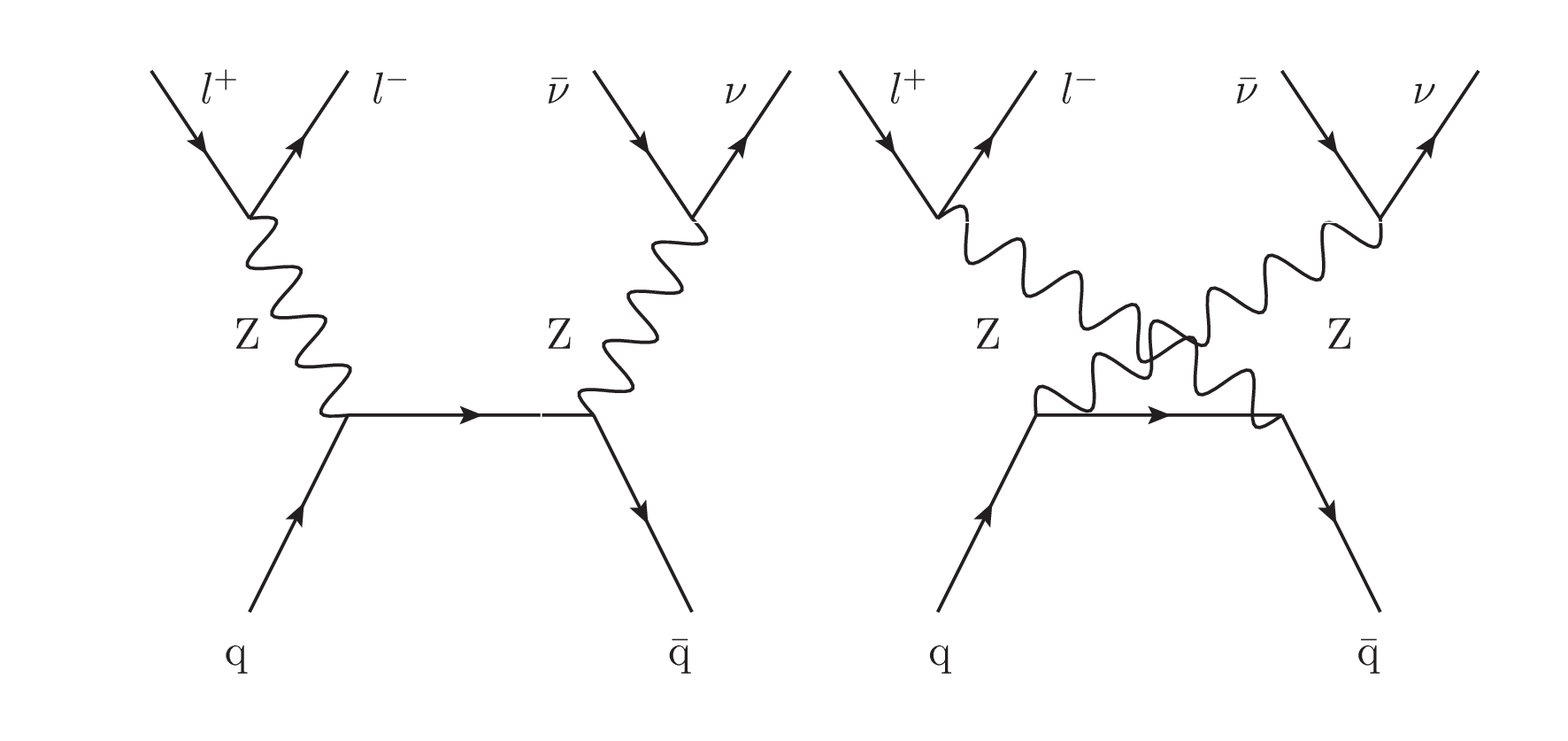}
  \caption{\label{fig:feynzz} Representative Feynman diagrams of the SM $\mathrm{Z Z}\to 2l2\nu$ production. }
\end{figure}

\begin{figure}[!h]
  \centering
  \includegraphics[width=4.5cm]{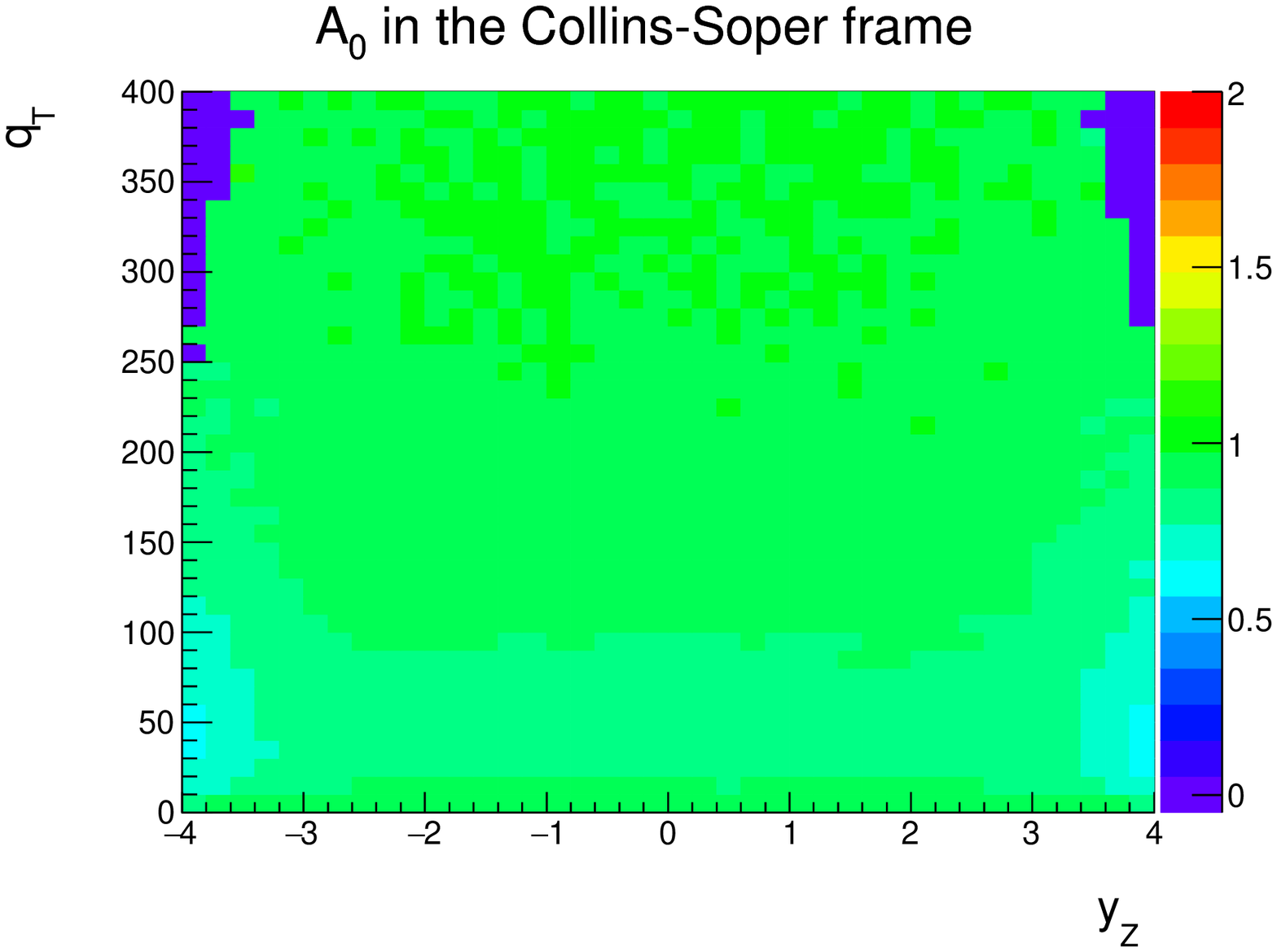}
  \includegraphics[width=4.5cm]{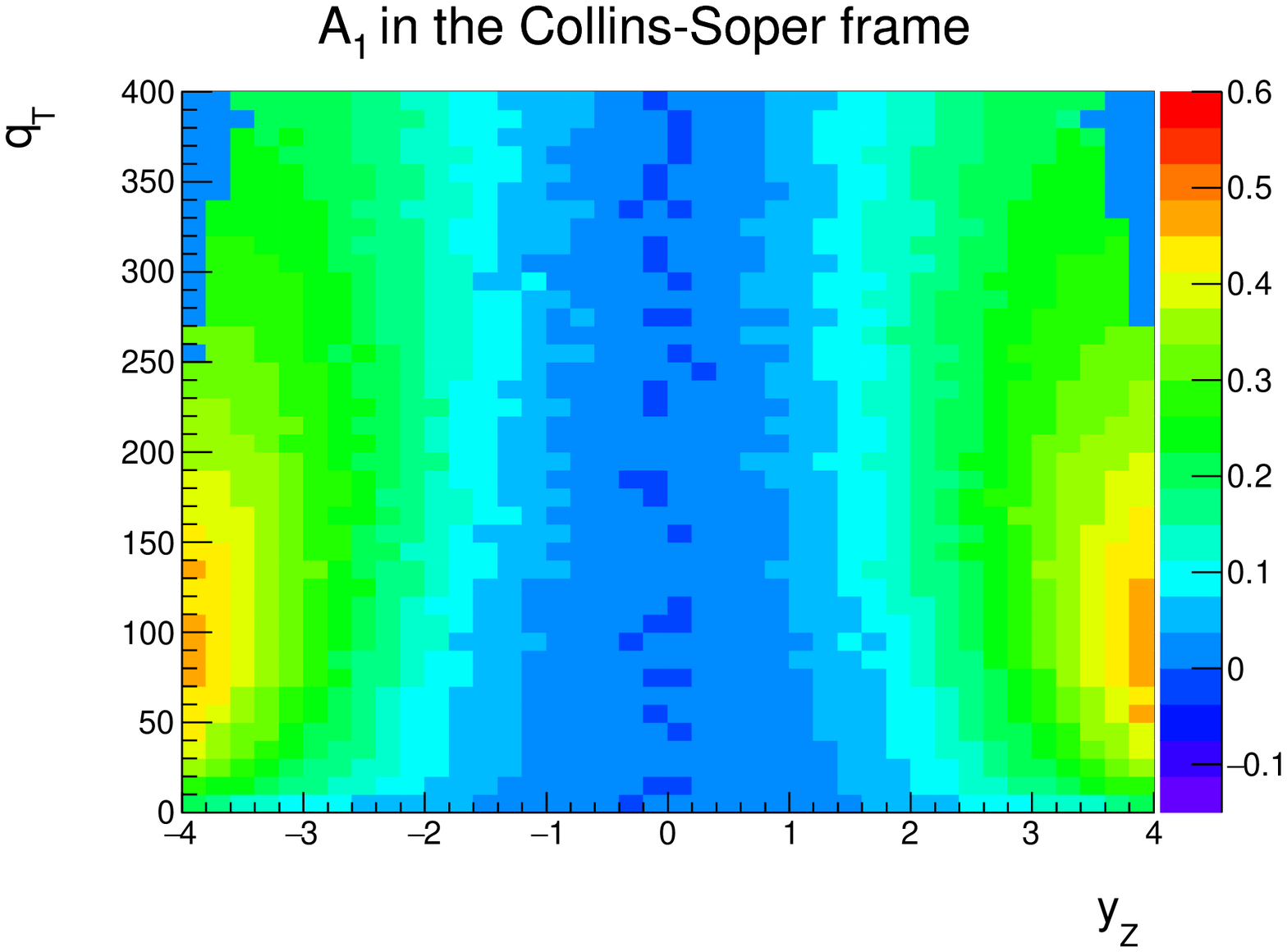} 
  \includegraphics[width=4.5cm]{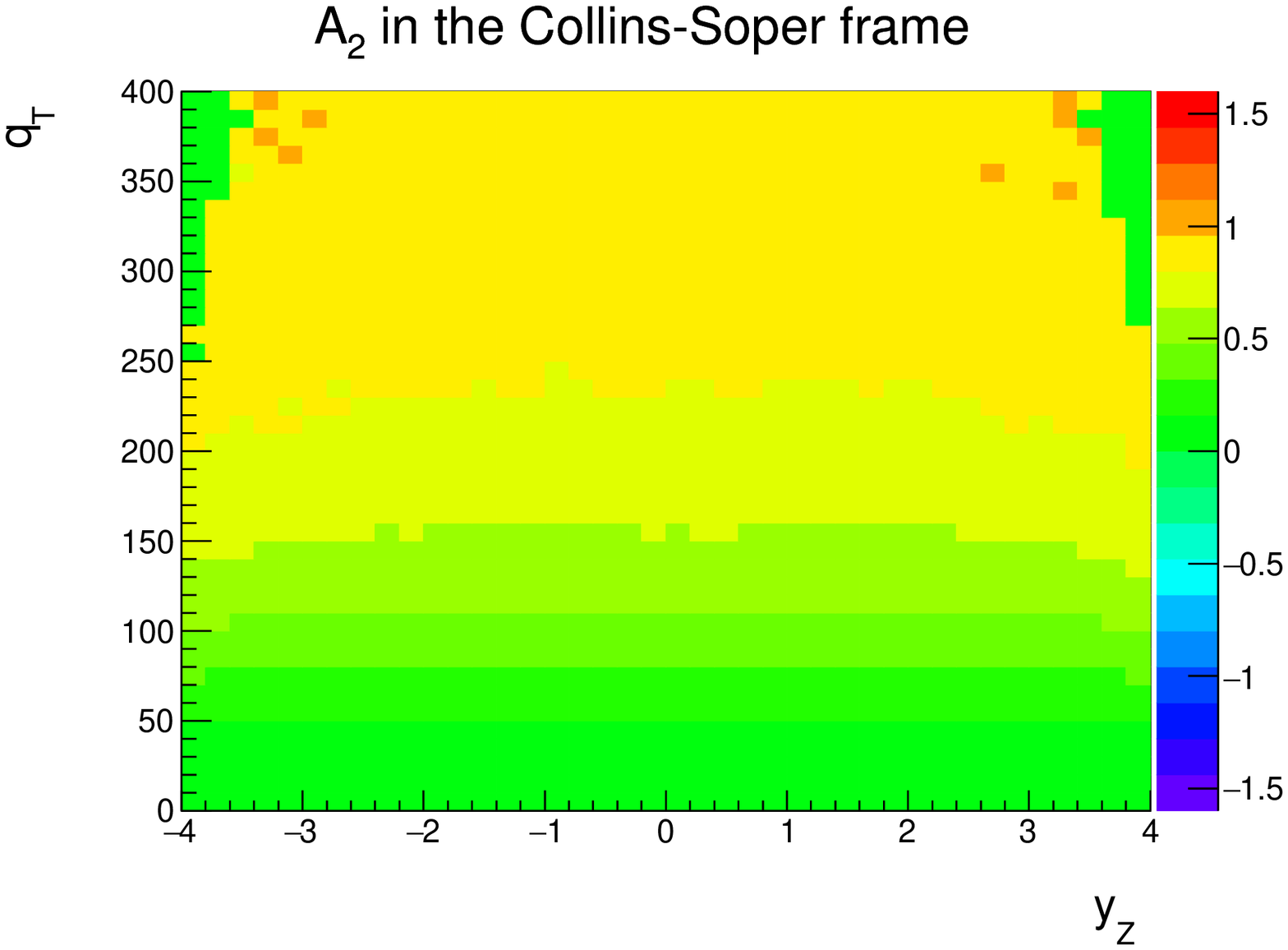} \\
  \includegraphics[width=4.5cm]{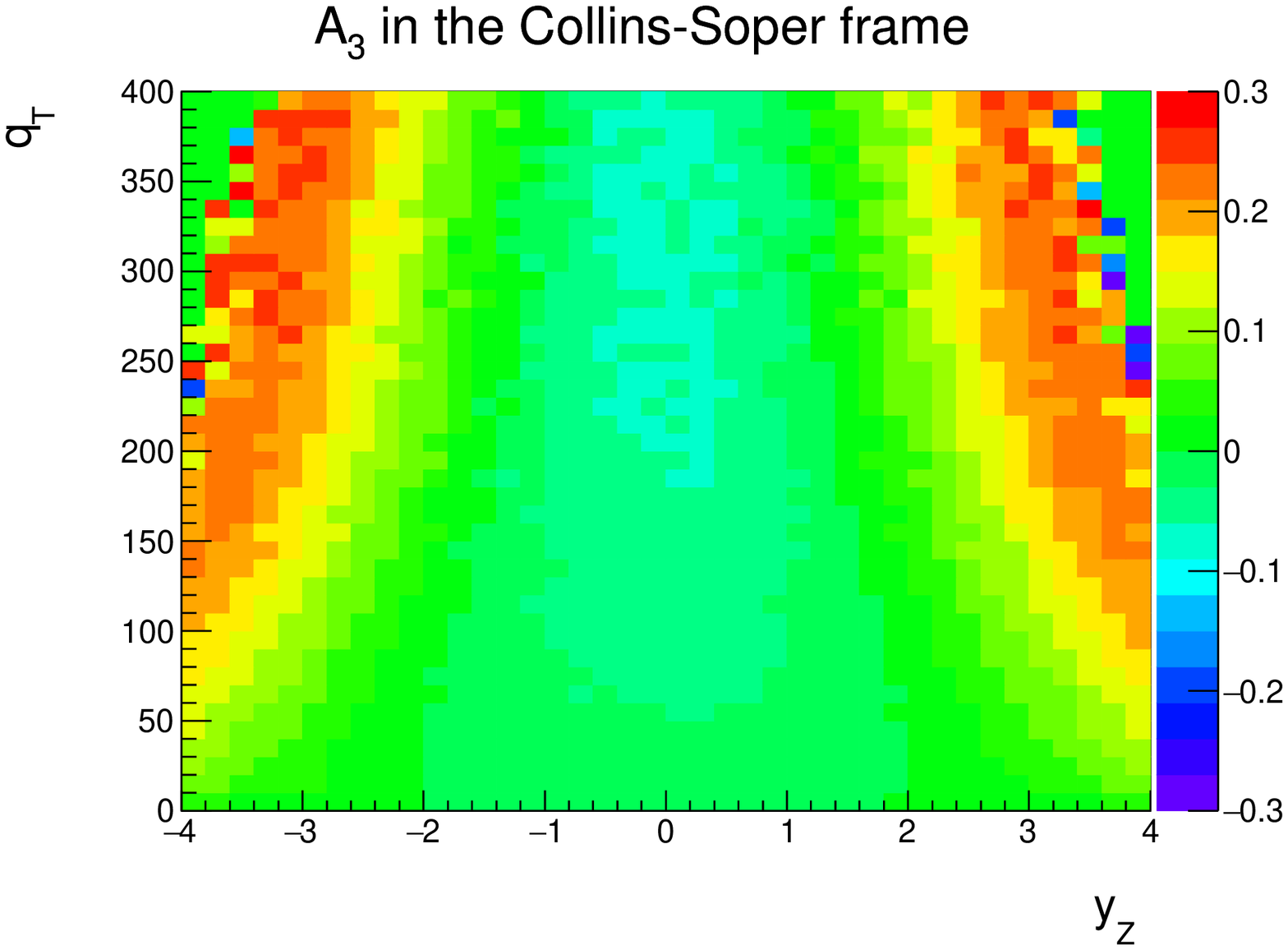} 
  \includegraphics[width=4.5cm]{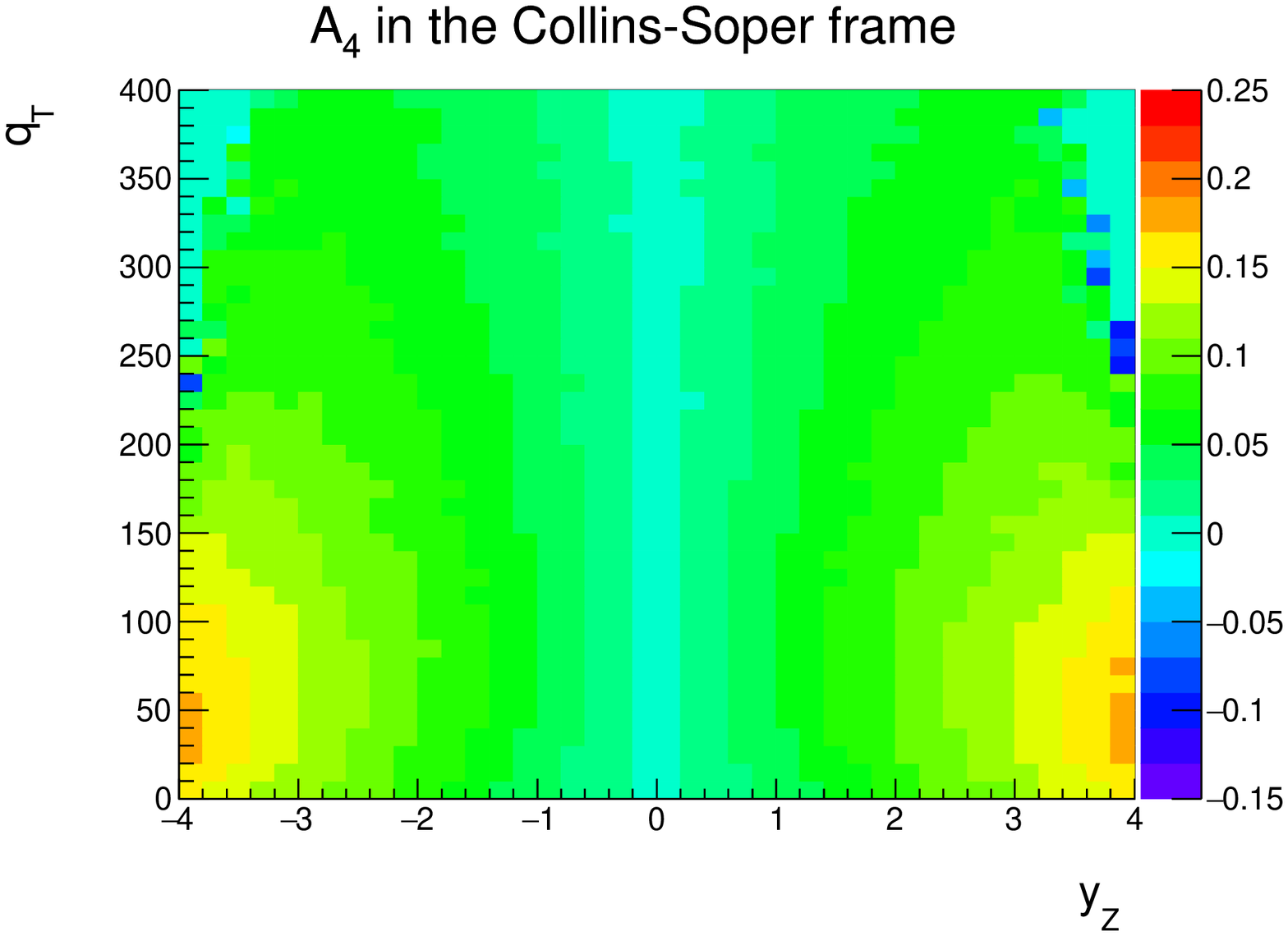}
  \includegraphics[width=4.5cm]{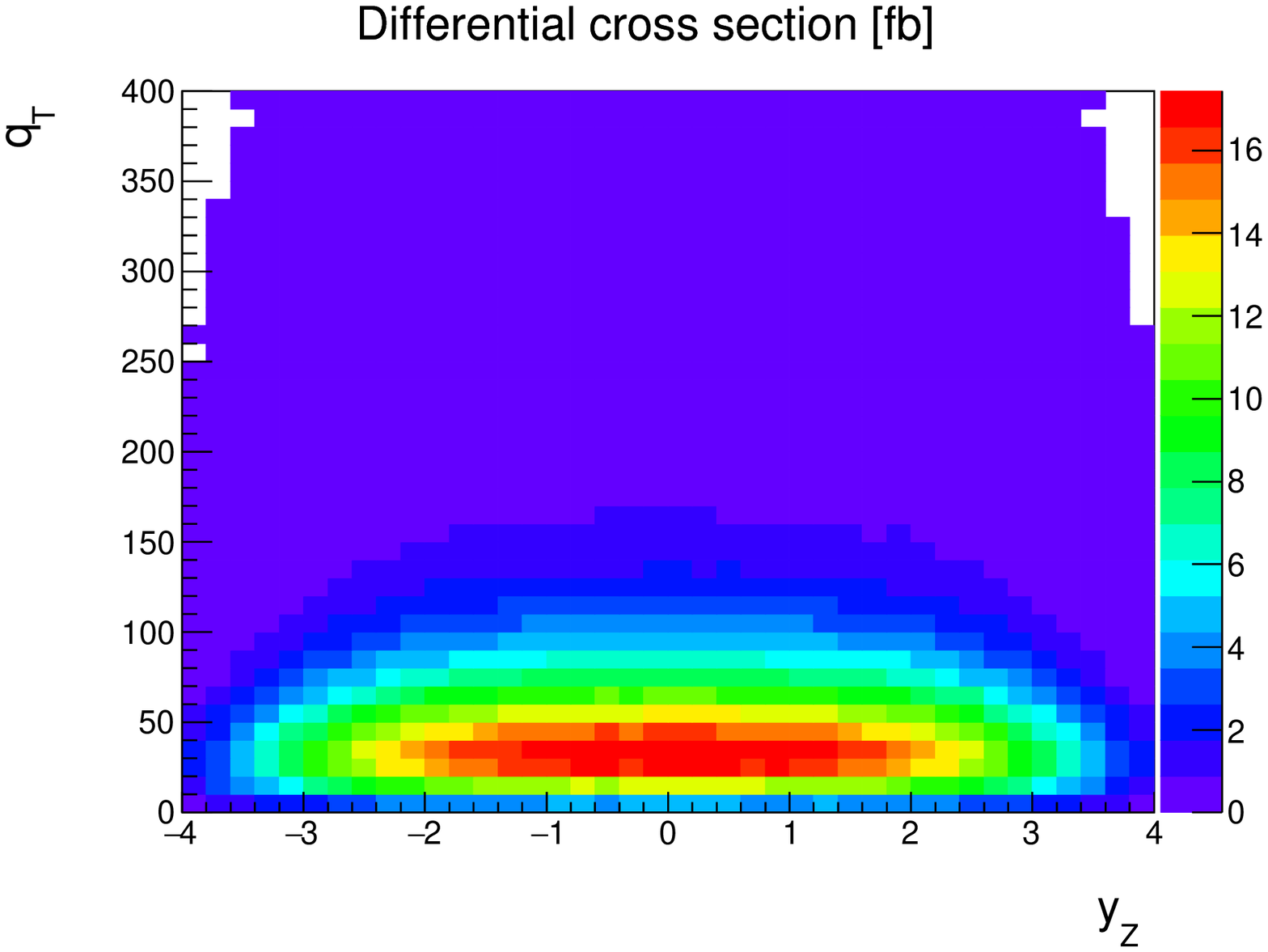}
  \caption{\label{fig:angzz} Angular coefficients $A_0-A_4$ and the $\yz-\qt$ differential cross section of the SM $\mathrm{Z Z}\to 2l2\nu$ process. }
\end{figure}

The SM $\mathrm{Z Z}\to 2l2\nu$ production is the major background of our DM search.
It has a similar final state signature as the signal process, as depicted in Fig.~\ref{fig:feynzz}.
Hence we first take a look at the Fig.~\ref{fig:angzz} for the angular coefficients of this process.
In general, the angular coefficient $A_0$ measures the difference between longitudinal and transverse polarizations, and it looks more longitudinal at high $\qt$.
The coefficient $A_4$ measures forward-backward asymmetry, the Z boson looks more like left-handed in the forward region.
The $A_2$ measures the interference between the transverse amplitudes and the $A_{1,3}$ measures the interference between transverse and longitudinal.

\clearpage

\subsection{Spin-0 mediator}

\begin{figure}[!h]
  \centering
  \includegraphics[width=5.0cm]{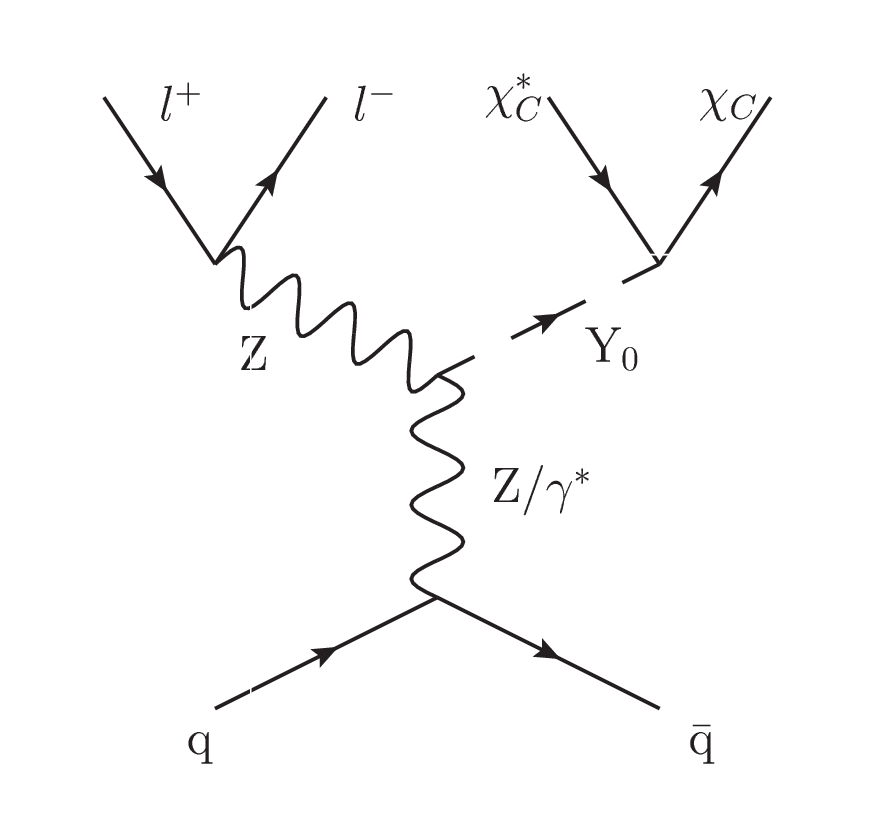}
  \caption{\label{fig:feyns0} Representative Feynman diagrams of the dark sector with a spin-0 mediator. 
           For the S0$_c$ model, there is no virtual photon propagator. }
\end{figure}

We consider a simplified model with a scalar s-channel mediator as described in Ref.~\cite{Neubert:2015fka}.
The dark sector model is constructed as follows:

\begin{eqnarray}
{\cal L}_{SM EW}^{Y_0} &=& \dfrac{1}{\Lambda} g^S_{h3} (D^{\mu}\phi)^{\dagger} (D_{\mu}\phi) Y_0 \\
                       & & + \dfrac{1}{\Lambda} B_{\mu\nu} \left( g^S_{B} B^{\mu\nu} + g^P_B \tilde{B}^{\mu\nu} \right)  Y_0 
                           + \dfrac{1}{\Lambda} W_{\mu\nu}^i \left( g^S_W W^{i,\mu\nu} + g^P_W \tilde{W}^{i,\mu\nu} \right) Y_0, \\
{\cal L}_{X}^{Y_0} &=&  \mathrm{m}_{\chi_C} g^S_{X_C} \chi^*_C \chi_C Y_0 + \bar{\chi}_D ( g^S_{X_D} + i g^P_{X_D} \gamma_5 ) \chi_{D} Y_0,
\end{eqnarray}

where $\tilde{V}^{\mu\nu}=\frac{1}{2}\epsilon_{\mu\nu\rho\sigma}V^{\rho\sigma}$ is the dual field strength tensor of $V$ field,
$\Lambda$ is a high energy scale. 
As discussed in Ref.~\cite{Neubert:2015fka}, this operator can be induced by a fermion loop graph with heavy fermion integrated out. 
Signature of this model is very different from the SM $\mathrm{Z Z}\to 2l2\nu$ process, 
the dark mediator is emitted from the SM gauge bosons as depicted in Fig.~\ref{fig:feyns0}.
We consider three benchmark scenarios of the parameters labeled by S0$_{a,b,c}$. 
As our angular distributions are more sensitive to changes in couplings, we fix
the mass of dark matter $\mchi=10$~GeV and the mass of the mediator $\mys=1000$~GeV. 
The angular distributions won't be changed drastically as long as $2\mchi$ is much smaller than $\mys$.
The parameter values and inclusive cross sections are listed in Table~\ref{tab:exhi-s0}. 

\begin{figure}[!h]
  \centering
  \includegraphics[width=4.5cm]{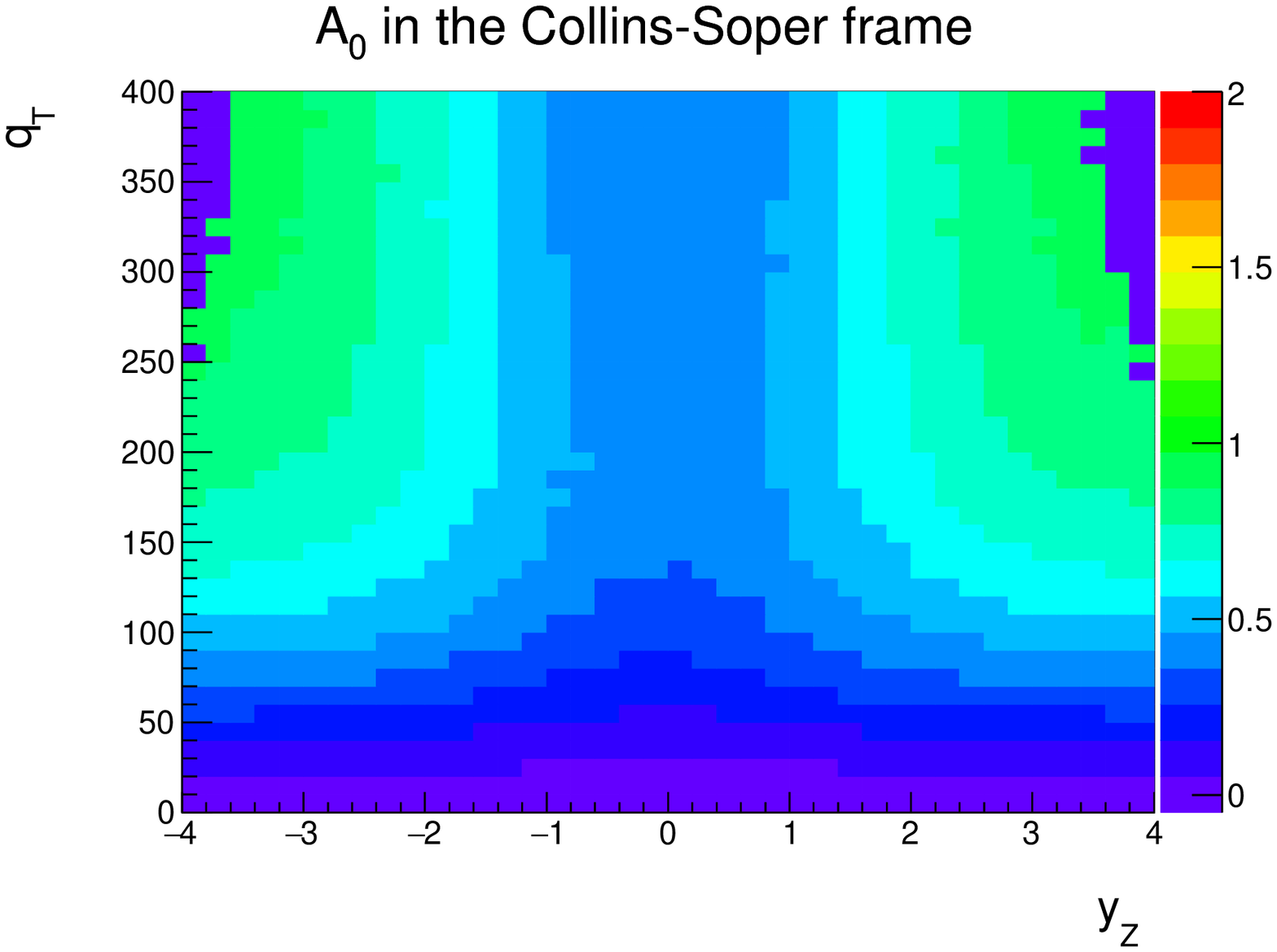}
  \includegraphics[width=4.5cm]{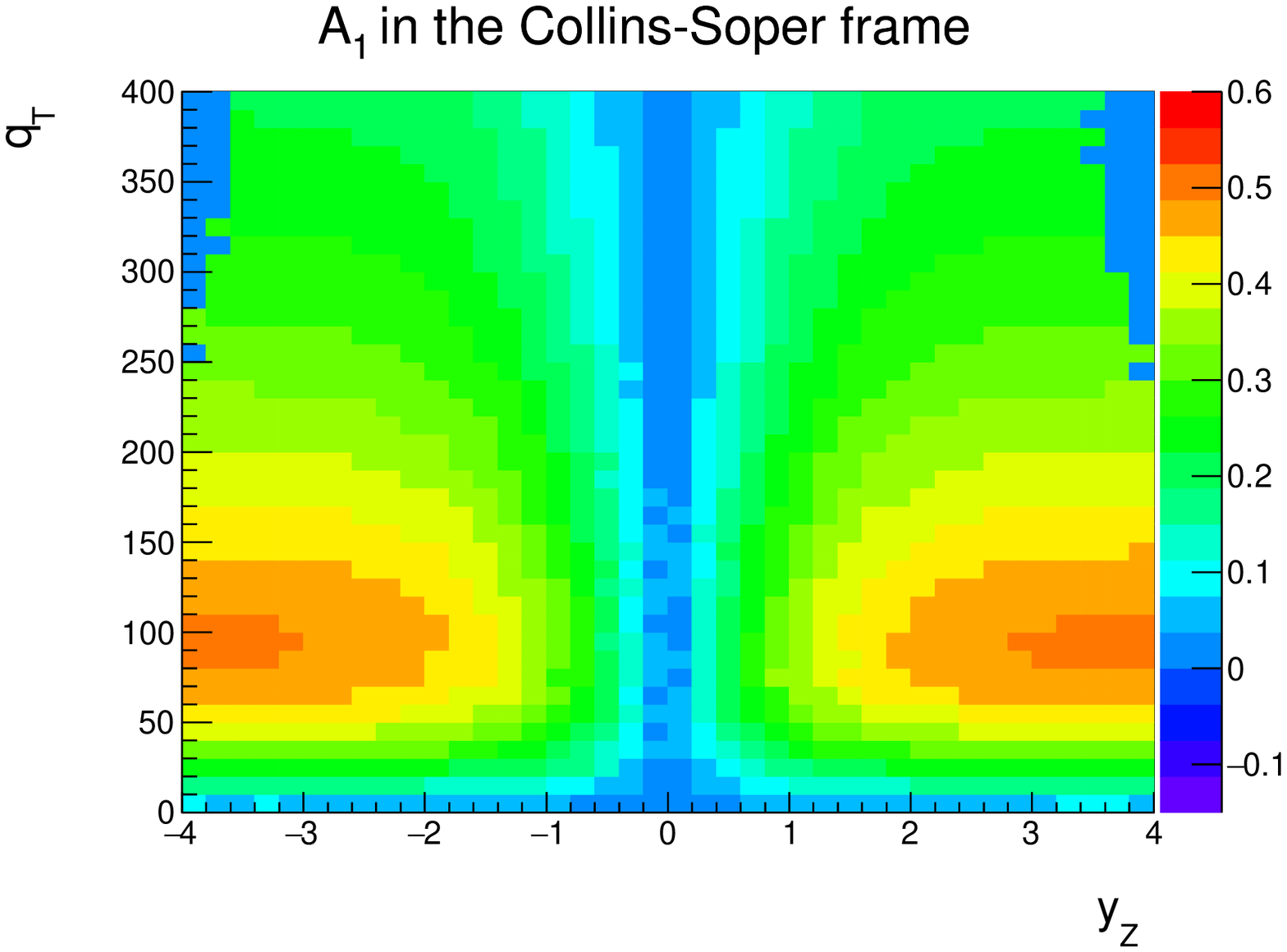} 
  \includegraphics[width=4.5cm]{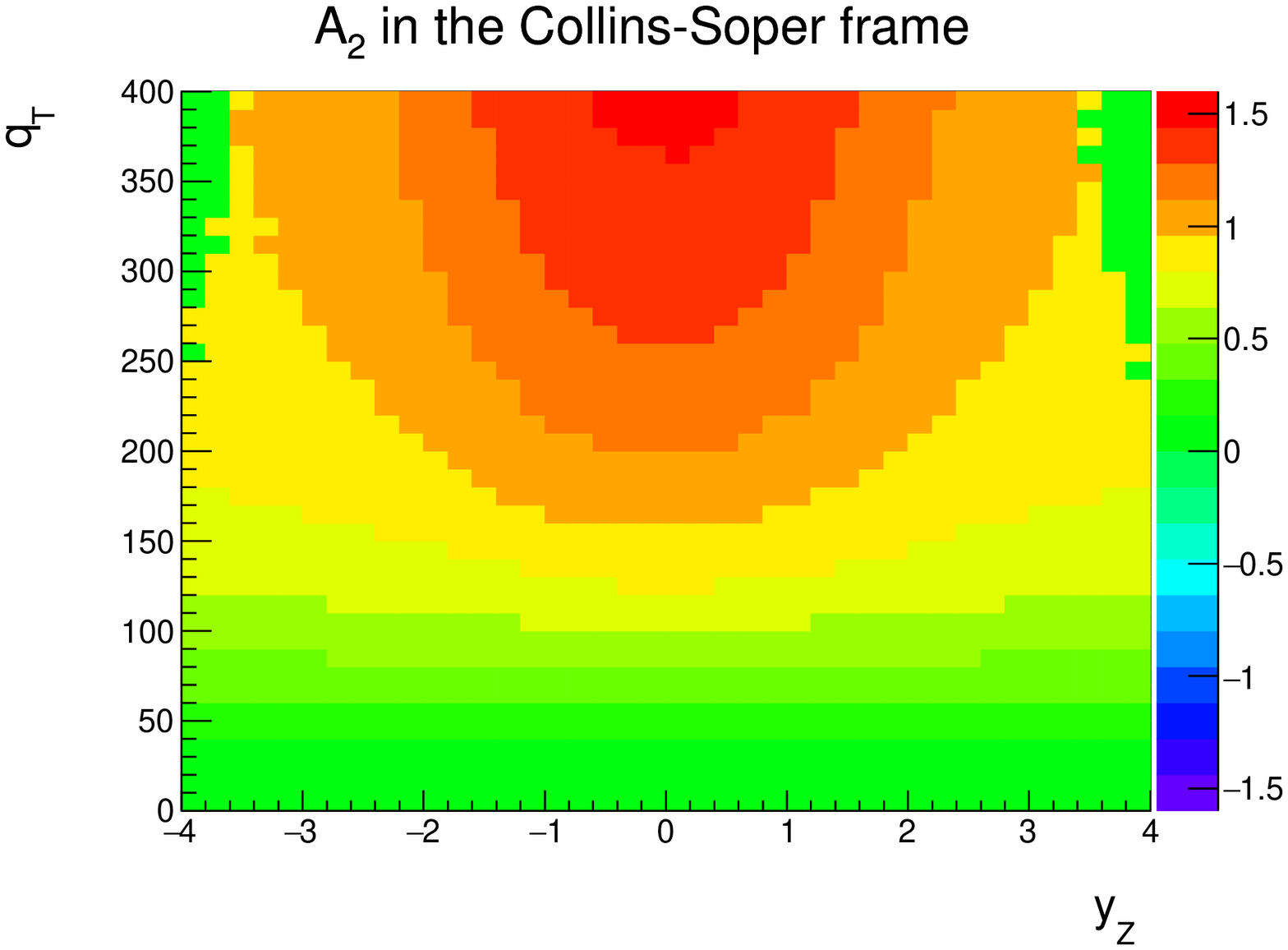} \\
  \includegraphics[width=4.5cm]{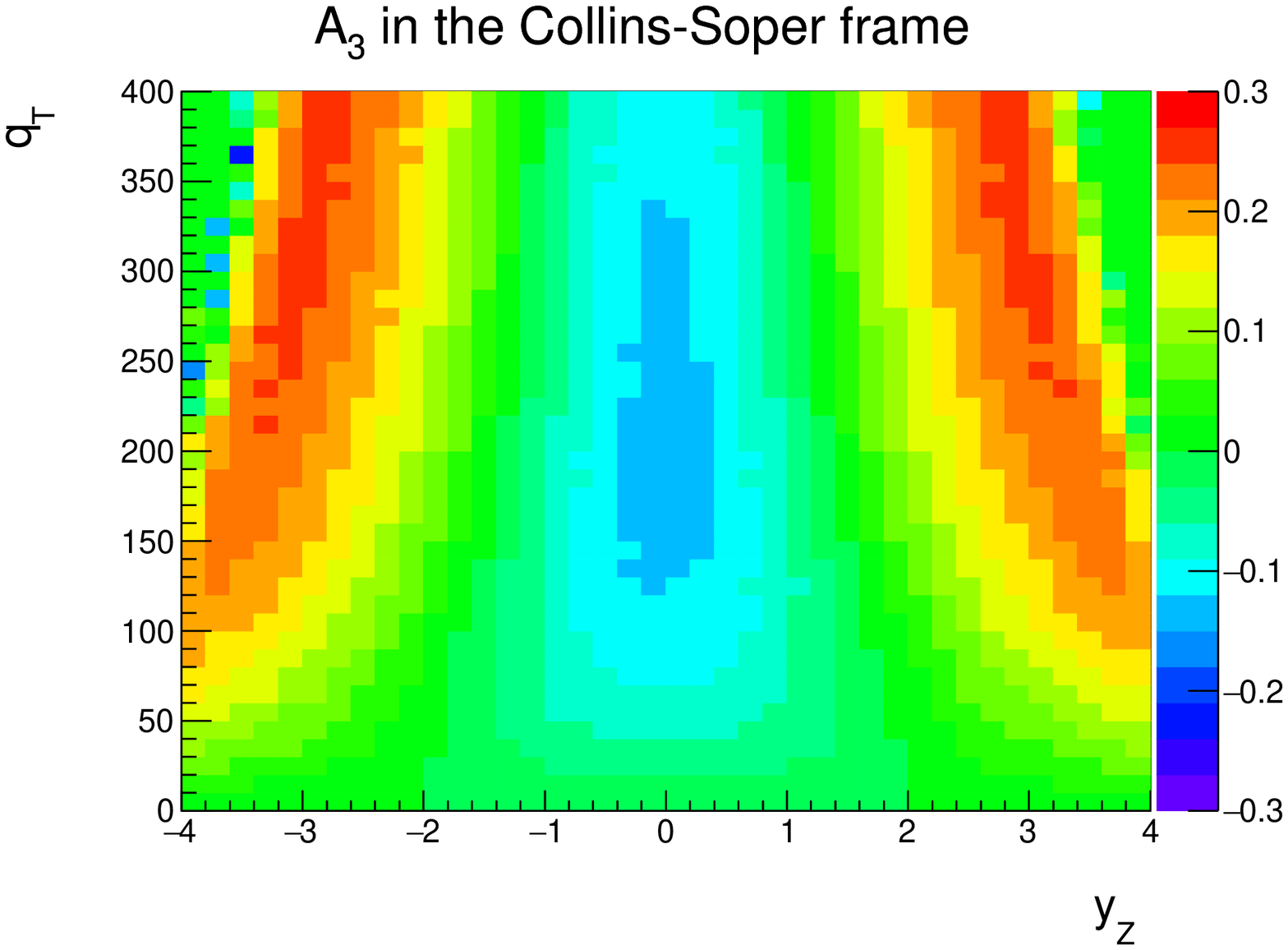} 
  \includegraphics[width=4.5cm]{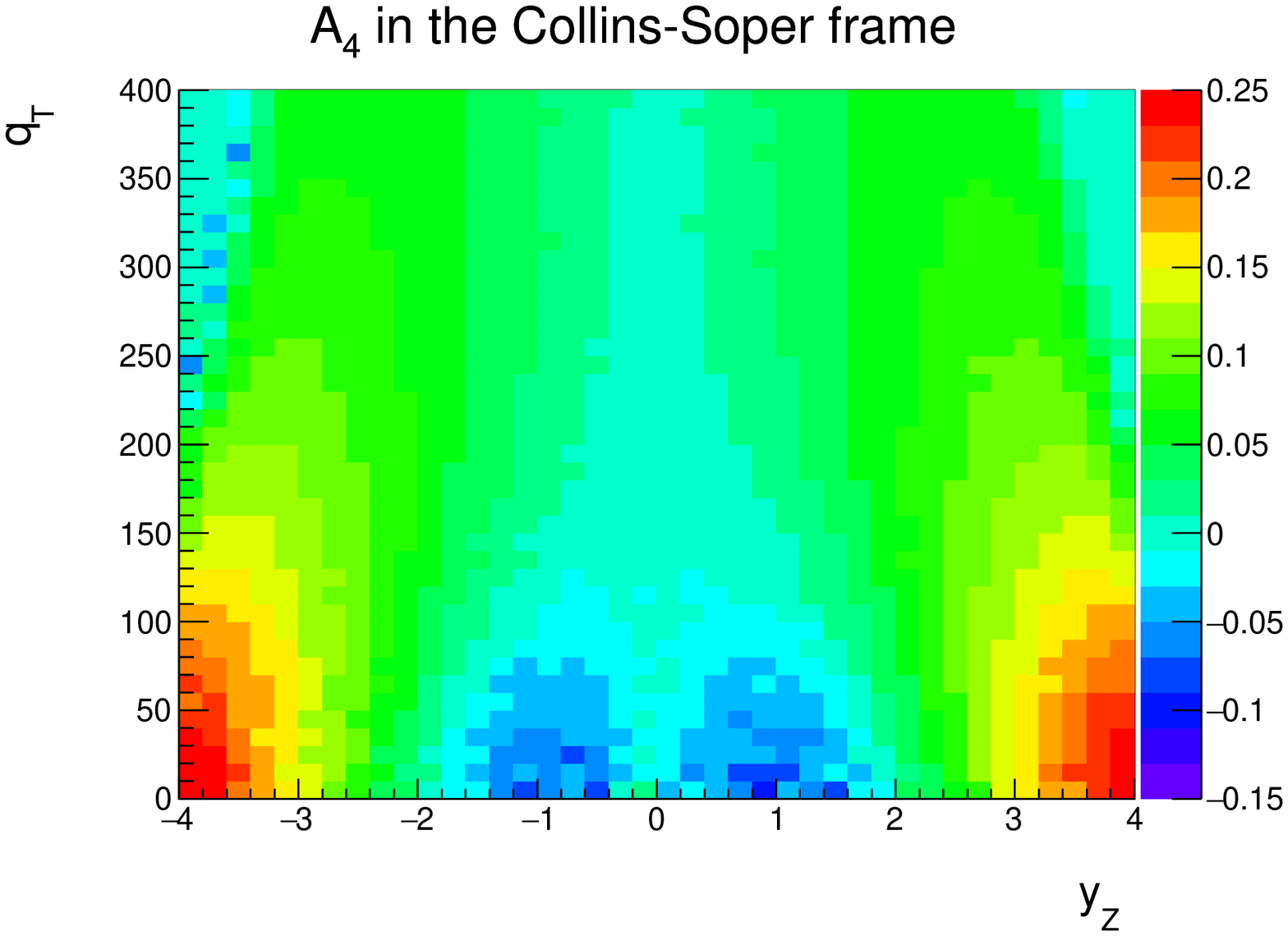}
  \includegraphics[width=4.5cm]{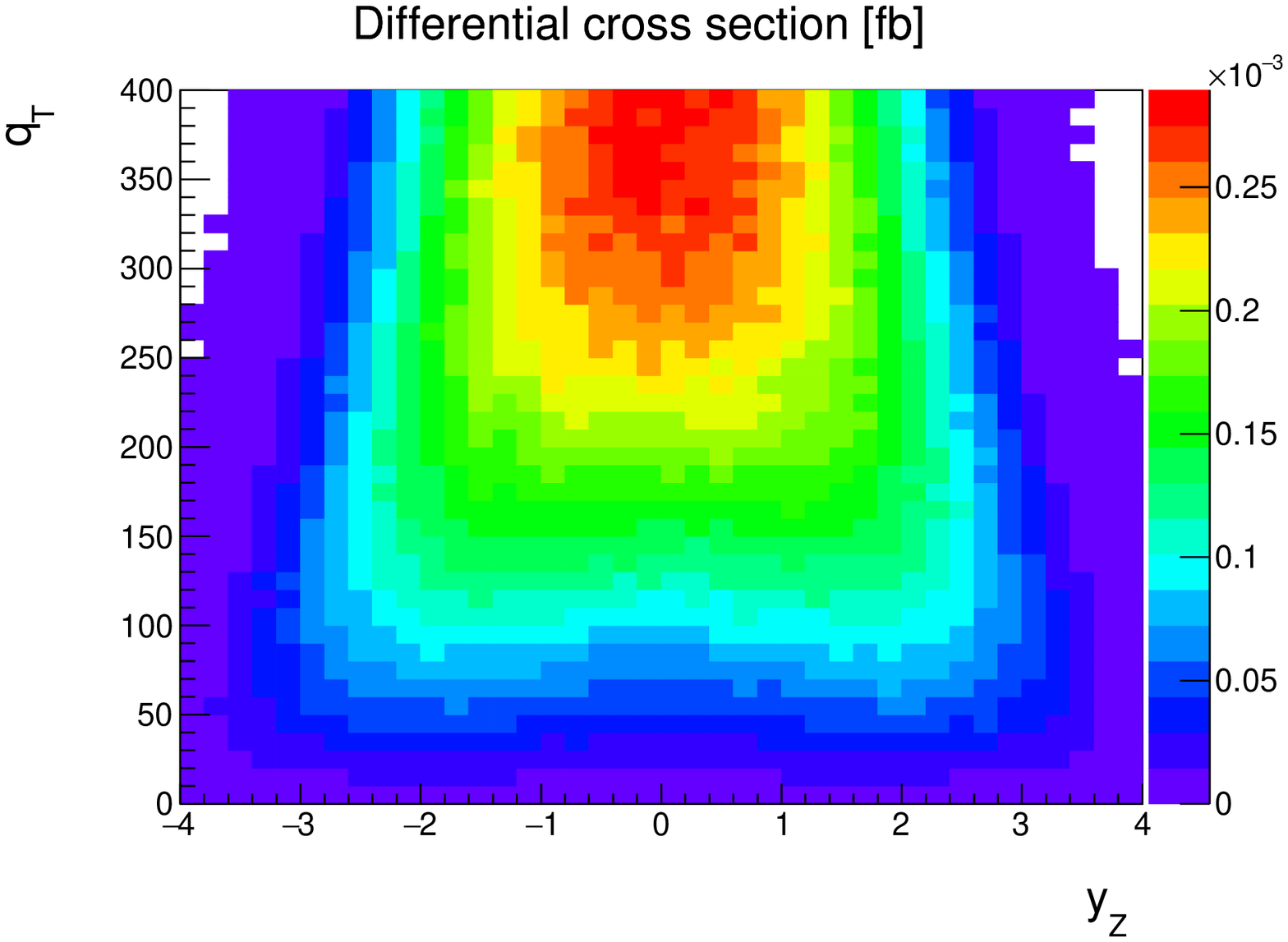} 
  \caption{\label{fig:angs0a} Angular coefficients $A_0-A_4$ and the $\yz-\qt$ differential cross section of the benchmark scenario S0$_a$. }
\end{figure}

\begin{figure}[!h]
  \centering
  \includegraphics[width=4.5cm]{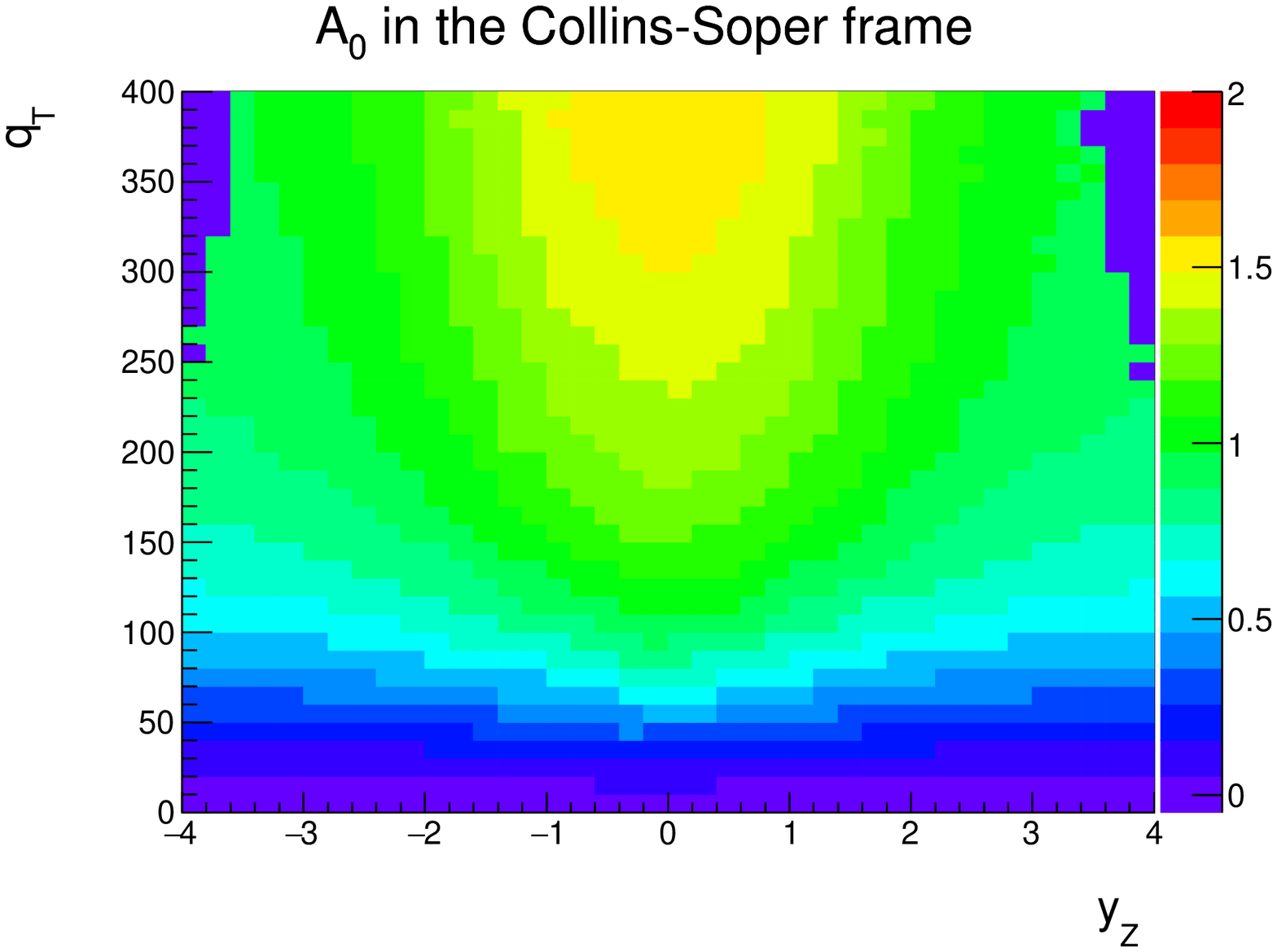}
  \includegraphics[width=4.5cm]{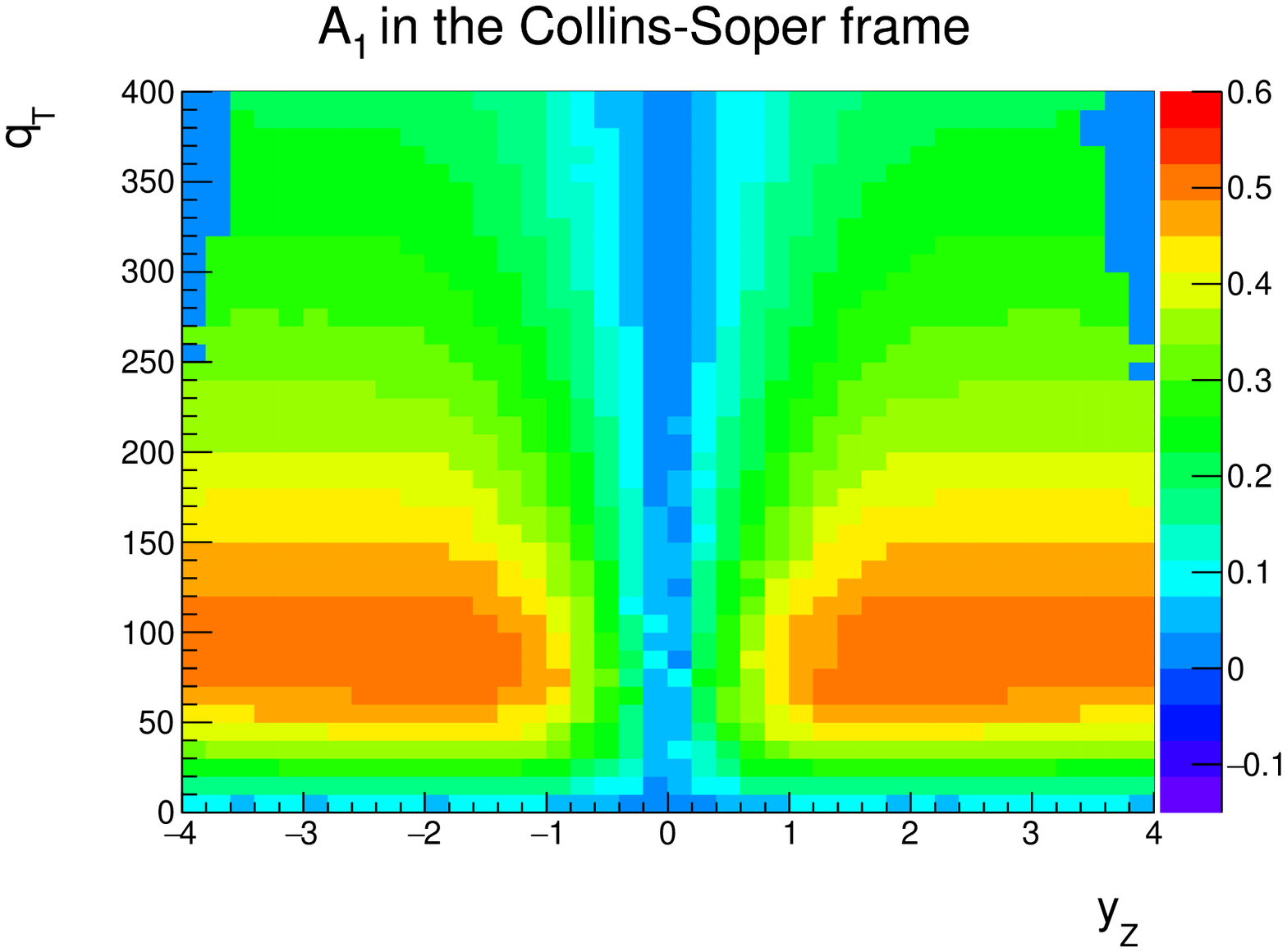} 
  \includegraphics[width=4.5cm]{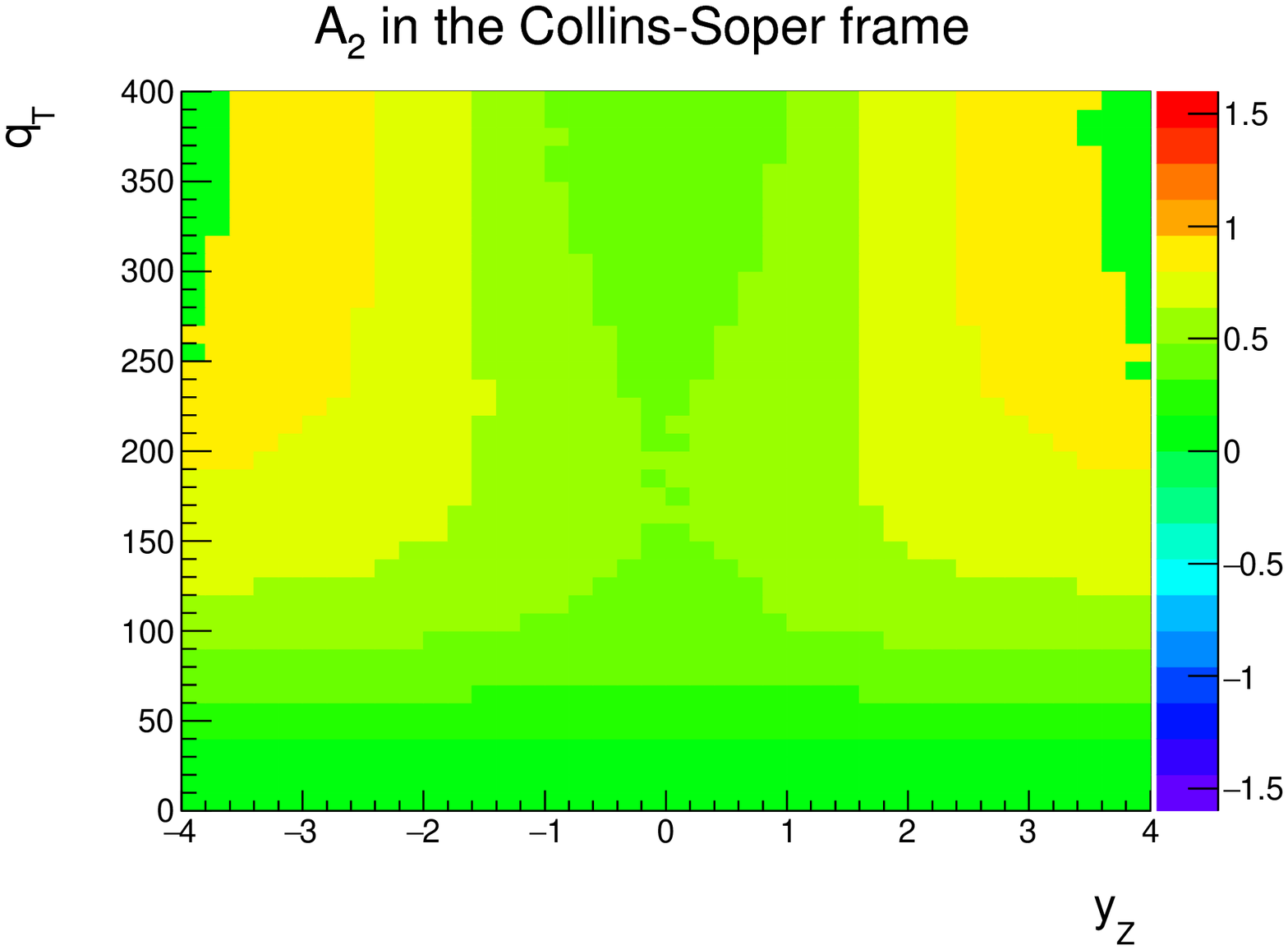} \\
  \includegraphics[width=4.5cm]{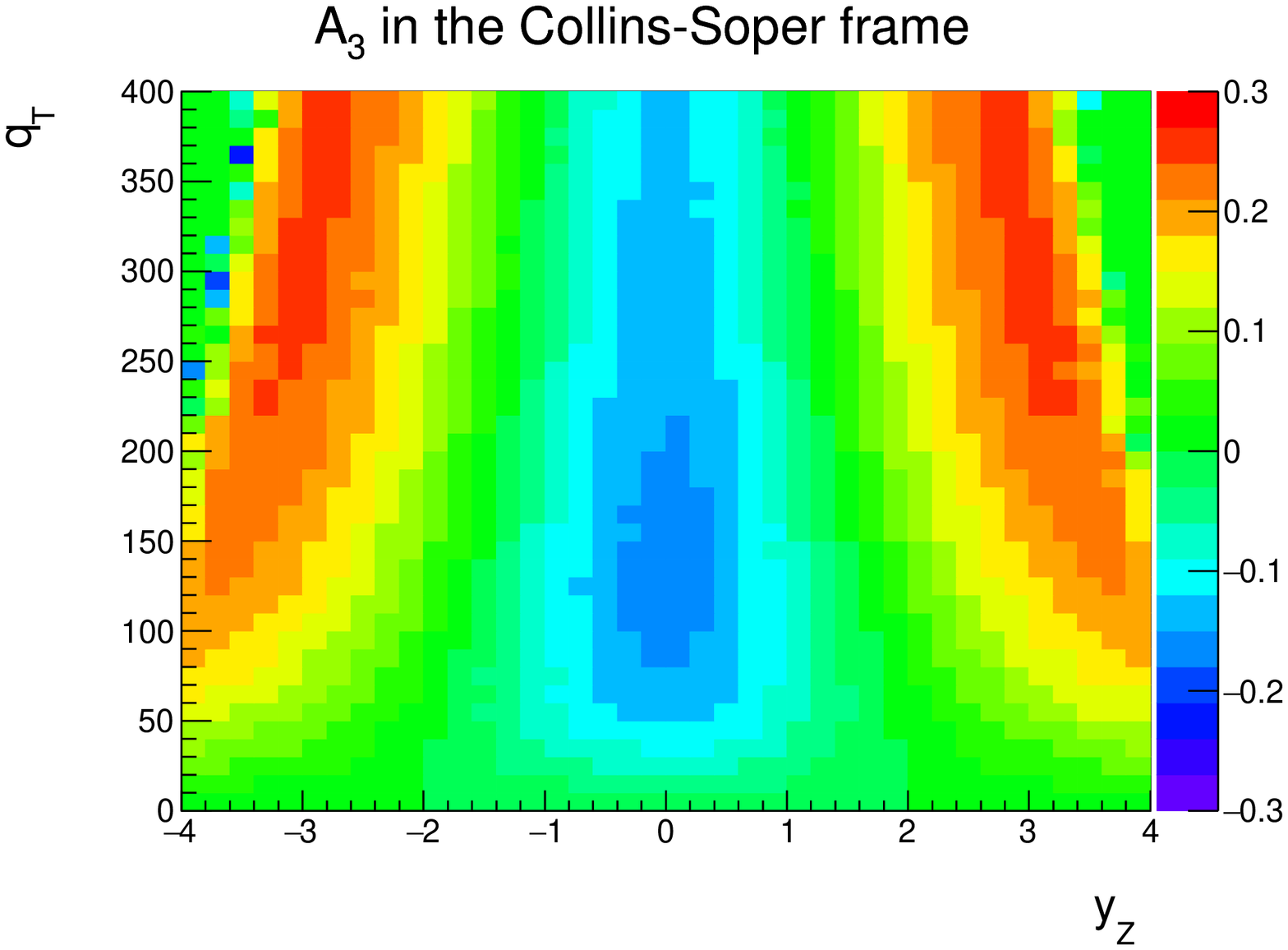} 
  \includegraphics[width=4.5cm]{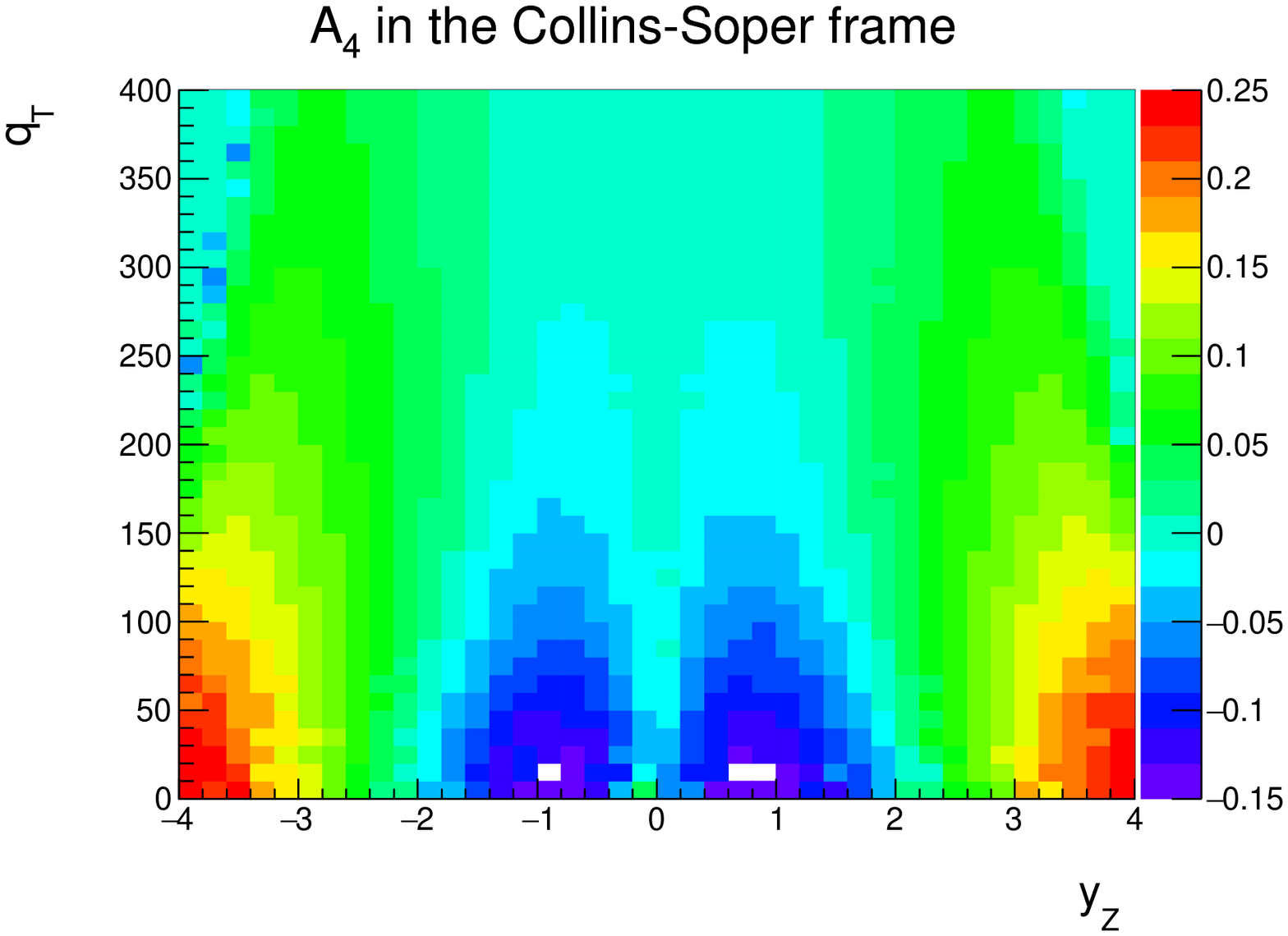}
  \includegraphics[width=4.5cm]{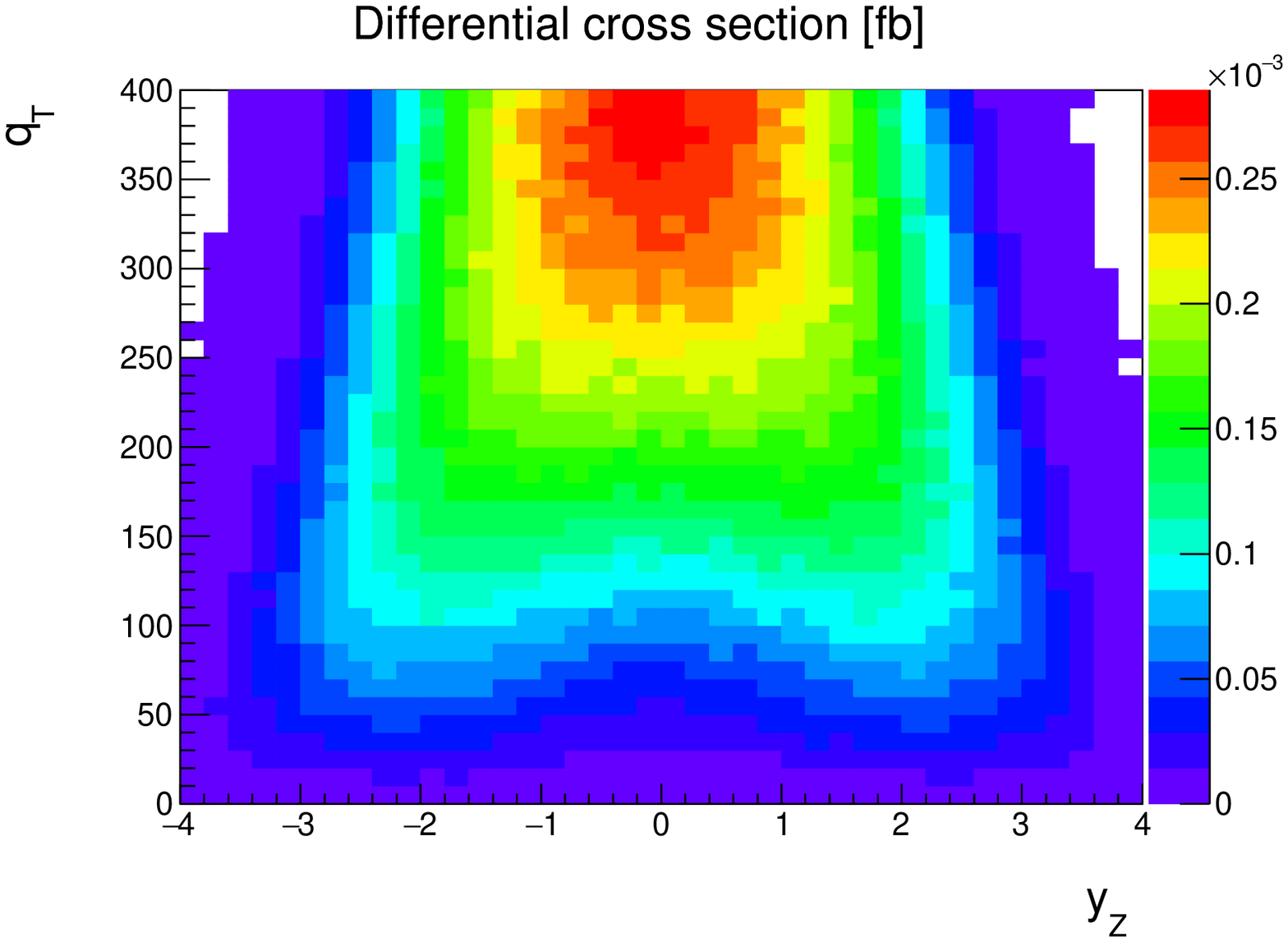}
  \caption{\label{fig:angs0b} Angular coefficients $A_0-A_4$ and the $\yz-\qt$ differential cross section of the benckmark scenario S0$_b$. }
\end{figure}

\begin{figure}[!h]
  \centering
  \includegraphics[width=4.5cm]{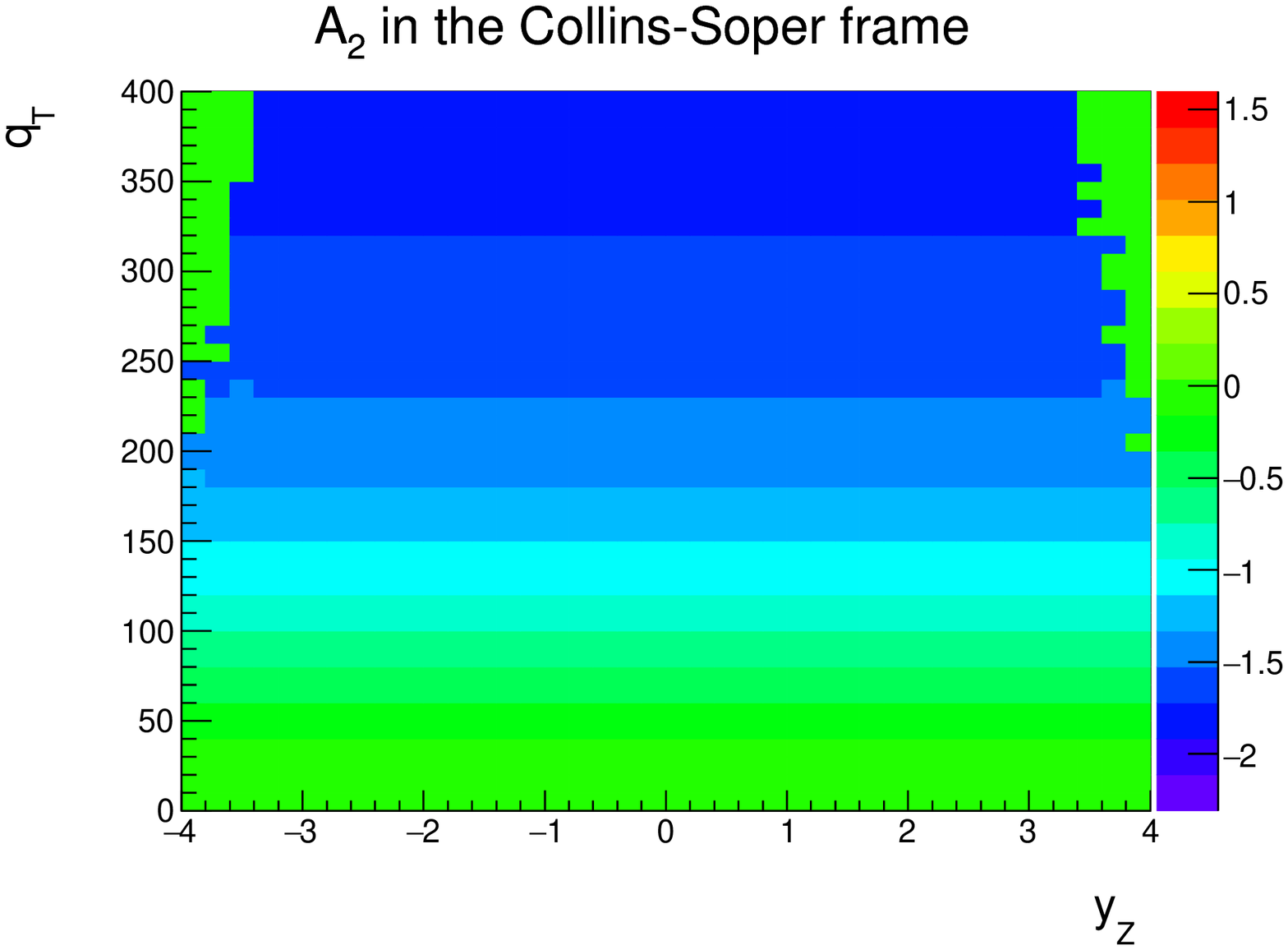} 
  \includegraphics[width=4.5cm]{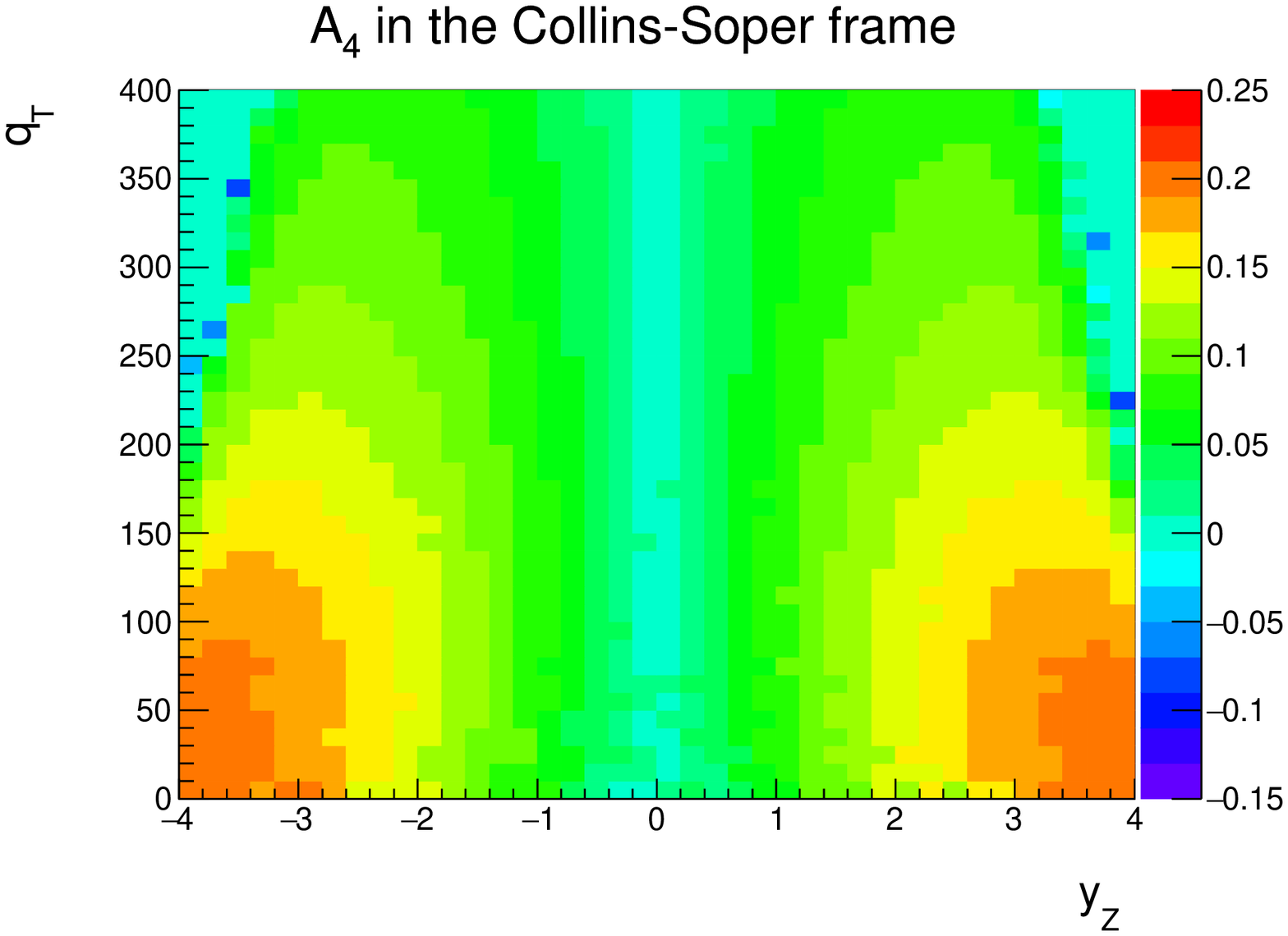}
  \includegraphics[width=4.5cm]{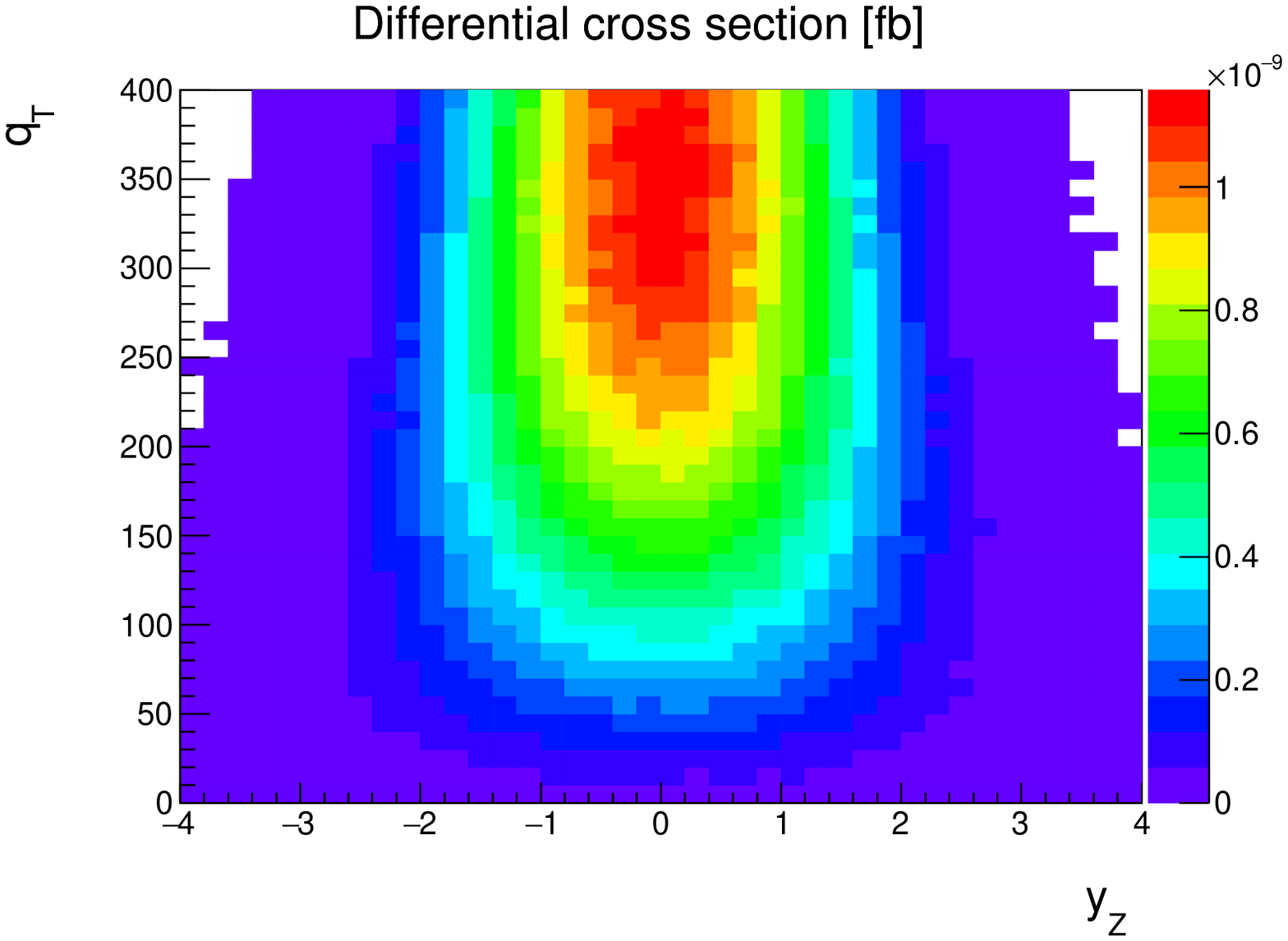}
  \caption{\label{fig:angs0c} Angular coefficients $A_0-A_4$ and the $\yz-\qt$ differential cross section of the benchmark scenario S0$_c$. 
            Comparing with other figures, we extended the range of the $A_2$ for a better demonstration. }
\end{figure}

Angular coefficients of the benchmark scenatios S0$_{a,b,c}$ are shown in Fig.~\ref{fig:angs0a}, Fig.~\ref{fig:angs0b} and Fig.\ref{fig:angs0c} respectively. 
Comparing to the SM $\mathrm{Z Z}\to 2l2\nu$, the dark matter signal is produced with much higher $\qt$ and have very different angular coefficients distributions, e.g., more transverse at low $\qt$.  
The S0$_a$ and S0$_b$ can be distinguished from $A_0,A_2$, where the $\yz$ dependences are very different.  
In the case of S0$_c$, $Y_0$ couples to weak bosons like a Higgs boson and cannot perturb the coupling structure with the Z boson production. 
Consequently, the $A_0$, $A_1$ and $A_3$ in the CS frame are all zero hence are not shown in the figure.

%\clearpage

\begin{table}[htb]
\centering
    \begin{tabular}{c|cccccccc}
      \hline
      \hline
  Benchmark         &       S0$_a$       &       S0$_b$     &       S0$_c$        \\ \hline
  $g^S_{X_D}$       &        1           &        0         &         0           \\
  $g^P_{X_D}$       &        0           &        1         &         0           \\
  $g^S_{X_C}$       &        0           &        0         &         1           \\
  \hline
  $g^S_{W}$         &       0.25         &        0         &         0           \\
  $g^P_{W}$         &       0            &       0.25       &         0           \\
  $g^S_{h3}$        &       0            &        0         &         1           \\
  $\Lambda$ (GeV)   &       3000         &       3000       &        3000         \\
  \hline
  Interaction       &     CP-even        &     CP-odd       &       CP-even       \\
  $\mchi$ (GeV)       &      10            &      10          &        10           \\
  $\mys$ (GeV)      &      1000          &      1000        &        1000         \\
  \hline
$\Gamma_{Y_0}$ (GeV)&      41.4          &      41.4        &        1.05         \\
Cross section (fb)  &      0.0103        &     0.00977      &       2.98e-08      \\
      \hline
      \hline
    \end{tabular}
  \caption{ Benchmark scenarios with a spin-0 mediator. }
  \label{tab:exhi-s0}
\end{table}

\subsection{Spin-1 mediator}

We consider the same dark sector with a spin-1 mediator as in the LHC experiment~\cite{Abercrombie:2015wmb} with the following interactions of the dark sector:

\begin{eqnarray}
{\cal L}_{X_D}^{Y_1} &=& \bar{\chi}_D \gamma_{\mu} \left( g^V_{X_D} + g^A_{X_D} \gamma_5 \right) \chi_D Y_1^{\mu} \\ \nonumber
{\cal L}_{SM}^{Y_1} &=& \bar{d}_i \left( g^V_{d_{ij}} + g^A_{d_{ij}}\gamma_5 \right) d_j Y_1^{\mu} + 
                        \bar{u}_i \left( g^V_{u_{ij}} + g^A_{u_{ij}}\gamma_5 \right) u_j Y_1^{\mu}  
\end{eqnarray}

\begin{figure}[htb]
  \centering
  \includegraphics[width=10.0cm]{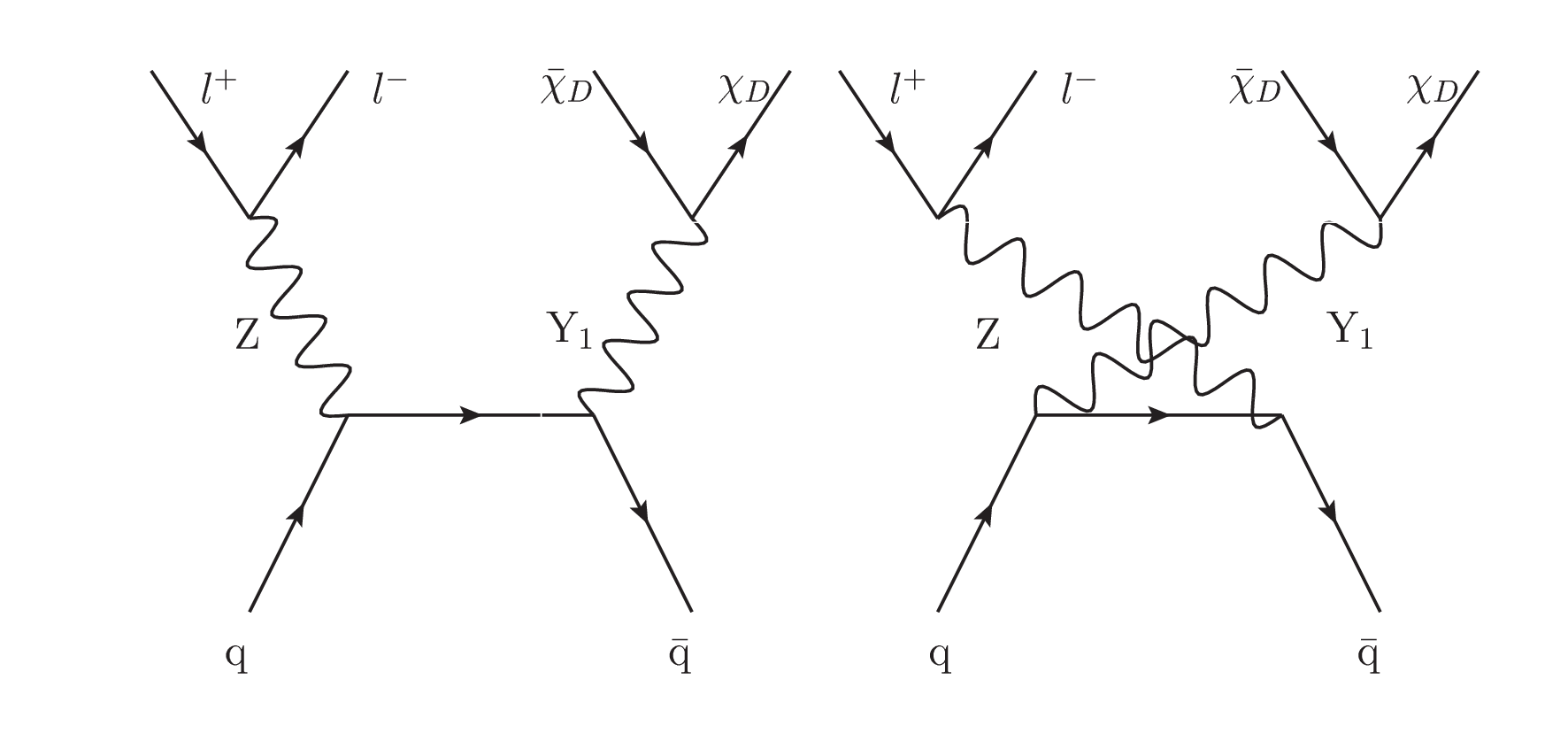}
  \caption{\label{fig:feyns1} Representative Feynman diagrams of the dark sector with a spin-1 mediator. }
\end{figure}

The masses of the dark matter and the mediator are chosen to be the same as in the spin-0 model.
A sound discussion of the impact of the choice of masses is available in the Ref.~\cite{Abercrombie:2015wmb}.
Since our analysis is more suitable for testing couplings, we consider benchmark scenarios as listed in Table~\ref{tab:exhi-s1}.
The signal signature is close to the SM $\mathrm{Z Z}\to 2l2\nu$ process, as shown in Fig.~\ref{fig:feyns1}, and we include here the
SM $\mathrm{Z Z}\to 2l2\nu$ as a special case with zero coupling for comparison.
The S1$_b$ and S1$_c$ project out the right- and left-handed part the Z-q-$\mathrm{\bar{q}}$ couplings.
Since the magnitude of the left-handed couplings are larger than the one of the right-handed,
cross section of the S1$_c$ scenario is found to be much larger than the S1$_b$ scenario.

\begin{figure}[htp]
  \centering
  \includegraphics[width=4.5cm]{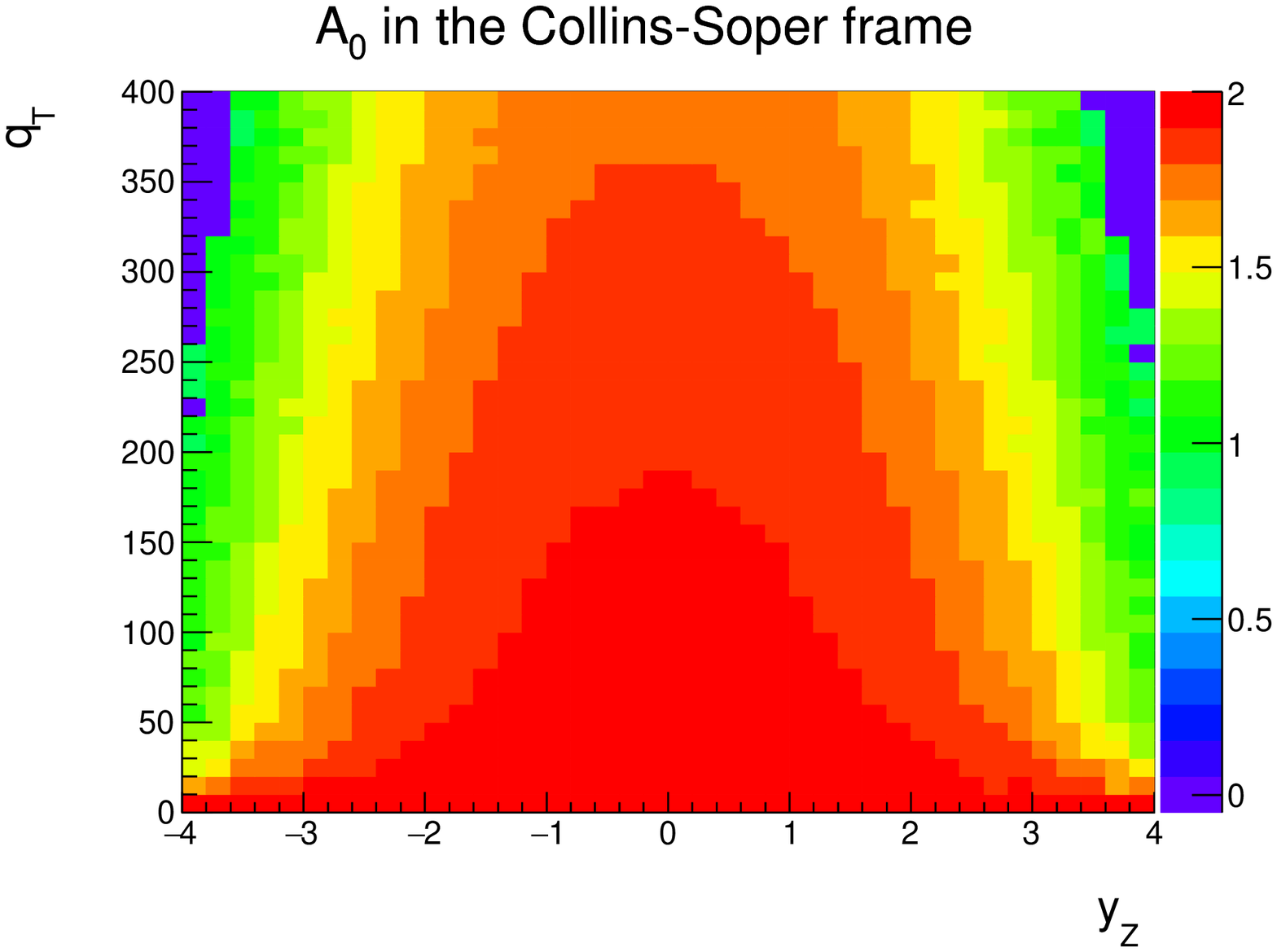}
  \includegraphics[width=4.5cm]{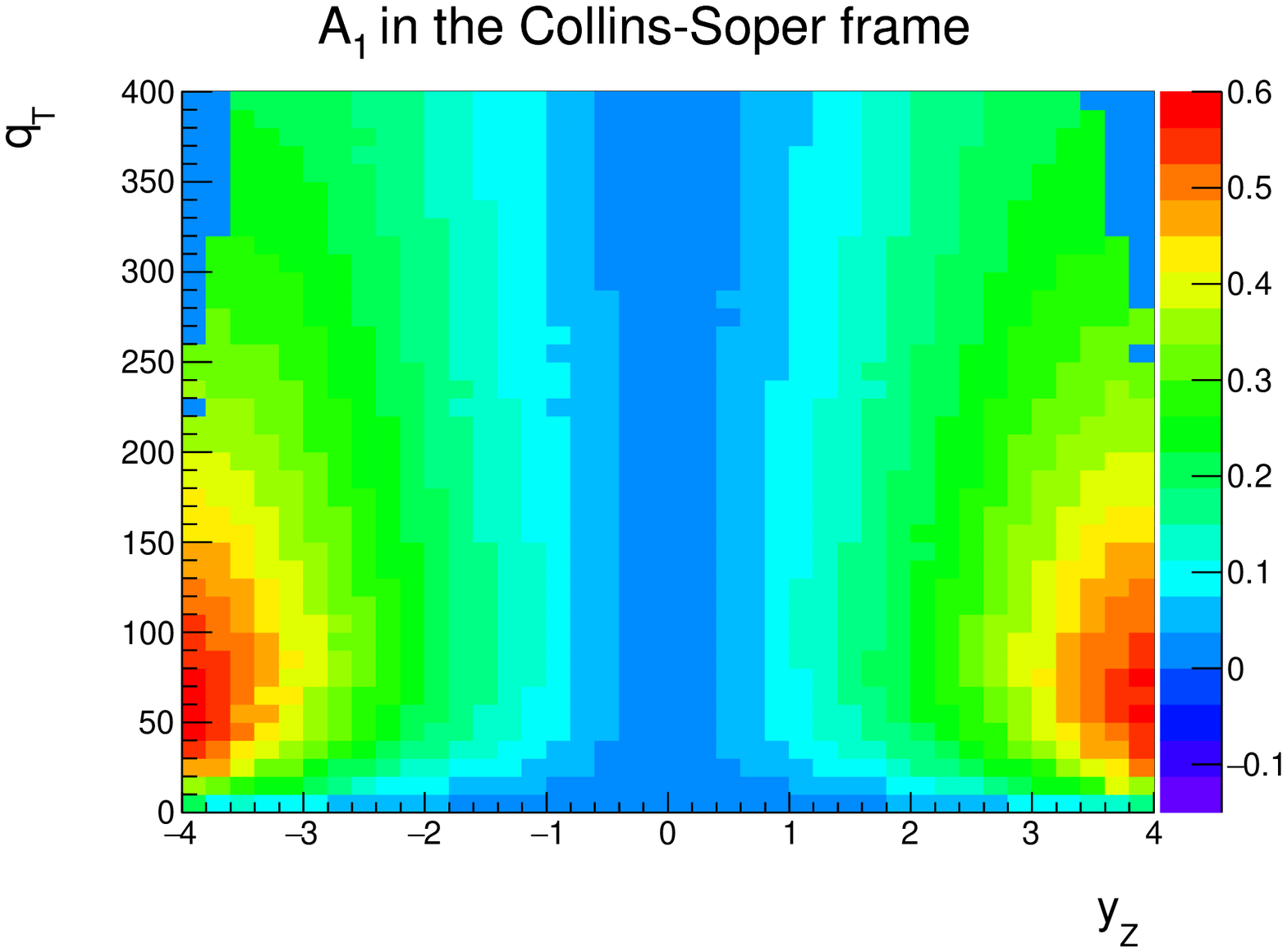} 
  \includegraphics[width=4.5cm]{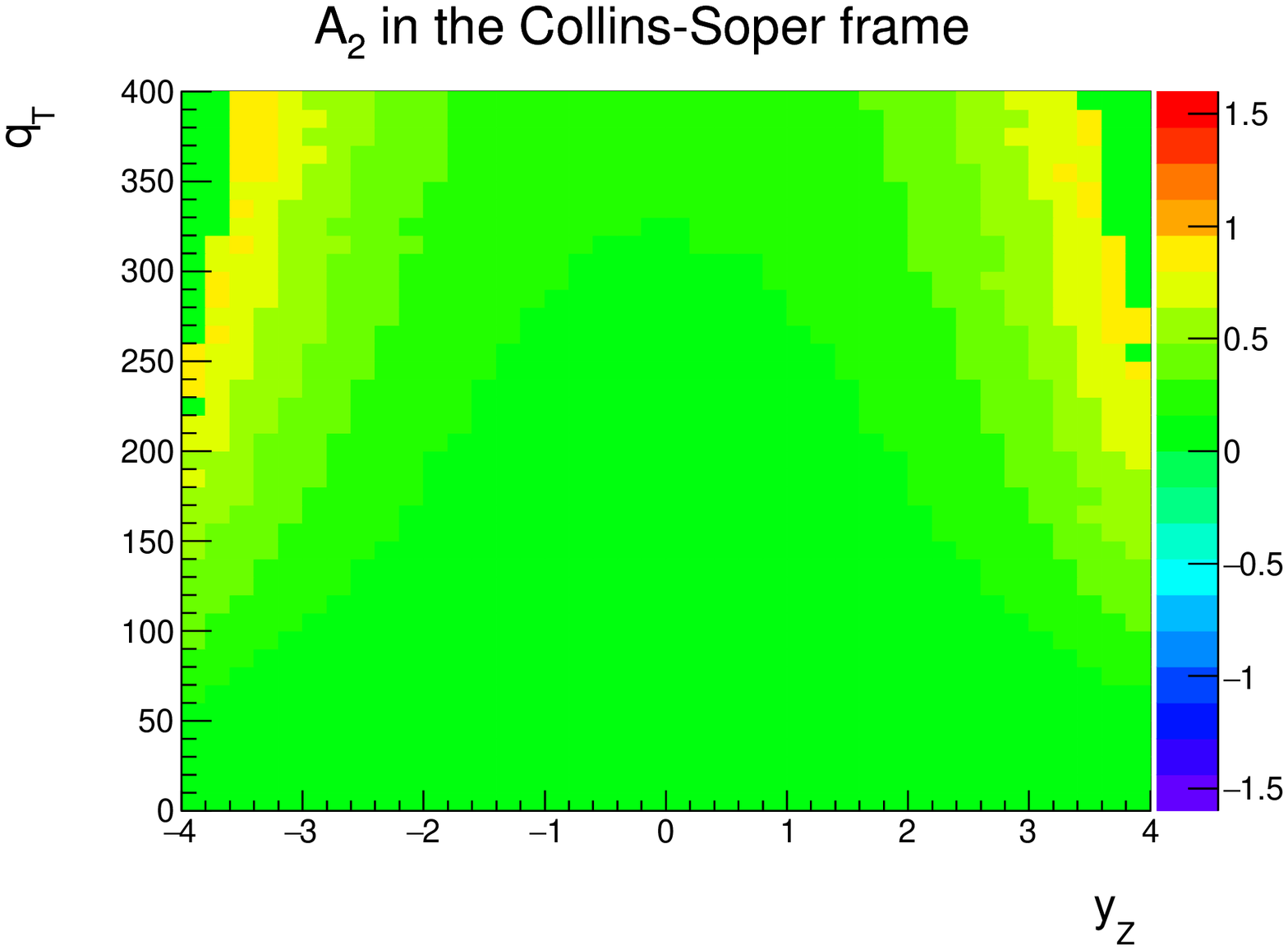} \\
  \includegraphics[width=4.5cm]{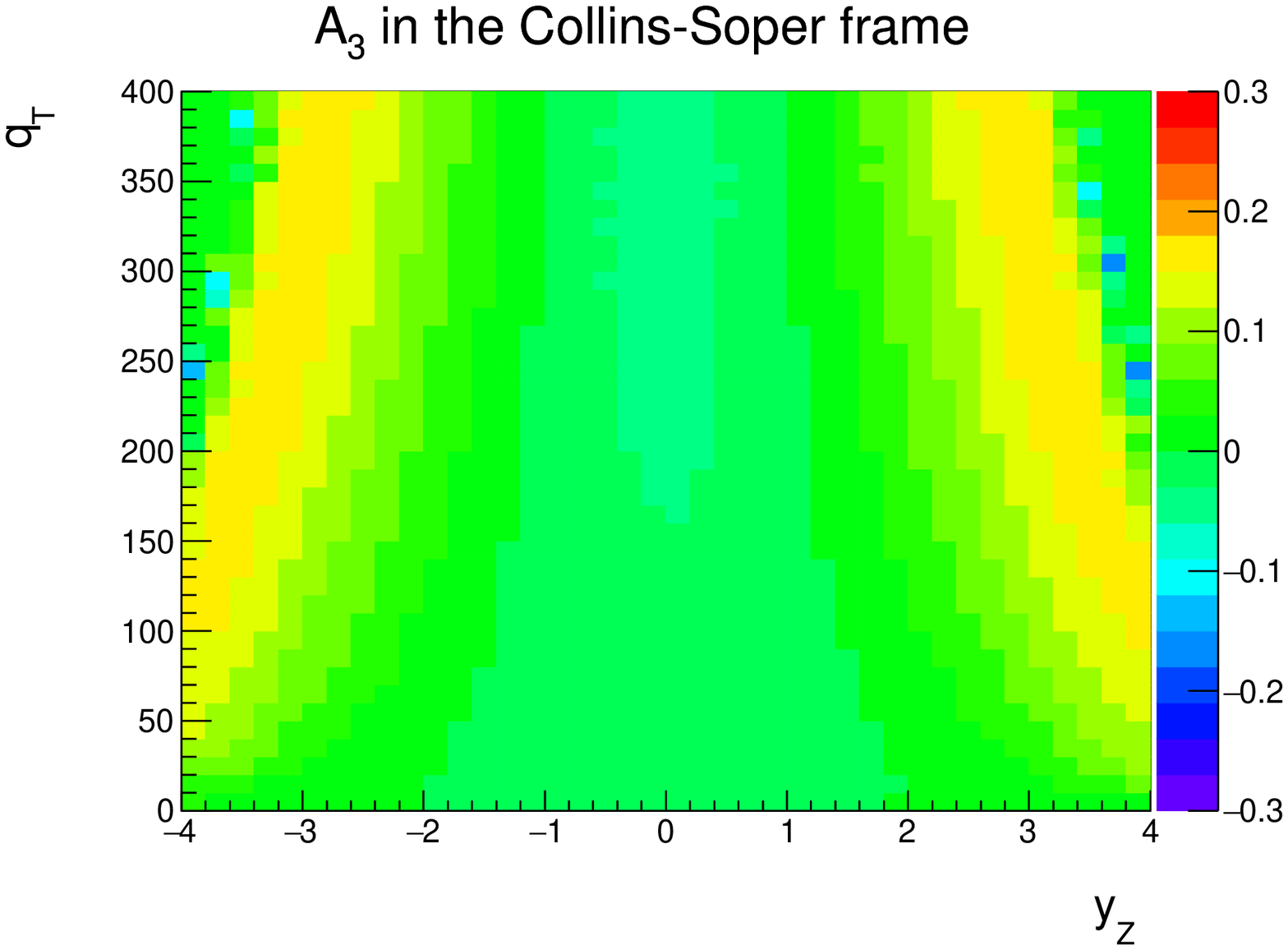} 
  \includegraphics[width=4.5cm]{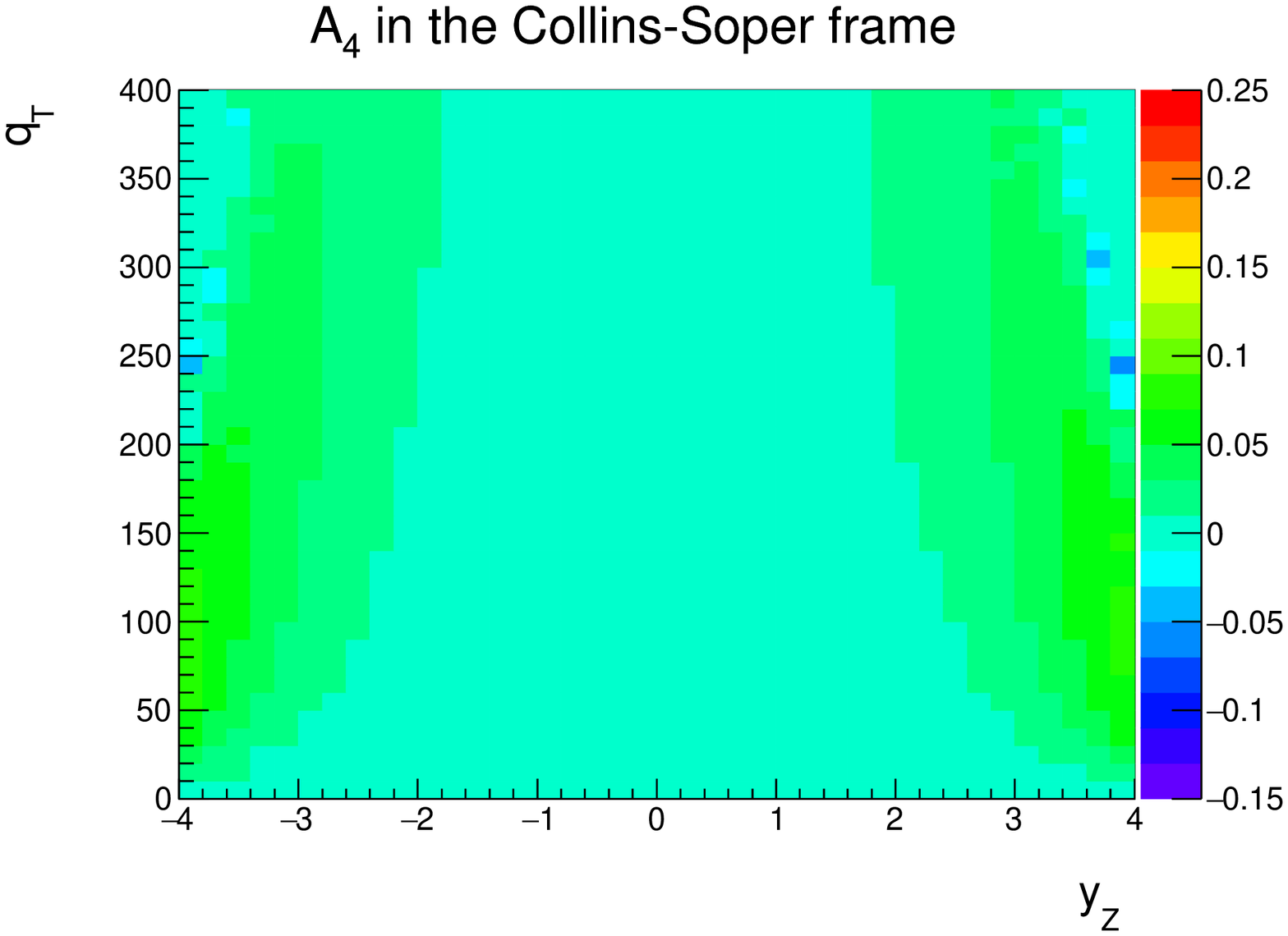}
  \includegraphics[width=4.5cm]{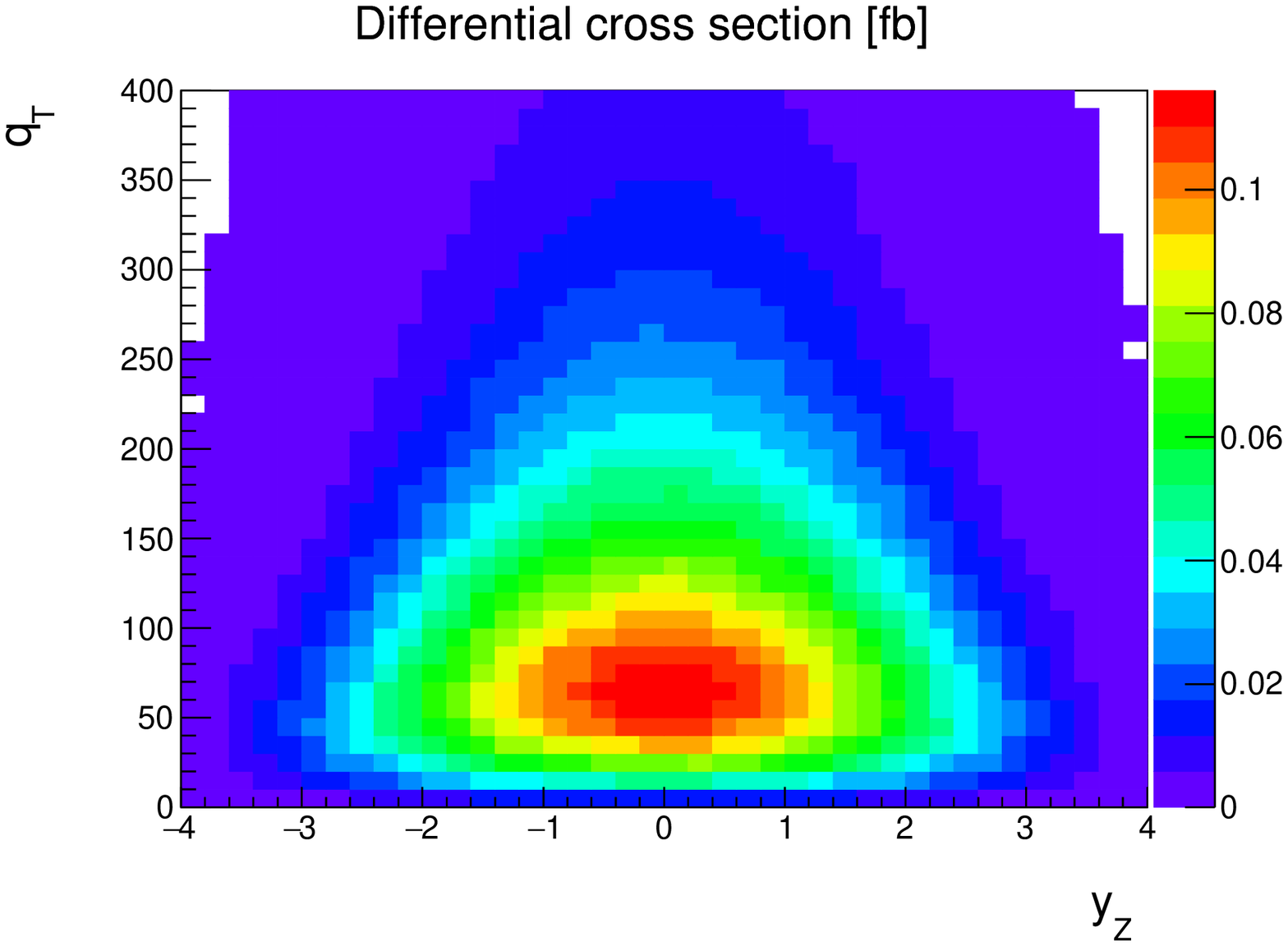}
  \caption{\label{fig:angs1a} Angular coefficients $A_0-A_4$ and the $\yz-\qt$ differential cross section of the benchmark scenario S1$_a$. }
\end{figure}

\begin{figure}[htp]
  \centering
  \includegraphics[width=4.5cm]{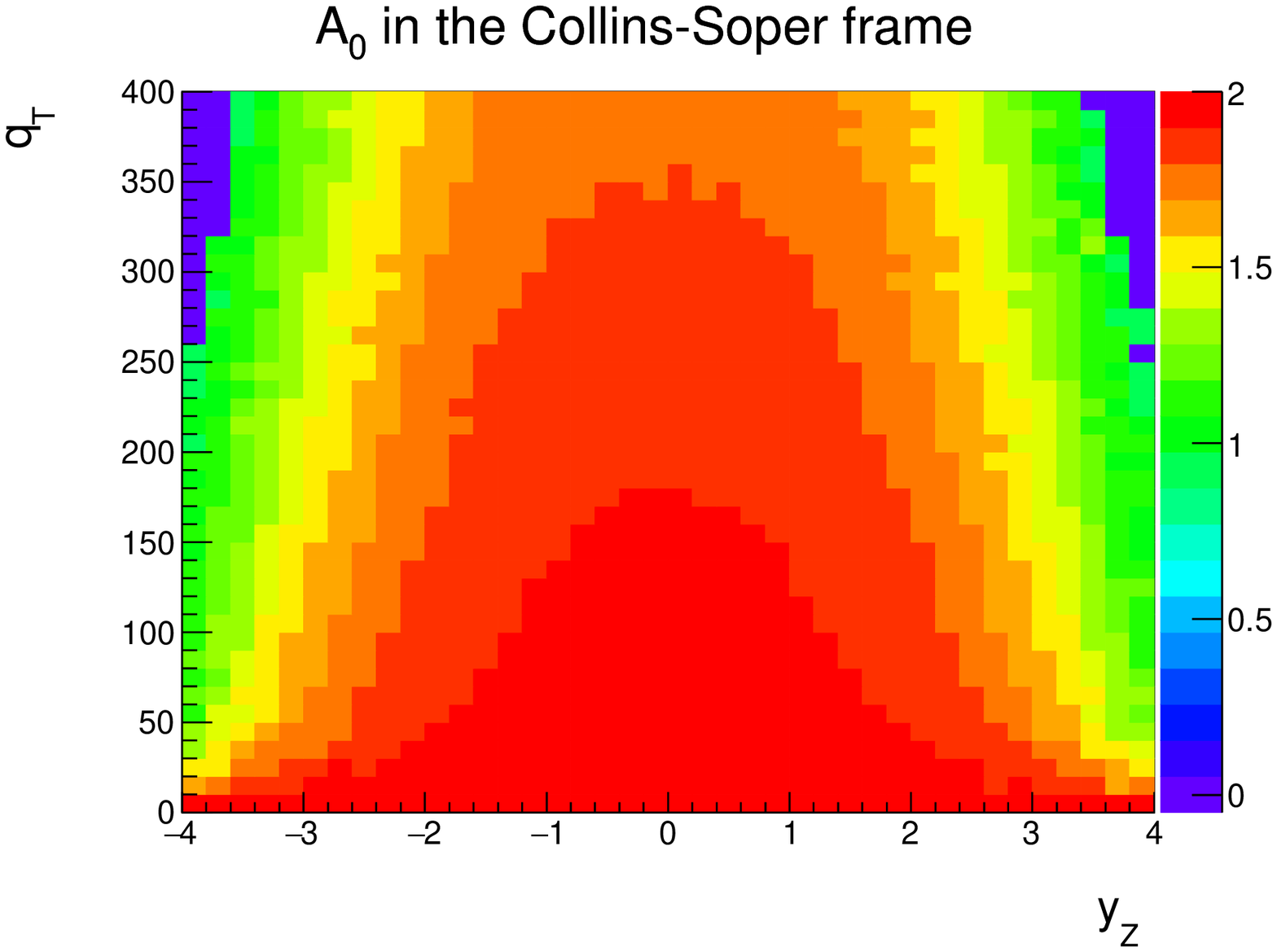}
  \includegraphics[width=4.5cm]{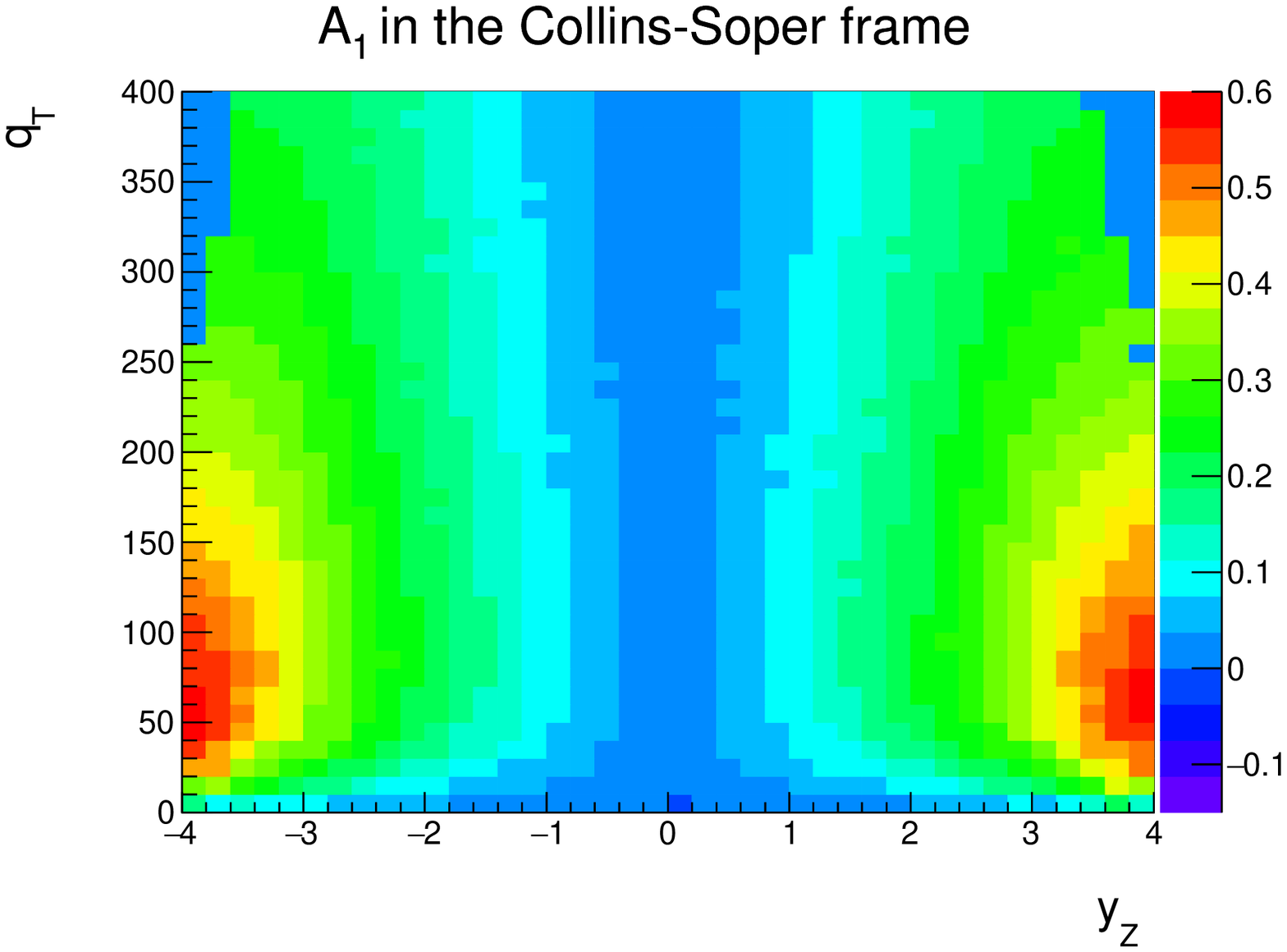} 
  \includegraphics[width=4.5cm]{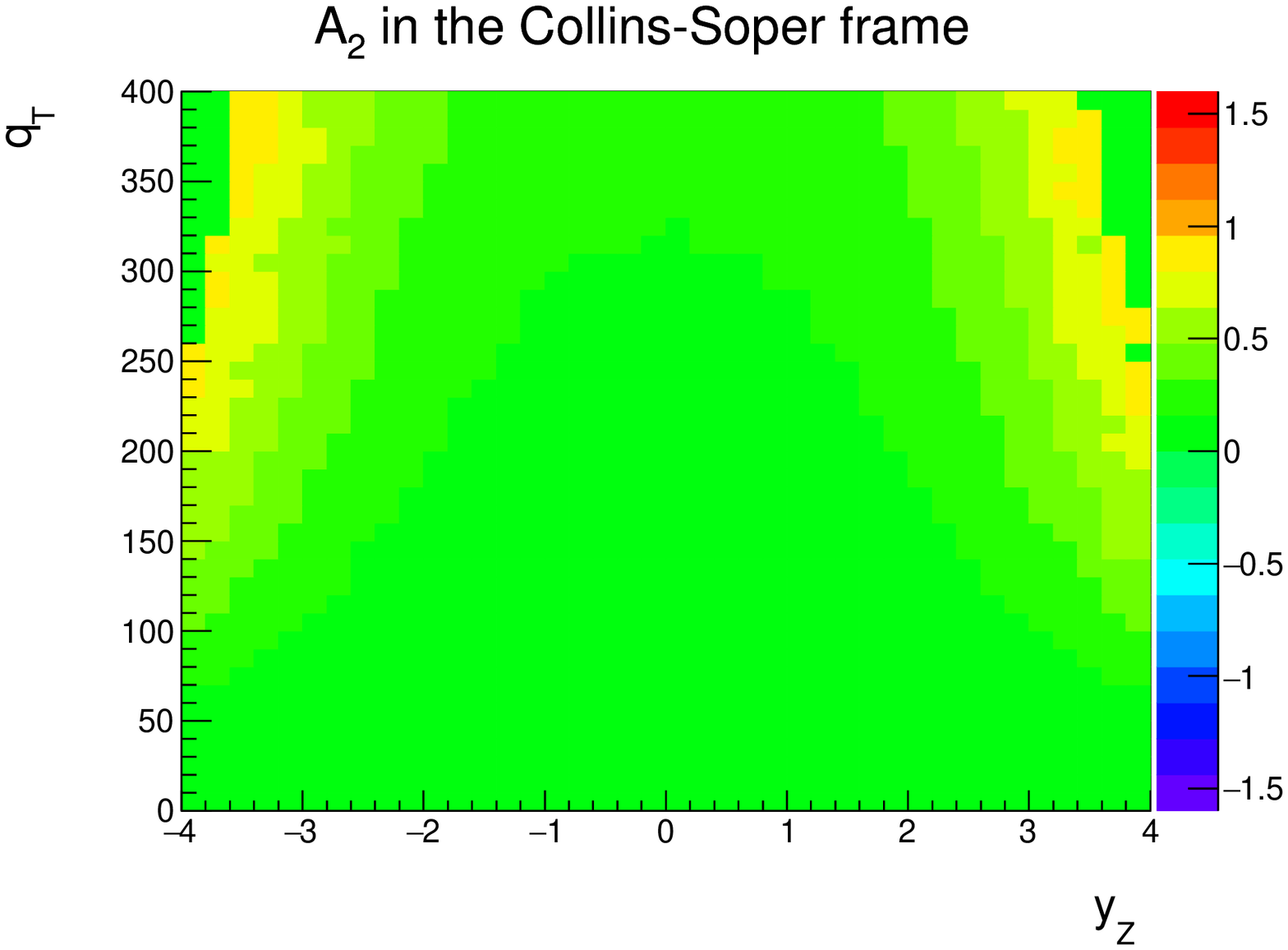} \\
  \includegraphics[width=4.5cm]{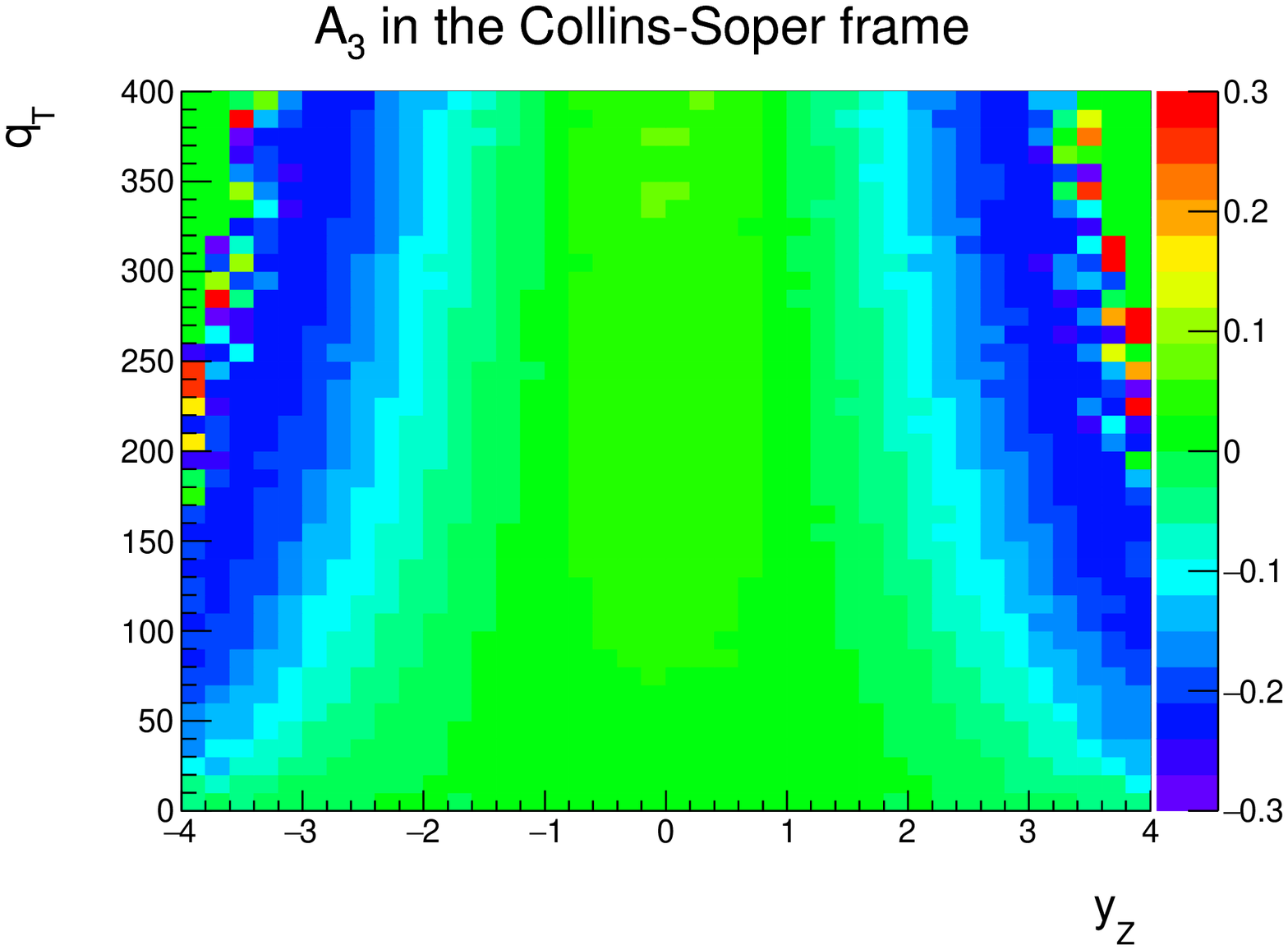} 
  \includegraphics[width=4.5cm]{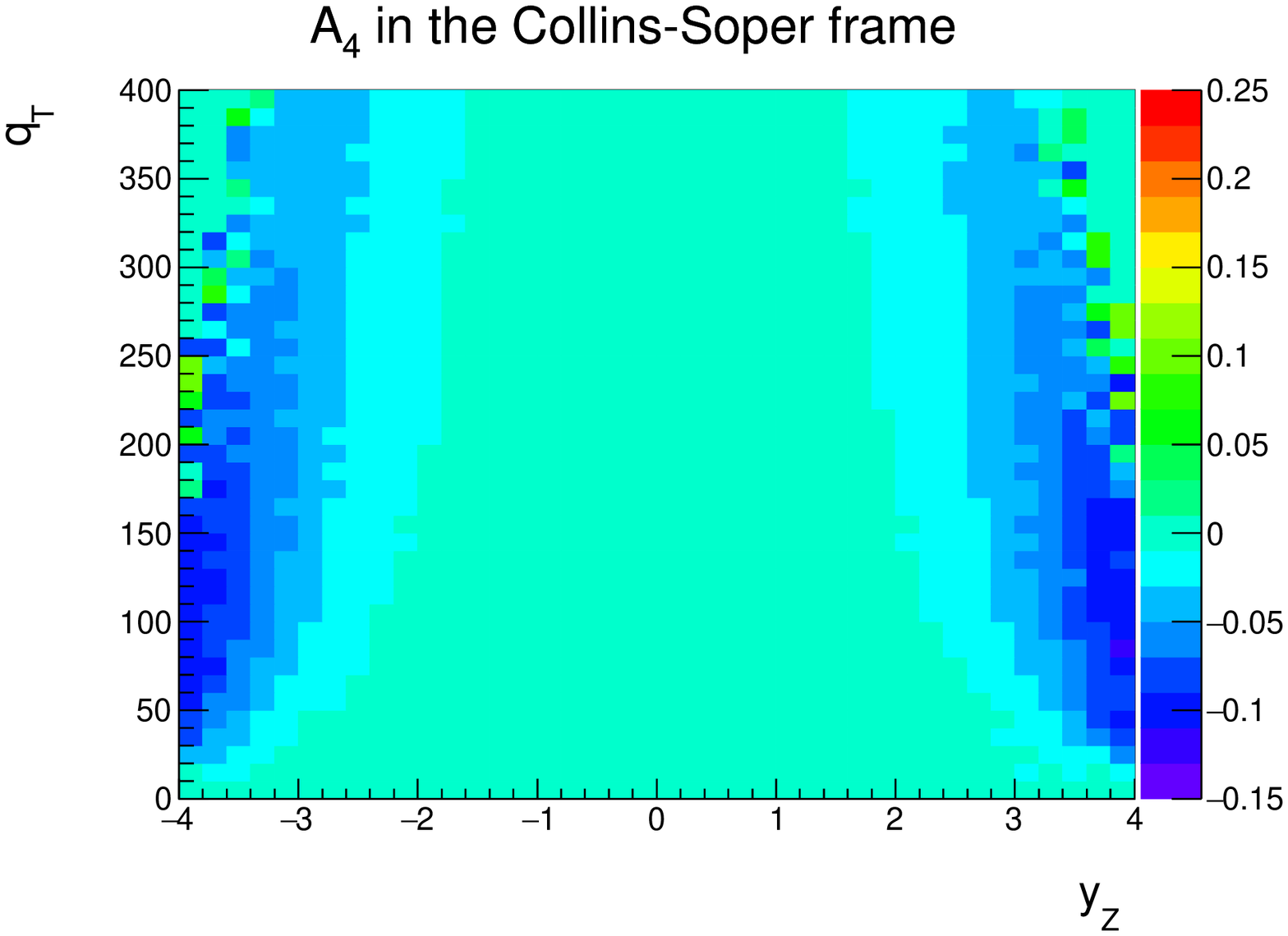}
  \includegraphics[width=4.5cm]{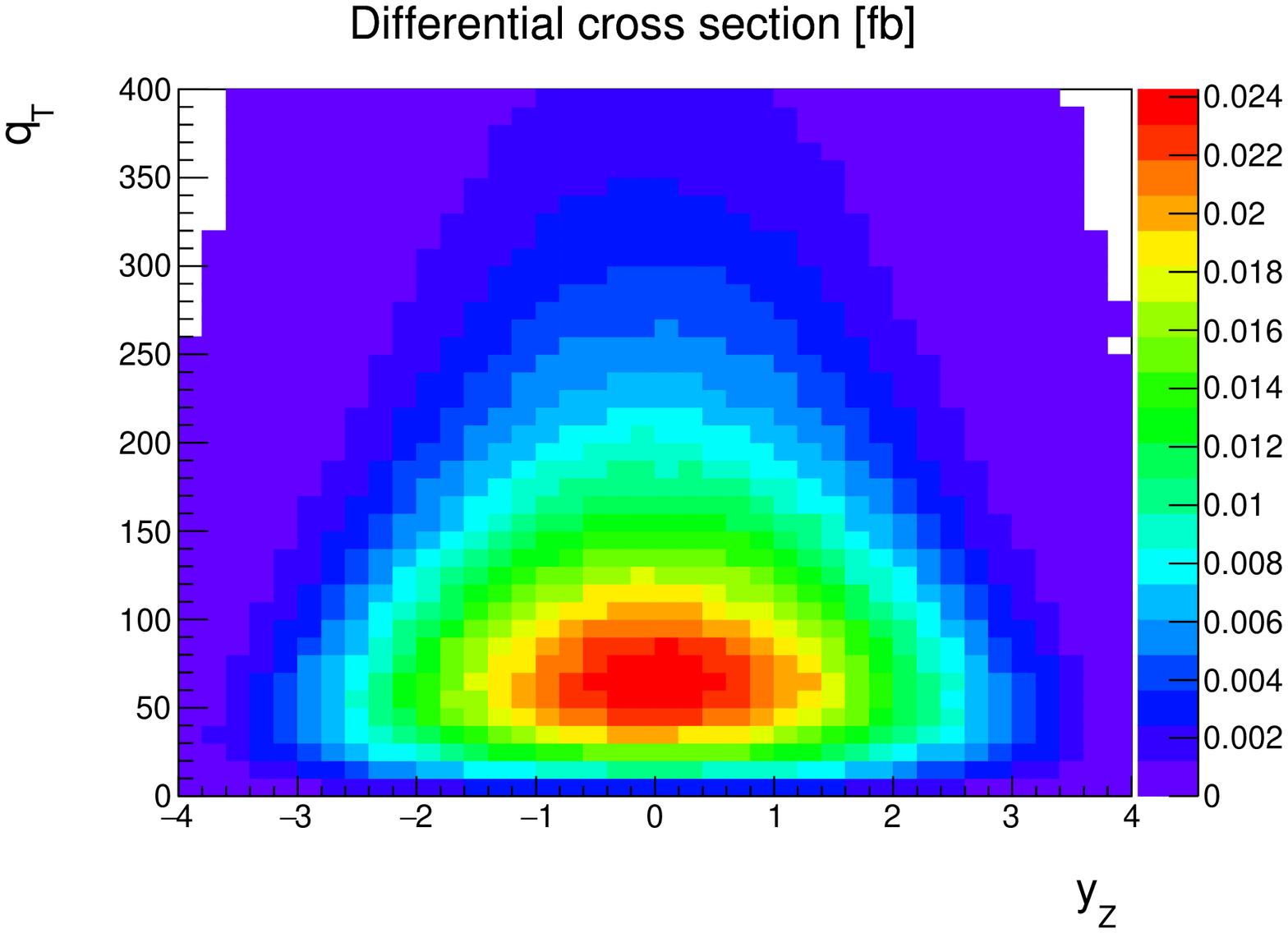}
  \caption{\label{fig:angs1b} Angular coefficients $A_0-A_4$ and the $\yz-\qt$ differential cross section of the benchmark scenario S1$_b$. }
\end{figure}

\begin{figure}[htp]
  \centering
  \includegraphics[width=4.5cm]{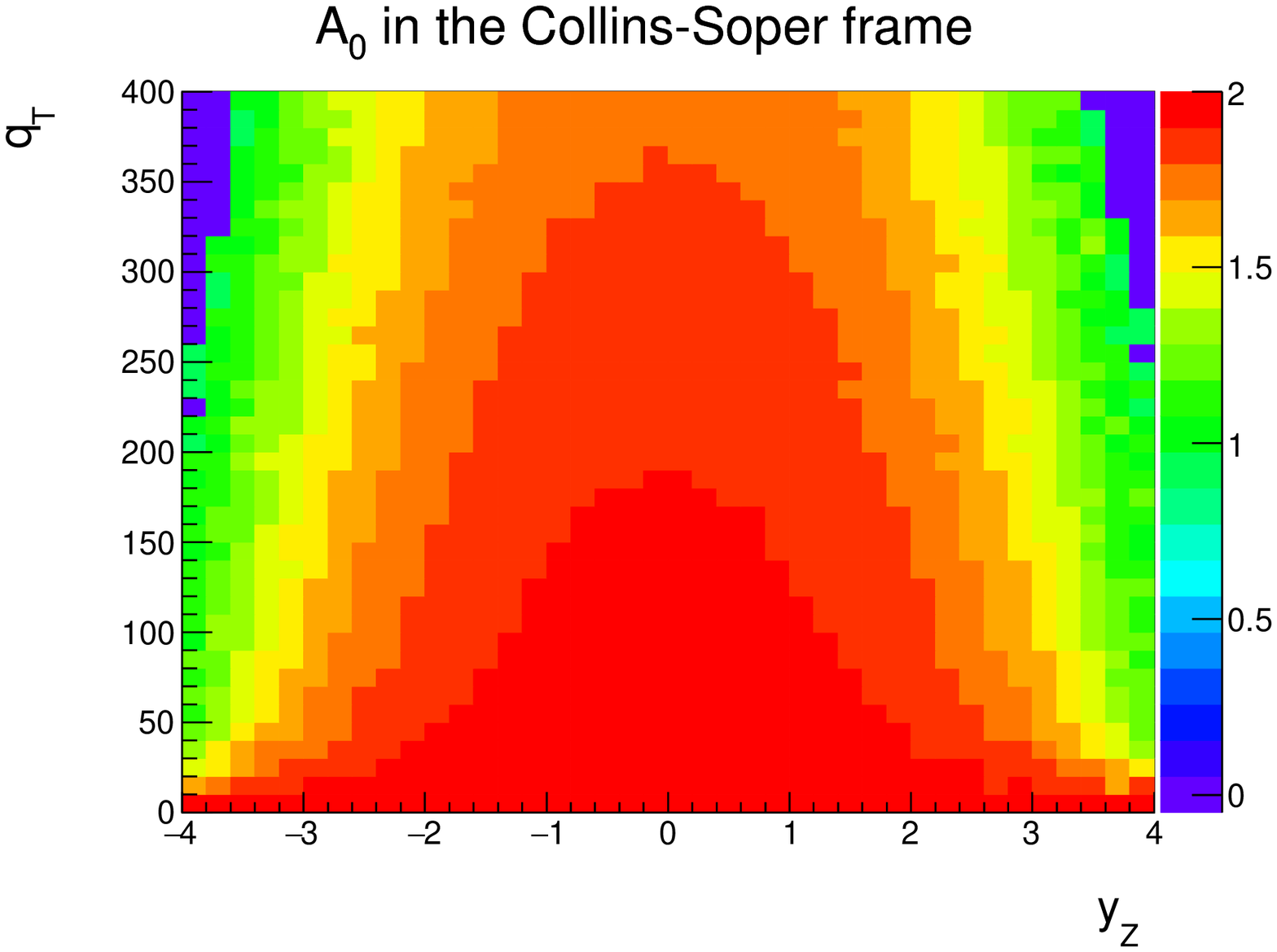}
  \includegraphics[width=4.5cm]{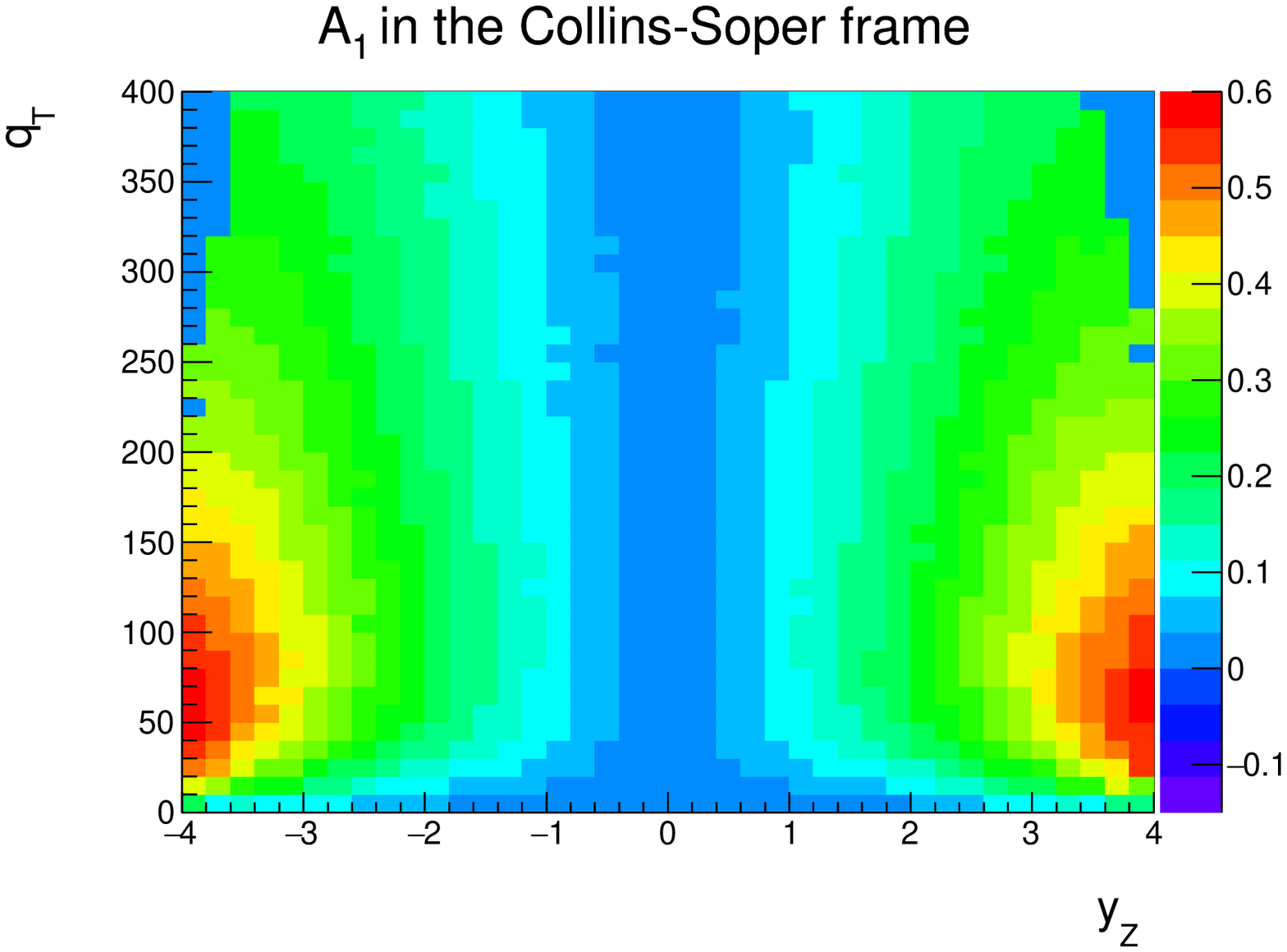} 
  \includegraphics[width=4.5cm]{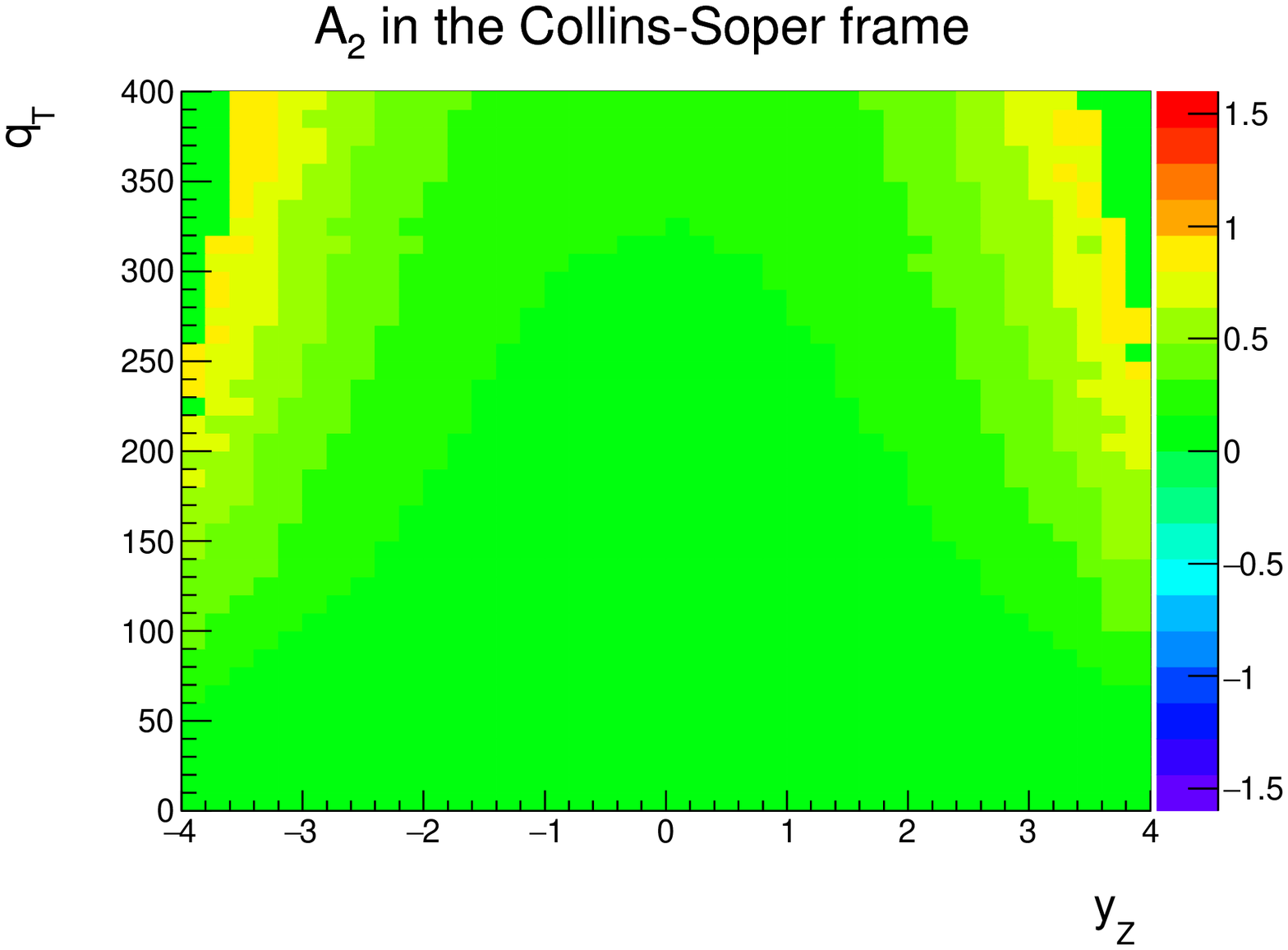} \\
  \includegraphics[width=4.5cm]{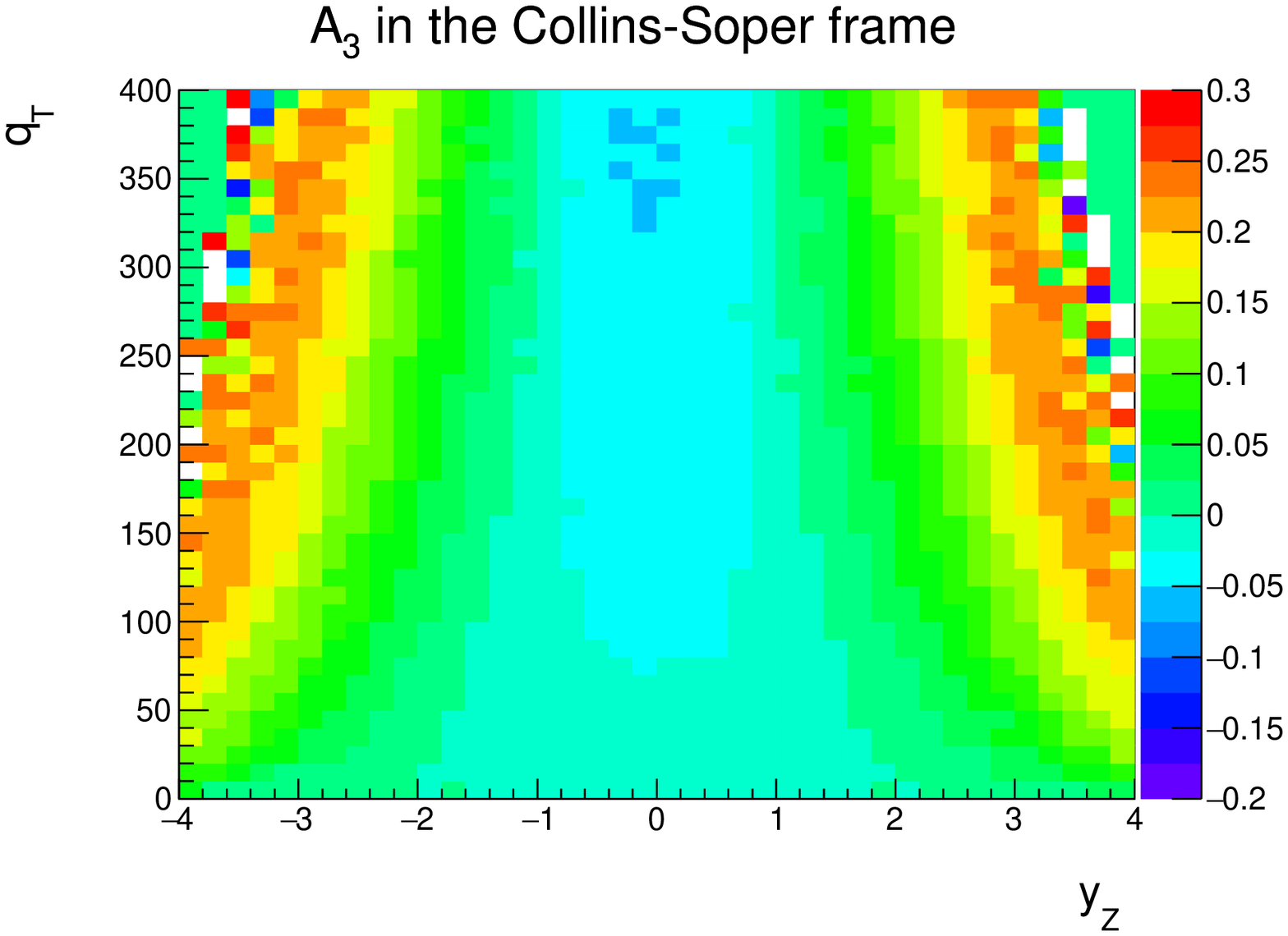} 
  \includegraphics[width=4.5cm]{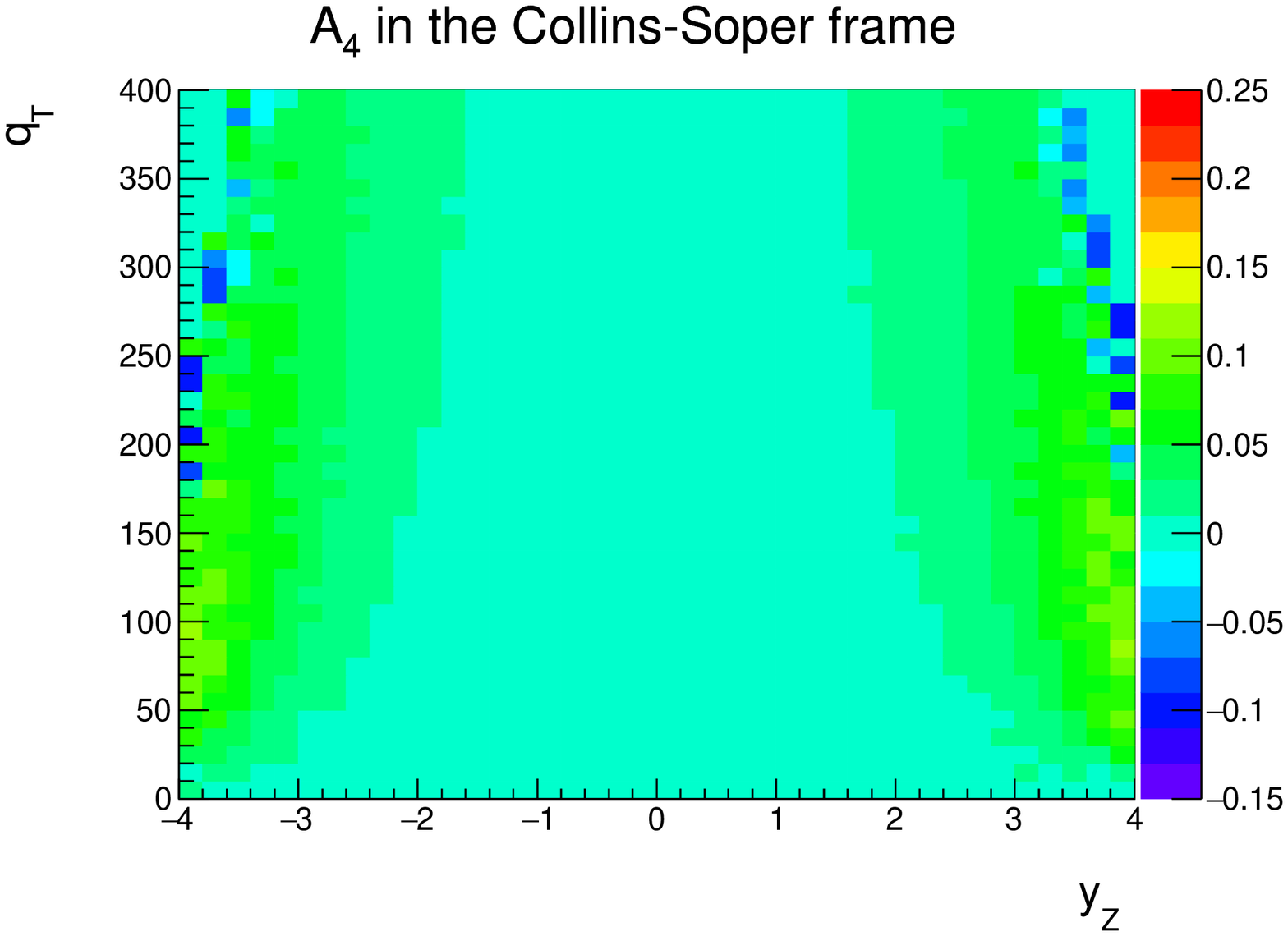}
  \includegraphics[width=4.5cm]{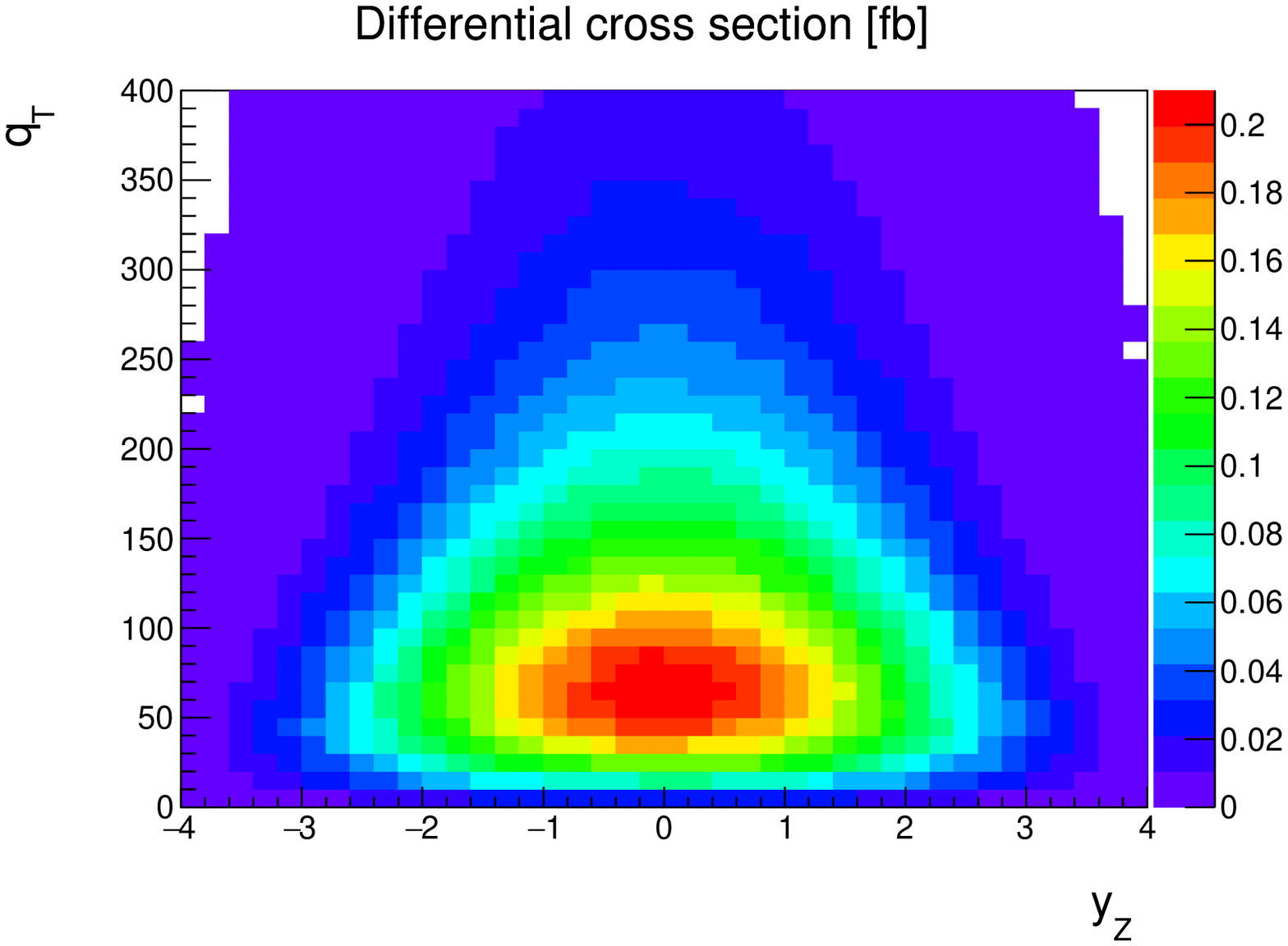}
  \caption{\label{fig:angs1c} Angular coefficients $A_0-A_4$ and the $\yz-\qt$ differential cross section of the benchmark scenario S1$_c$. }
\end{figure}

Angular coefficients of the benchmark scenarios S1$_{a,b,c}$ are shown in Fig.~\ref{fig:angs1a}, Fig.~\ref{fig:angs1b} and Fig.~\ref{fig:angs1c} respectively.
Comparing with the SM $\mathrm{Z Z}\to 2l2\nu$ and spin-0 dark sector models, 
$A_0$ of the spin-1 models are found to be very significant.
Among the three scenarios, most signatures look similar, 
but $A_3$ and $A_4$ take different signs between the S1$_b$ and S1$_c$. 
Hence the $A_3$ and $A_4$ can be used to quantify the parity violation of the dark sector.

\begin{table}[htb]
\centering
    \begin{tabular}{c|ccccccc}
      \hline
      \hline
  Benchmark         &       S1$_a$           &       S1$_b$                &       S1$_c$            &        S1$_0$                \\
                    &   Spin independent     &   Right handed              &  Left handed            &   SM ($\mathrm{Z Z}\to2l2\nu$)  \\ \hline
  $g^V_{X_D}$       &      1                 &        $1/\sqrt{2}$         &    $1/\sqrt{2}$         &           -                  \\
  $g^A_{X_D}$       &      0                 &        $1/\sqrt{2}$         &   -$1/\sqrt{2}$         &           -                  \\
  $g^V_{X_C}$       &      0                 &          0                  &         0               &           -                  \\
  \hline
  $g^V_{u}$         &     0.25               &        $\sqrt{2}/8$         &    $\sqrt{2}/8$         &           -                  \\
  $g^A_{u}$         &      0                 &        $\sqrt{2}/8$         &   -$\sqrt{2}/8$         &           -                  \\
  $g^V_{d}$         &     0.25               &        $\sqrt{2}/8$         &    $\sqrt{2}/8$         &           -                  \\
  $g^A_{d}$         &      0                 &        $\sqrt{2}/8$         &   -$\sqrt{2}/8$         &           -                  \\ 
  \hline
  $\mchi$ (GeV)     &      10                &          10                 &       10                &           -                  \\
  $\myv$ (GeV)      &      1000              &          1000               &       1000              &           -                  \\
  \hline
$\Gamma_{Y_1}$ (GeV)&     56.3               &         55.9                &       55.9              &           -                  \\
Cross section (fb)  &      2.50              &         0.533               &       4.50              &         239                  \\
      \hline
      \hline
    \end{tabular}
  \caption{ Benchmark scenarios with a spin-1 mediator. }
  \label{tab:exhi-s1}
\end{table}

\clearpage

\subsection{Spin-2 mediator}

\begin{figure}[htp]
  \centering
  \includegraphics[width=4.5cm]{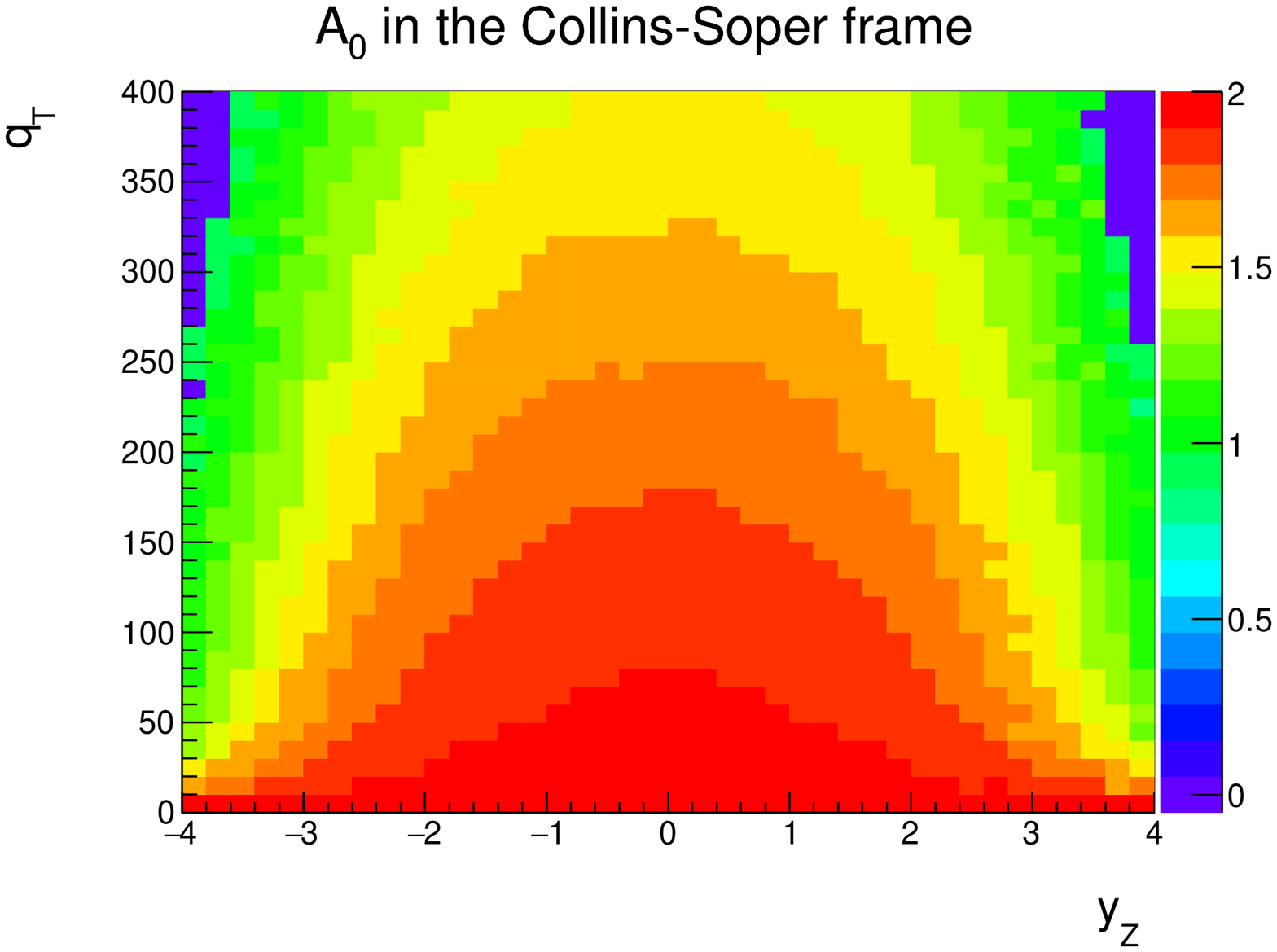}
  \includegraphics[width=4.5cm]{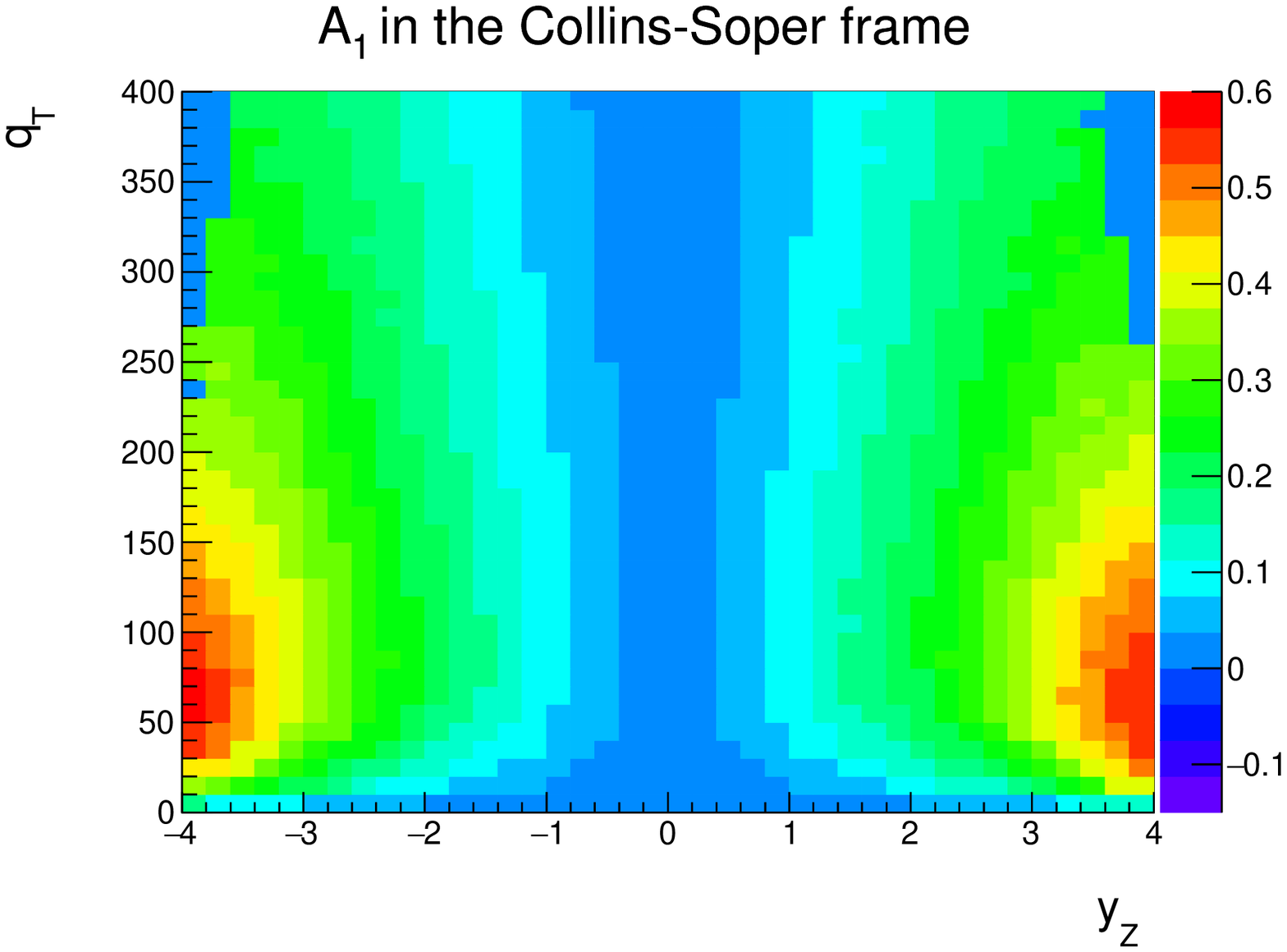} 
  \includegraphics[width=4.5cm]{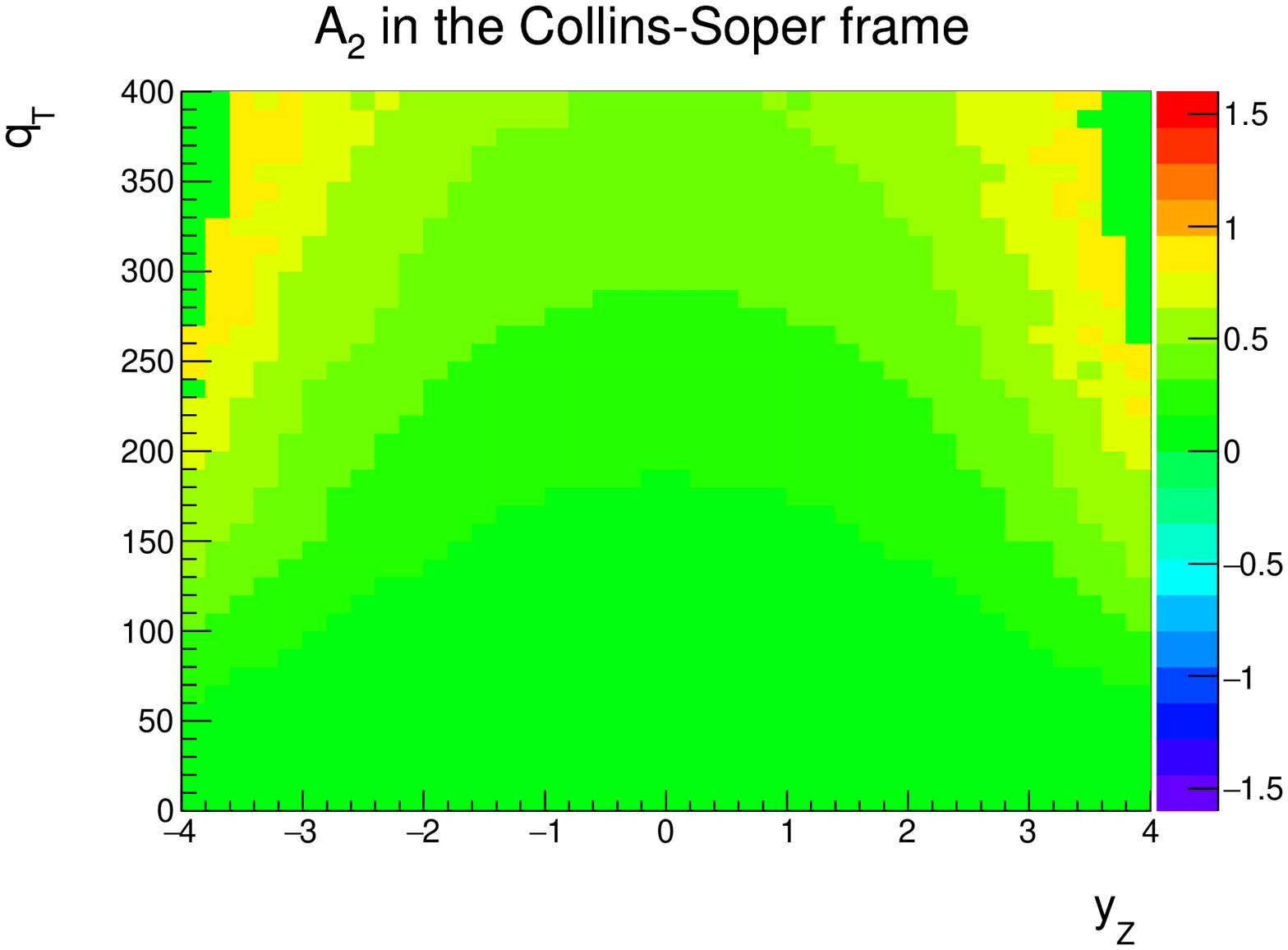} \\
  \includegraphics[width=4.5cm]{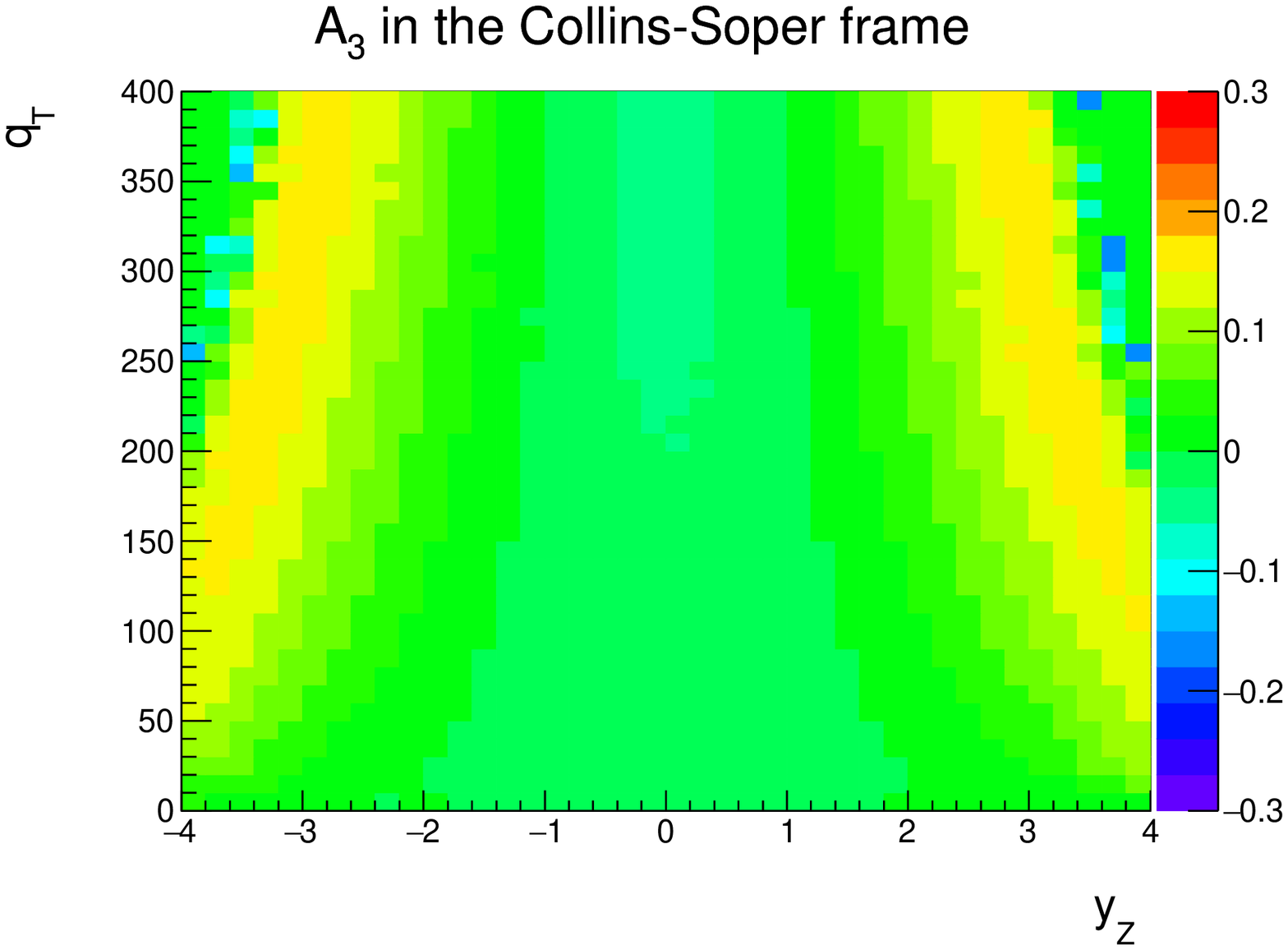} 
  \includegraphics[width=4.5cm]{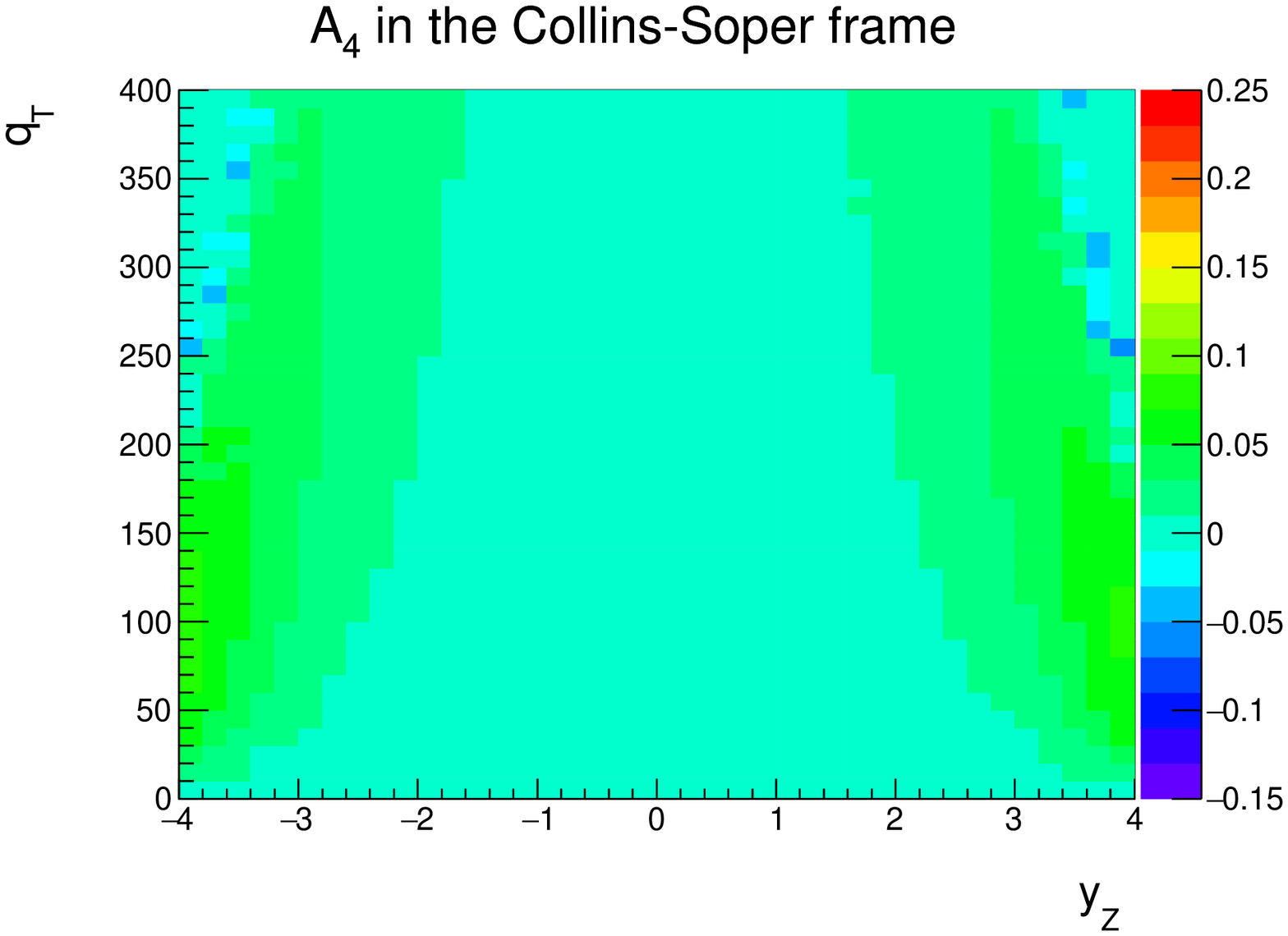}
  \includegraphics[width=4.5cm]{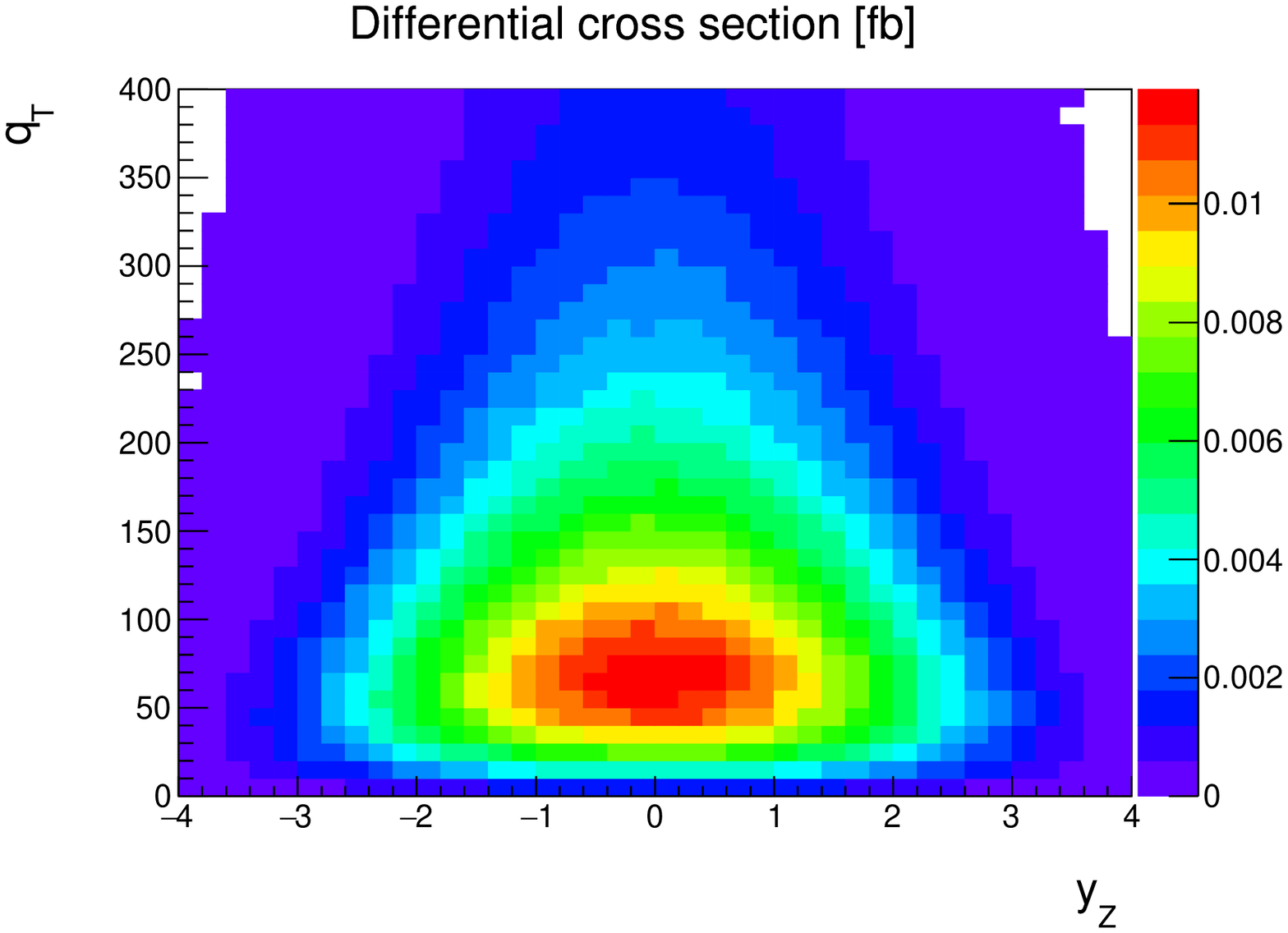}
  \caption{\label{fig:angs2a} Angular coefficients and the $\yz-\qt$ differential cross section of the benchmark scenario S2$_a$. }
\end{figure}

The dark sector with a spin-2 mediator is also tested.
We consider a model as described in the Ref.\cite{Kraml:2017atm}, with benchmark scenarios listed in Table~\ref{tab:exhi-s2}.
The masses of the dark matter and the mediator are also chosen to be the same as in the spin-0 model.
Despite an increase of complexity in the computation, we found the angular coefficients look similar to the benchmark scenario S1$_a$.
We show only the angular coefficients of the benchmark scenario S2$_a$ in Fig.~\ref{fig:angs2a}. 
Some visible differences from the S1$_a$ can be observed from the $A_0$ and $A_2$ distributions.
Since we do not measure the DM, the angular coefficients of S2$_{b,c}$ are found to be very close to the ones of S2$_a$.

\begin{table}[htb]
\centering
    \begin{tabular}{c|cccccccc}
      \hline
      \hline
  Benchmark         &       S2$_a$       &       S2$_b$     &       S2$_c$        \\ \hline
  $g^T_{X_D}$       &        1           &        0         &         0           \\
  $g^T_{X_R}$       &        0           &        1         &         0           \\
  $g^T_{X_V}$       &        0           &        0         &         1           \\
  $g^T_{SM}$        &        1           &        1         &         1           \\
  \hline
  $\mchi$ (GeV)     &      10            &      10          &        10           \\
  $\myt$ (GeV)      &      1000          &      1000        &        1000         \\
  $\Lambda$         &      3000          &      3000        &        3000         \\
  \hline
$\Gamma_{Y_2}$ (GeV)&      95.3          &      93.7        &        97.7         \\
Cross section (fb)  &      2.73          &     0.0462       &       0.578         \\
      \hline
      \hline
    \end{tabular}
  \caption{ Benchmark scenarios with a spin-2 mediator. Angular coefficients of the three scenarios look all the same. }
  \label{tab:exhi-s2}
\end{table}

\clearpage

%\clearpage

\section{Setting limits on the coupling strength parameters of dark sector models}
\label{sec:limits}

In Section~\ref{sec:angC}, we have shown that angular coefficients of the benchmark dark sector models can have distinct signatures from the 
SM $\mathrm{Z Z}\to 2l2\nu$ background process in the $\yz-\qt$ plane. 
In this section, we take advantage of these signatures and set limit on the coupling strength parameter $\lambda$ of each dark sector model,
based on observables $\mathbf{x}=(\yz,\qt,\cos\theta_{CS},\phi_{CS})$. 
The invisible part $(\yy,\sy,\cos\theta_{\chi},\phi_{\chi})$ was integrated out to construct pdfs, as described in Section~\ref{sec:param}.

\subsection{Statistical method}

With the pdfs of the signal and background processes obtained through MEM, 
one can construct an unbinned likelihood function over N events in the data sample~\cite{Barlow:1990vc}:

\begin{eqnarray}
{\cal L}(\text{data}|\lambda,\boldsymbol{\theta}) &=& \text{Poisson}(N|S(\lambda,\boldsymbol{\theta})+B(\boldsymbol{\theta})) \rho(\boldsymbol{\theta}) \prod_i \rho(\mathbf{x}^i|\lambda,\boldsymbol{\theta}), \\ 
\rho(\mathbf{x}|\lambda,\boldsymbol{\theta}) &=& \dfrac{ S(\lambda,\boldsymbol{\theta}) \rho_s(\mathbf{x}^i,\lambda) + B(\boldsymbol{\theta}) \rho_b(\mathbf{x}^i) }{ S(\lambda,\boldsymbol{\theta}) + B(\boldsymbol{\theta}) }, 
\end{eqnarray}

where $\rho_s(\mathbf{x},\lambda)$ and $\rho_b(\mathbf{x})$ represent pdfs of the signal and background, 
$S(\lambda,\boldsymbol{\theta})$ and $B(\boldsymbol{\theta})$ corresponding to the expected signal and background yields. 
The $\boldsymbol{\theta}$ represents the full set of nuisance parameters with pdf $\rho(\boldsymbol{\theta})$, which are designed to incorporate systematic uncertainties.

To set limits on the parameters $\lambda$, we compare the compatibility of the data with the $\lambda$ fixed and $\lambda$ floated hypotheses and construct a test statistic based on the profile likelihood ratio:

\begin{eqnarray}
t_{\lambda} = -2\ln\dfrac{ {\cal L}(\text{data}|\lambda,\boldsymbol{\hat{\theta}}_{\lambda}) }{ {\cal L}(\text{data}|\hat{\lambda},\boldsymbol{\hat{\theta}}) }.
\end{eqnarray}

According to the Wilk's theorem, this test statistic satisfies the $\chi^2$ distribution of the same degrees of freedom as $\lambda$ in the large sample limit~\cite{wilks1938}. 
One can, therefore, set limits on the $\lambda$ through a parameter space scan and cut on the $-2\ln\Delta{\cal L}$ values. 

Neglecting pdf of the nuisance parameters, it follows that

\begin{eqnarray}
t_{\lambda} = -2\ln\dfrac{\text{Poisson}(N|S(\lambda)+B)}{\text{Poisson}(N|S(\hat{\lambda})+B)} 
 -2 \sum_i \ln \dfrac{\rho(\mathbf{x}^i|\lambda)}{\rho(\mathbf{x}^i|\hat{\lambda})}
\end{eqnarray}

For setting limits on $\lambda$, we assume that there is a single dataset in agreement with $\lambda=0$. 
In the large sample limit, we have:

\begin{eqnarray}
t_{\lambda} &\xrightarrow{N\to\infty}& -2\ln\dfrac{\text{Poisson}(N|S(\lambda)+B)}{\text{Poisson}(N|B)} 
 +2 N \int \d \mathbf{x} \rho(\mathbf{x}|\lambda=0) \ln \dfrac{\rho(\mathbf{x}|\lambda=0)}{\rho(\mathbf{x}|\lambda)} \\ \nonumber
           &=& -2\ln\dfrac{\text{Poisson}(N|S(\lambda)+B)}{\text{Poisson}(N|B)} + 2 N \cdot D(\rho(\mathbf{x}|\lambda=0) || \rho(\mathbf{x}|\lambda)). 
\end{eqnarray}

where the first term is a test statistic for simple counting experiment and the second term is proportional to $N$ and a KL-divergence~\cite{kullback1951}.
As the KL-divergence measures the difference of the pdfs $\rho(\mathbf{x}|\lambda)$ and $\rho(\mathbf{x}|\lambda=0)$, it quantifies the powerfulness of the MEM. 
For simplicity, we will call the first term as normalization term and the second one as 
KL-divergence term. 

In our study, the likelihood function is prepared by {\sc BASES} numerical integration with HELAS subroutines for the helicity amplitudes. 
The evaluation of the KL-divergence term is performed using a plain integration provided by the {\sc GNU Scientific Library}. 
We validate our program by checking the normalizations of all the constructed pdfs and by comparing the angular coefficients and cross sections of all 
involved processes with the {\sc MG5}.
See more information in Appendix~\ref{app:valid}.

\subsection{Background modeling and event selections}

To make our limits more realistic, we consider a few selections -- marked as BL selections -- as listed in Table~\ref{tab:selections} to capture major detector acceptance effects for the processes involved. 
The values of these selections are set refering to recent 13 TeV LHC measurements~\cite{Sirunyan:2017onm,Aaboud:2017bja}.
There are several additional selections considered in experiments to improve the signal feasibility,
e.g., jet counting, $3^{rd}$-lepton veto, top quark veto, and $\Delta\phi_{ll,\ptvecmiss}$, $|\etmiss-\ptll|/\ptll$ for momentum balance~\cite{Sirunyan:2017onm}.
These selections reject most background from misidentification but lead to different acceptance efficiencies for different processes. 
Without detector simulation, we determine the event rate according to the CMS results 
(Table 3 of Ref.~\cite{Sirunyan:2017onm}), with an ancillary $A\cdot \epsilon$ incorporating the additional selections in the experiment and a scale factor normalizing to 150~$\fbinv$ data. 
The signal dark matter processes are assumed to have the same ancillary $A\cdot \epsilon$ as the SM ZZ$\to 2l 2\nu$ process.

\begin{table}[htb]
\centering
    \begin{tabular}{c|c}
      \hline
      \hline
   Variable             &          Requirements    \\ \hline
$p^l_{\mathrm{T}}$      &          $>20$~GeV       \\
    \sz                 &          \text{NWA}      \\
%$|\mathrm{m}_{ll}-mz|$  &          $<10$~GeV       \\
$E^{\text{miss}}_{\mathrm{T}}$ &   $>80$~GeV       \\
$|\eta_{l}|$            &          $<2.4$          \\
$\Delta R_{ll}$         &          $>0.4$          \\
$|\yz|$                 &          $<2.5$          \\
%$\qt$                  &          2580            \\ 
      \hline
      \hline
    \end{tabular}
  \caption{ Selections considered in our computations (BL-selections), where $l=e,\mu$. Additional selection requirements are considered in experiments to improve the signal feasibility.
            Their effects are included through an Ancillary $A\cdot \epsilon$. }
  \label{tab:selections}
\end{table}

Our background pdf is constructed based on components summarized in Table~\ref{tab:bkgevt}.
Apart from the non-resonant-$ll$ background, which is constructed using only the phase space,
other components are built using matrix elements.
The WZ$\to 3l\nu$ matrix element assumes W$\to e \nu$, where the electron is not identified by a detector.
The Z/$\gamma^*\to l^+l^-$ is estimated with matrix element of the Z$\to l^+l^-$ plus one jet production,
phase space of this process reduces to three final state particles.

\begin{table}[htb]
\centering
    \begin{tabular}{c|c|c|c}
      \hline
      \hline
  Process                &  Cross section with BL-selections (fb)   &       Ancillary $A\cdot \epsilon$     &       Events        \\ \hline
  ZZ$\to 2l 2\nu$        &                27.7                      &            0.488                      &        2028         \\ 
  Non-resonant-$ll$      &               1.57$\times 10^{3}$        &            5.80$\times 10^{-3}$       &        1370         \\
  WZ($\to e\nu 2l$)      &                17.05                     &            0.296                      &         757         \\
  Z/$\gamma^*\to l^+l^-$ &               3.61$\times 10^{4}$        &            1.23$\times 10^{-4}$       &         665         \\
      \hline
      \hline
    \end{tabular}
  \caption{ Background estimation with cross sections calculated in a phase space with BL-selections and ancillary $A\cdot \epsilon$ to obtain the same event rate as in Table 3 of Ref.~\cite{Sirunyan:2017onm}.
The number of events has been translated into 150~$\fbinv$ data.  }
  \label{tab:bkgevt}
\end{table}

\begin{figure}[htbp]
  \centering
  \includegraphics[width=4.5cm]{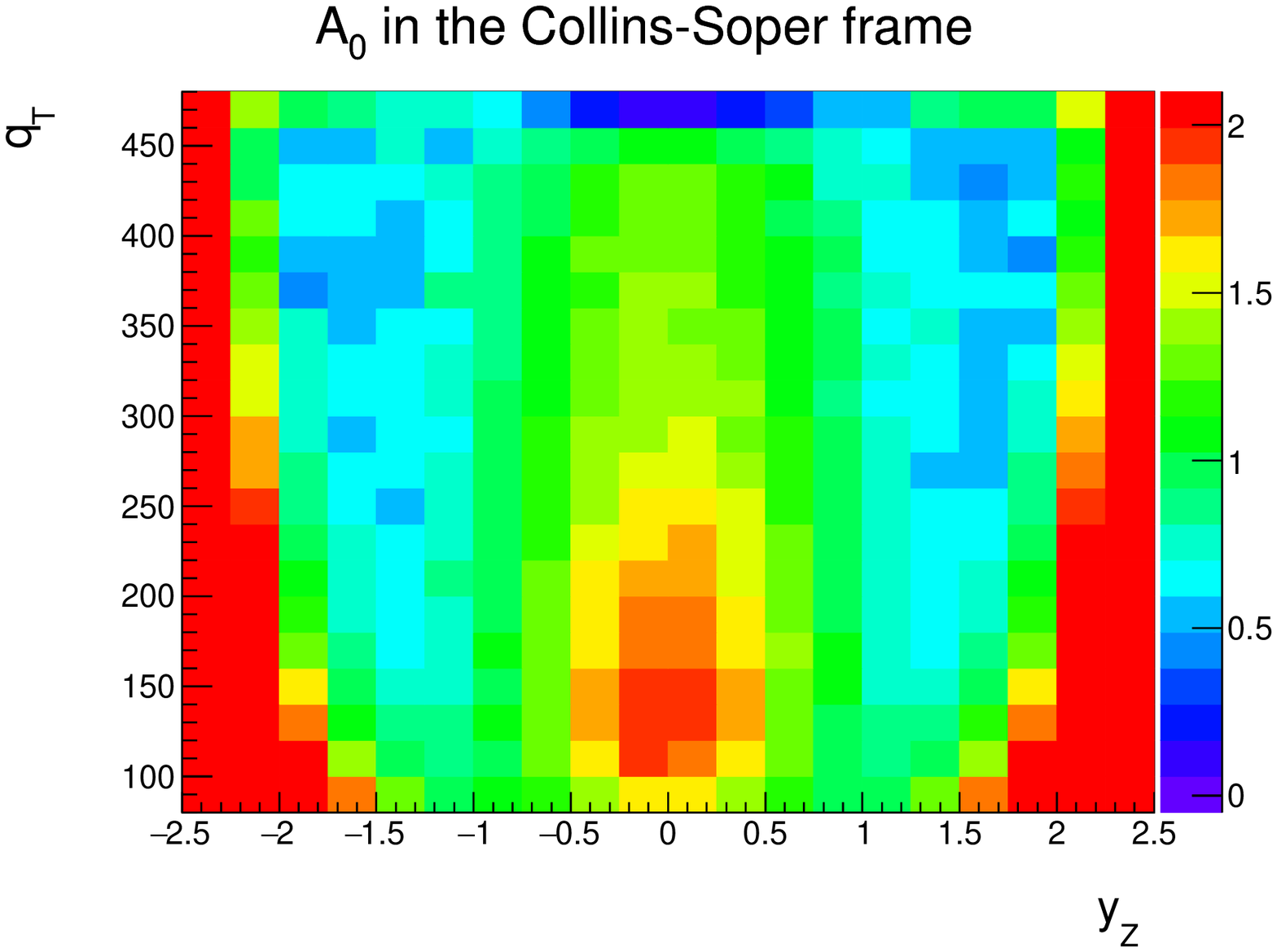}
  \includegraphics[width=4.5cm]{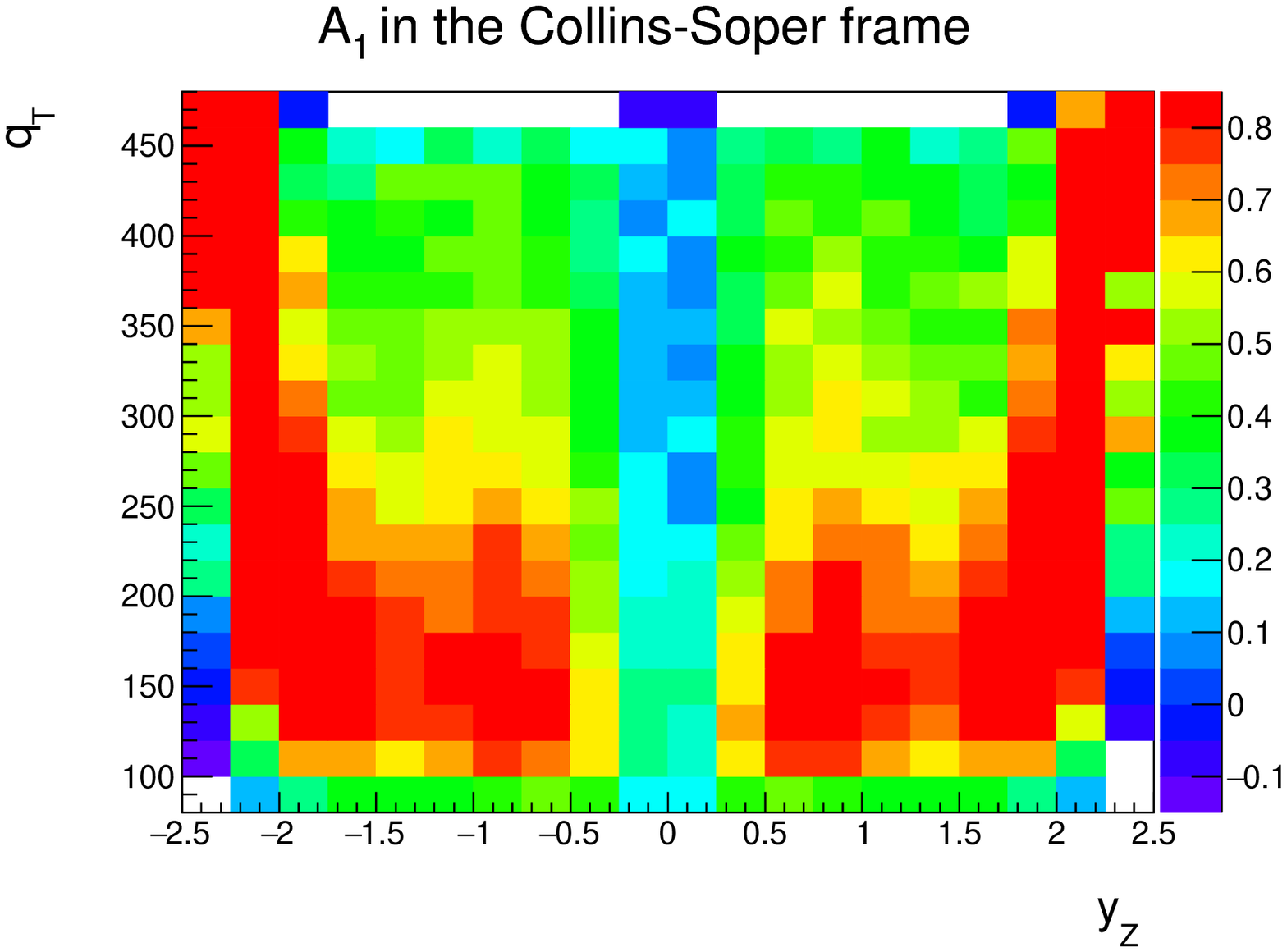}
  \includegraphics[width=4.5cm]{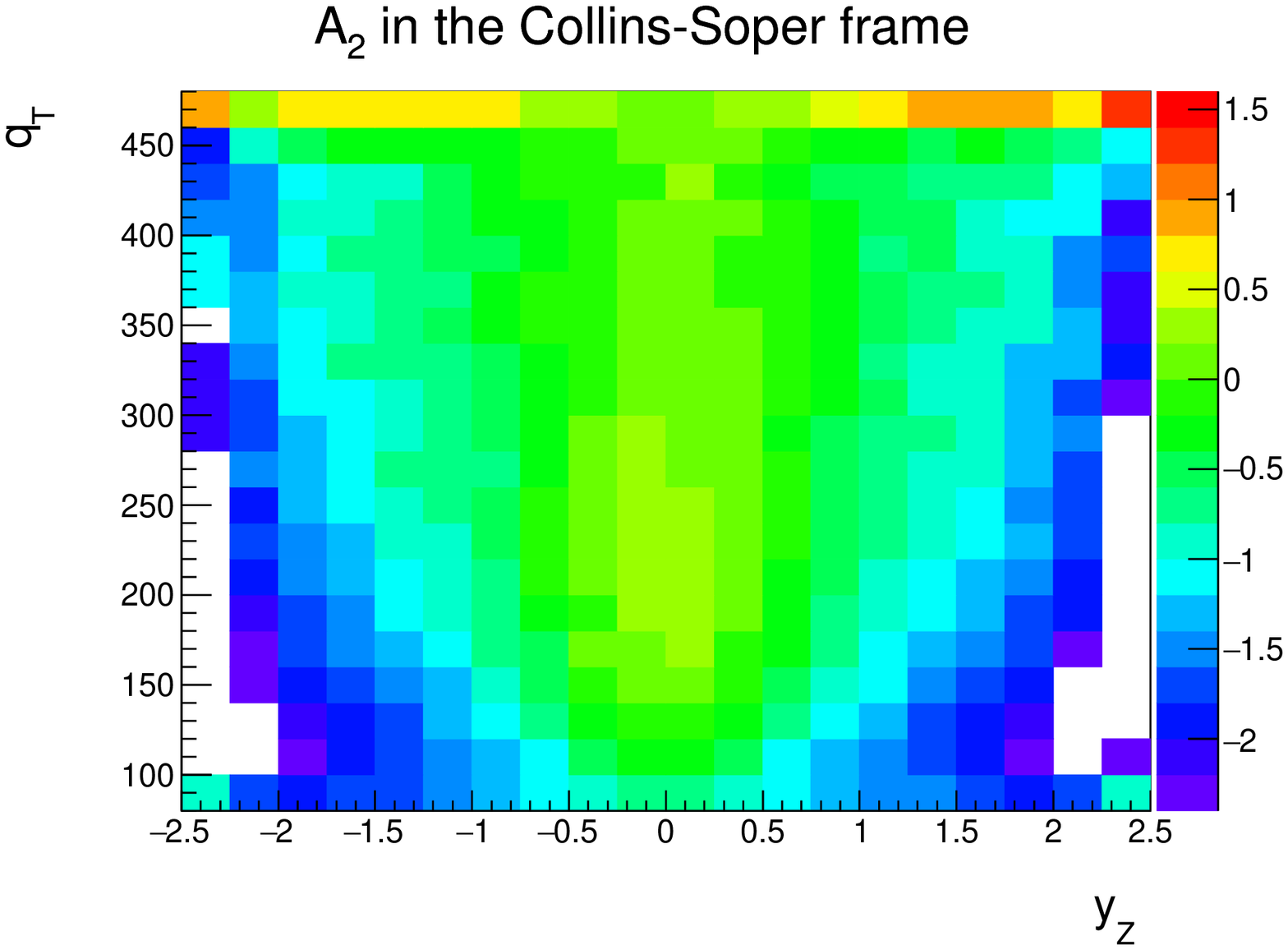} \\
  \includegraphics[width=4.5cm]{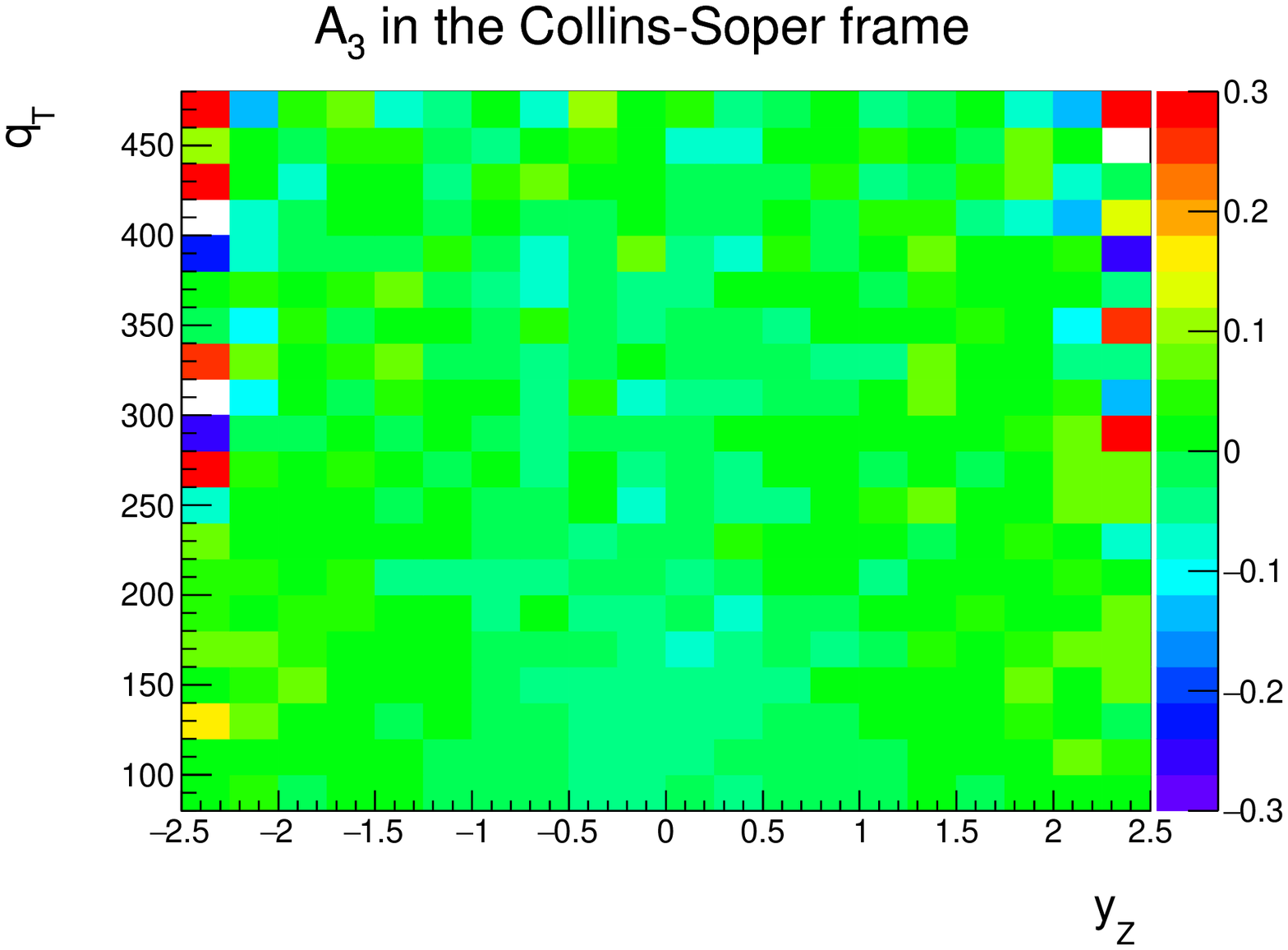}
  \includegraphics[width=4.5cm]{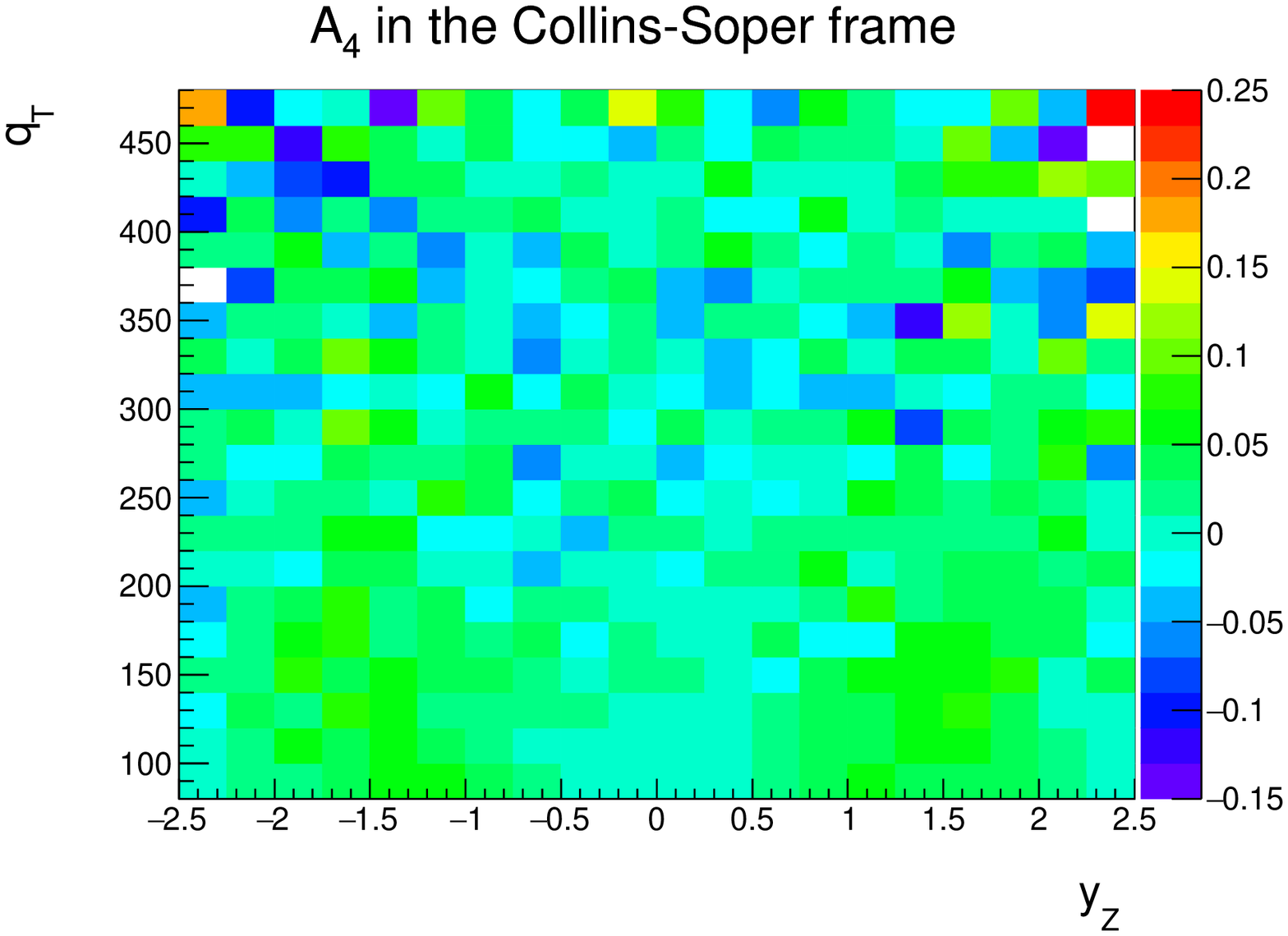}
  \includegraphics[width=4.5cm]{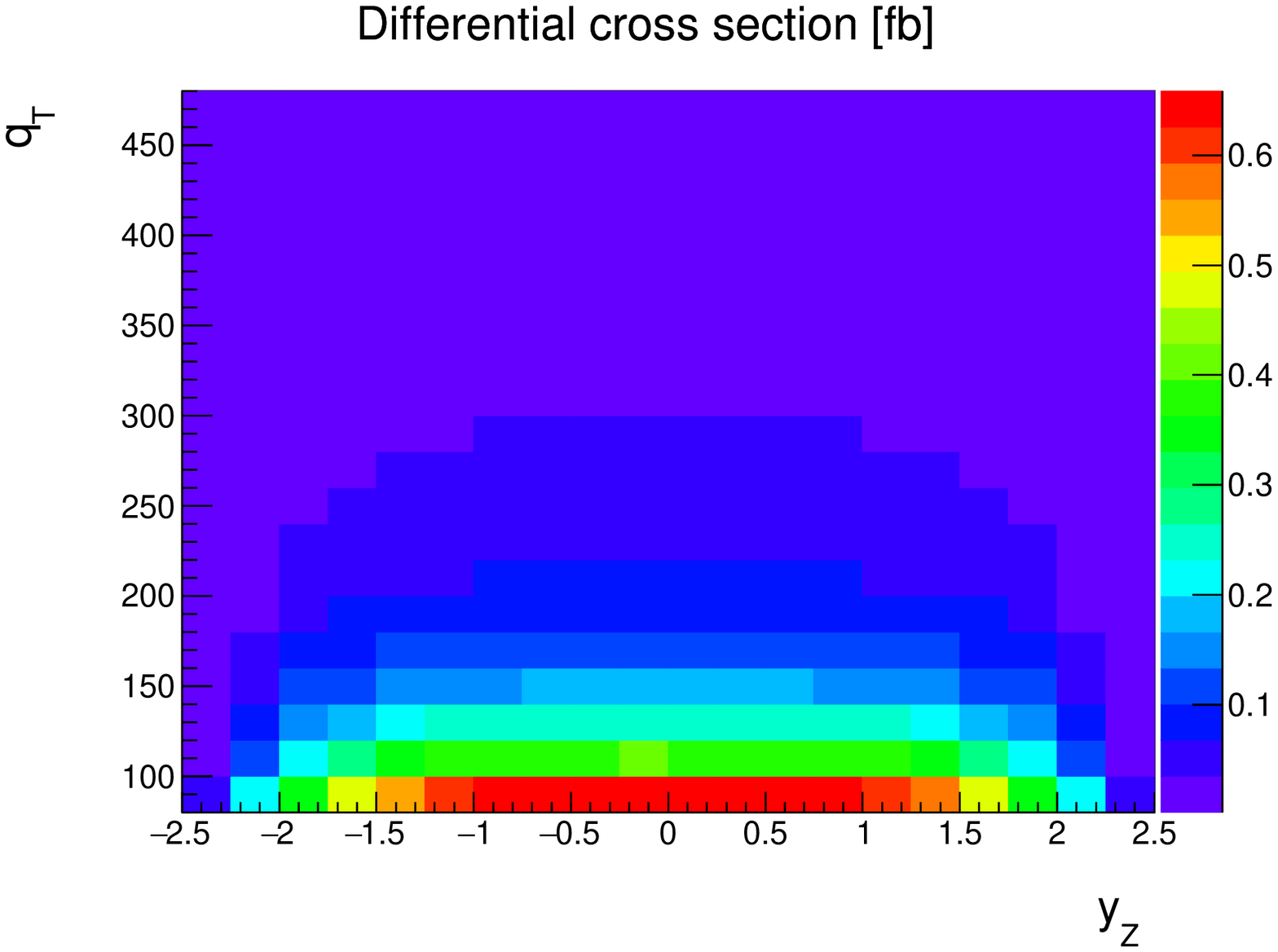}
  \caption{\label{fig:bkgak} Angular coefficients $A_0-A_4$ in the CS frame and $\yz-\qt$ differential cross section 
                          for background only hypothesis.  
                          Selections in Table~\ref{tab:selections} have been applied and cause irregular shapes in kinematic boundaries.}
\end{figure}

In the presence of selections, angular coefficients can be distorted. 
Fig.~\ref{fig:bkgak} shows the angular coefficients $A_0-A_4$ for the background only hypothesis. 
Irregular distributions on the boundaries are mainly caused by the selections on $|\eta_{l}|$ and $\Delta R_{ll}$. 
With the coupling strength at our expected limit, the presence of signal can only perturb the shapes of the background only ones. 
%and Fig.~\ref{fig:bkgakwithS0} shows those coefficients for signal plus background hypothesis considering benchmark model S0$_{a}$ with $\lambda=3.5$.

%\begin{figure}[htbp]
%  \centering
%  \includegraphics[width=4.5cm]{figs/fig_A0_bkgak_withS0a_yz.eps}
%  \includegraphics[width=4.5cm]{figs/fig_A1_bkgak_withS0a_yz.eps}
%  \includegraphics[width=4.5cm]{figs/fig_A2_bkgak_withS0a_yz.eps} \\
%  \includegraphics[width=4.5cm]{figs/fig_A3_bkgak_withS0a_yz.eps}
%  \includegraphics[width=4.5cm]{figs/fig_A4_bkgak_withS0a_yz.eps}
%  \includegraphics[width=4.5cm]{figs/fig_hrate_bkgak_withS0a_yz.eps}
%  \caption{\label{fig:bkgakwithS0} Angular coefficients $A_0-A_4$ in the CS frame and $\yz-\qt$ differential cross section
%                          for signal plus background hypothesis considering benchmark model S0$_{a}$ with $\lambda=3.5$.  
%                          Selections in Table~\ref{tab:selections} have been applied and cause irregular shapes in kinematic boundaries.}
%\end{figure}

\subsection{Limits on the coupling strength parameters of the dark sector models}

\begin{figure}[htbp]
  \centering
  \includegraphics[width=4.5cm]{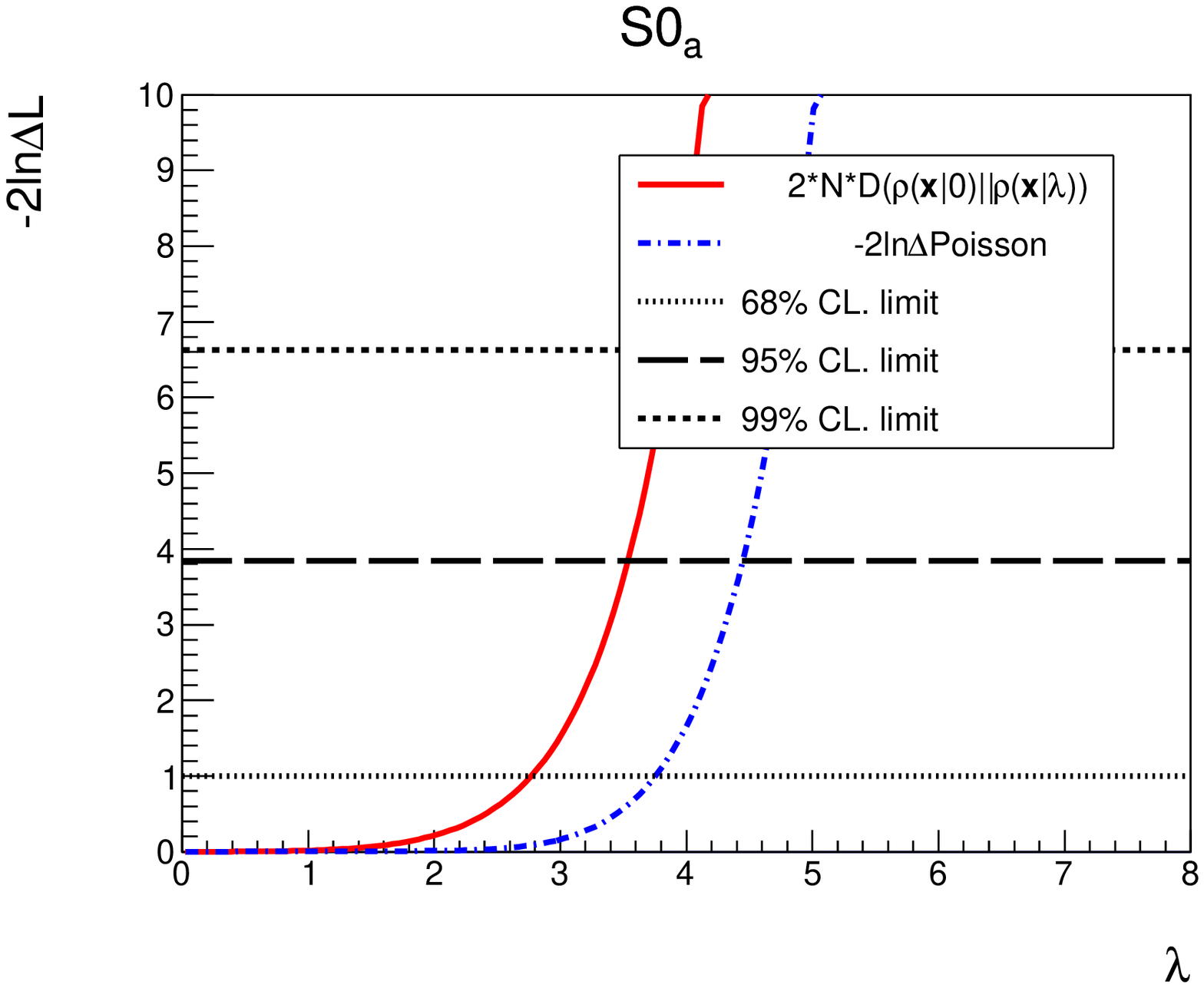}
  \includegraphics[width=4.5cm]{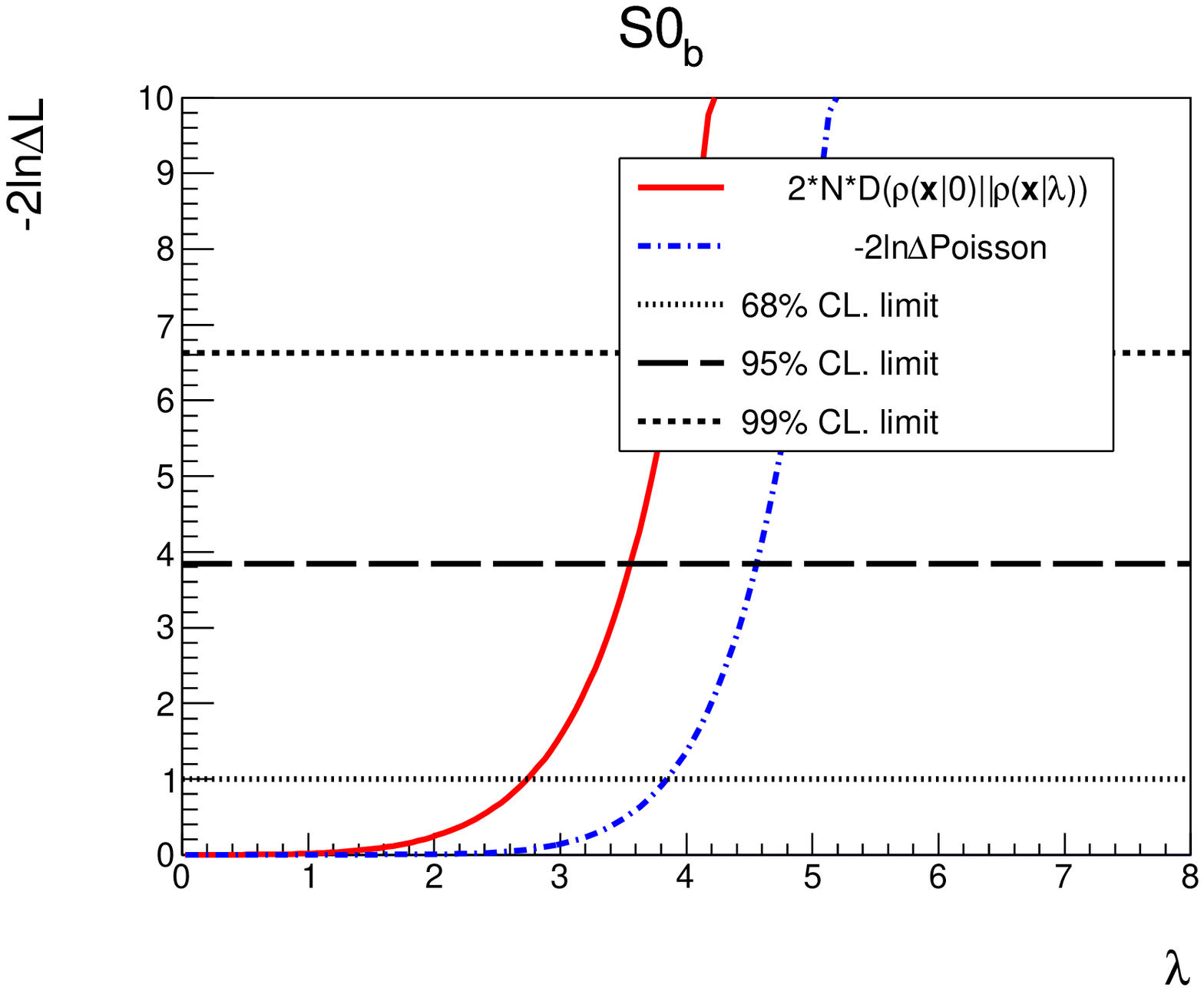}
  \includegraphics[width=4.5cm]{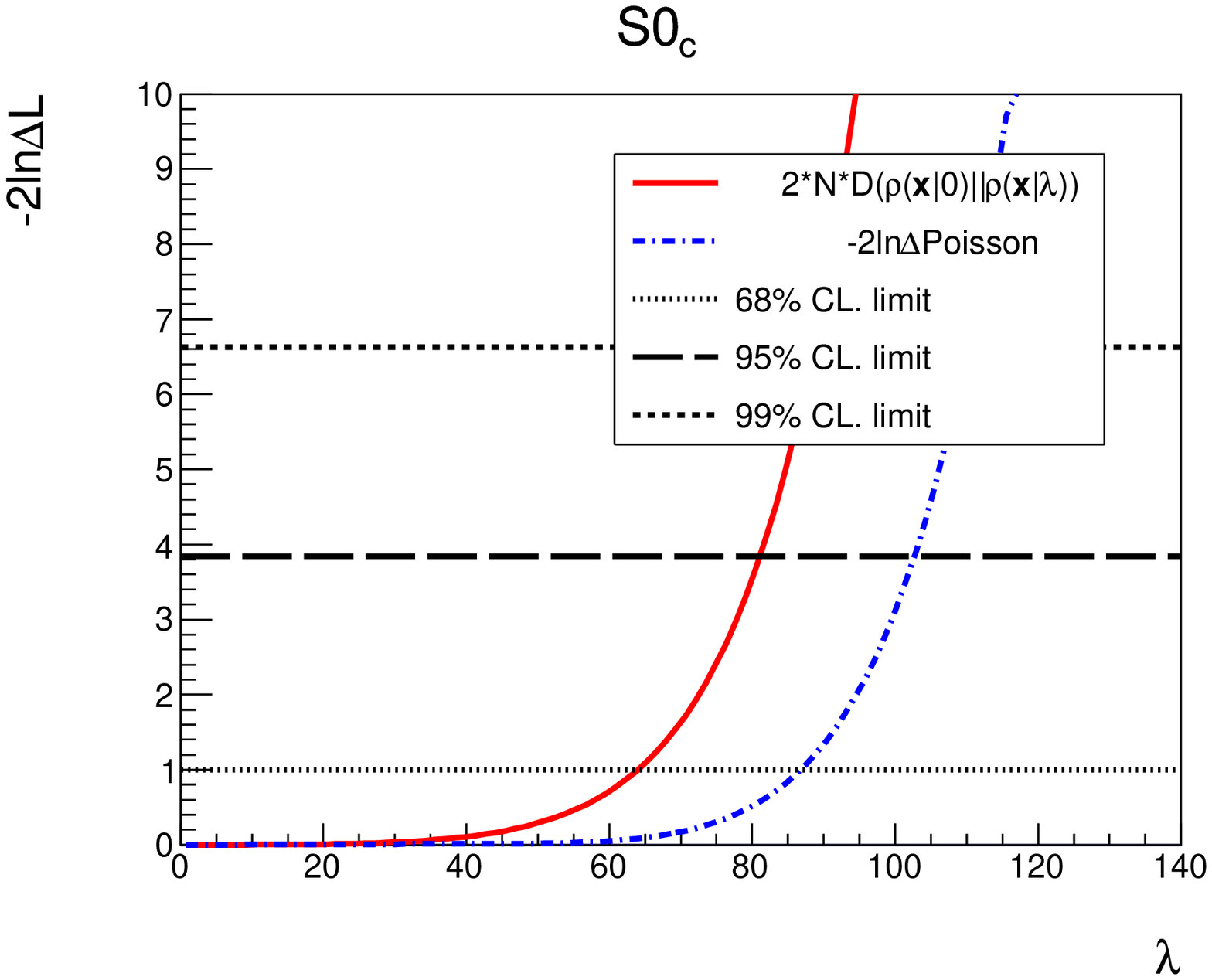}
  \caption{\label{fig:lims0} Upper limits on the coupling strength parameters of the S0 benchmark scenarios. }
\end{figure}

\begin{figure}[htbp]
  \centering
  \includegraphics[width=4.5cm]{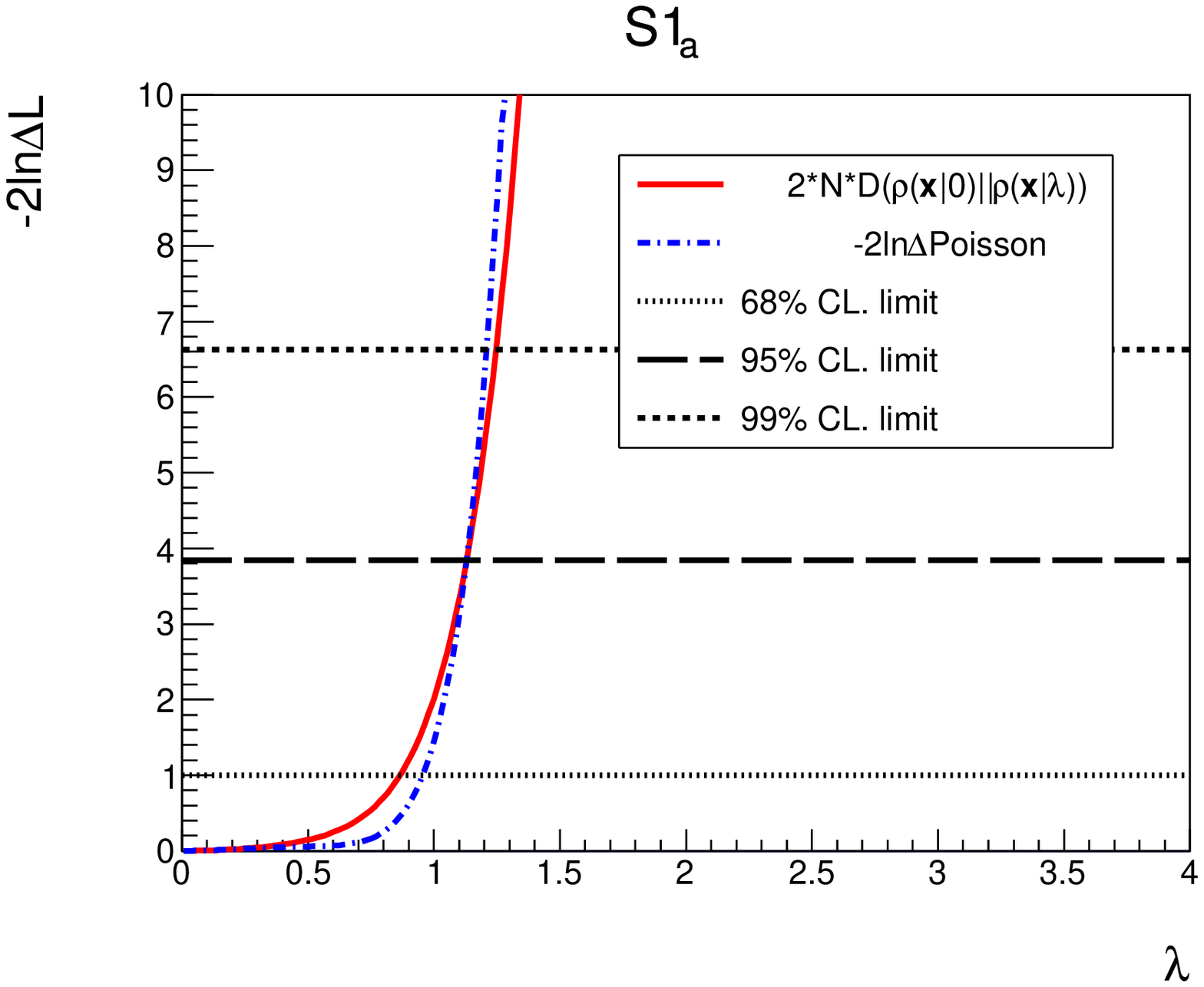}
  \includegraphics[width=4.5cm]{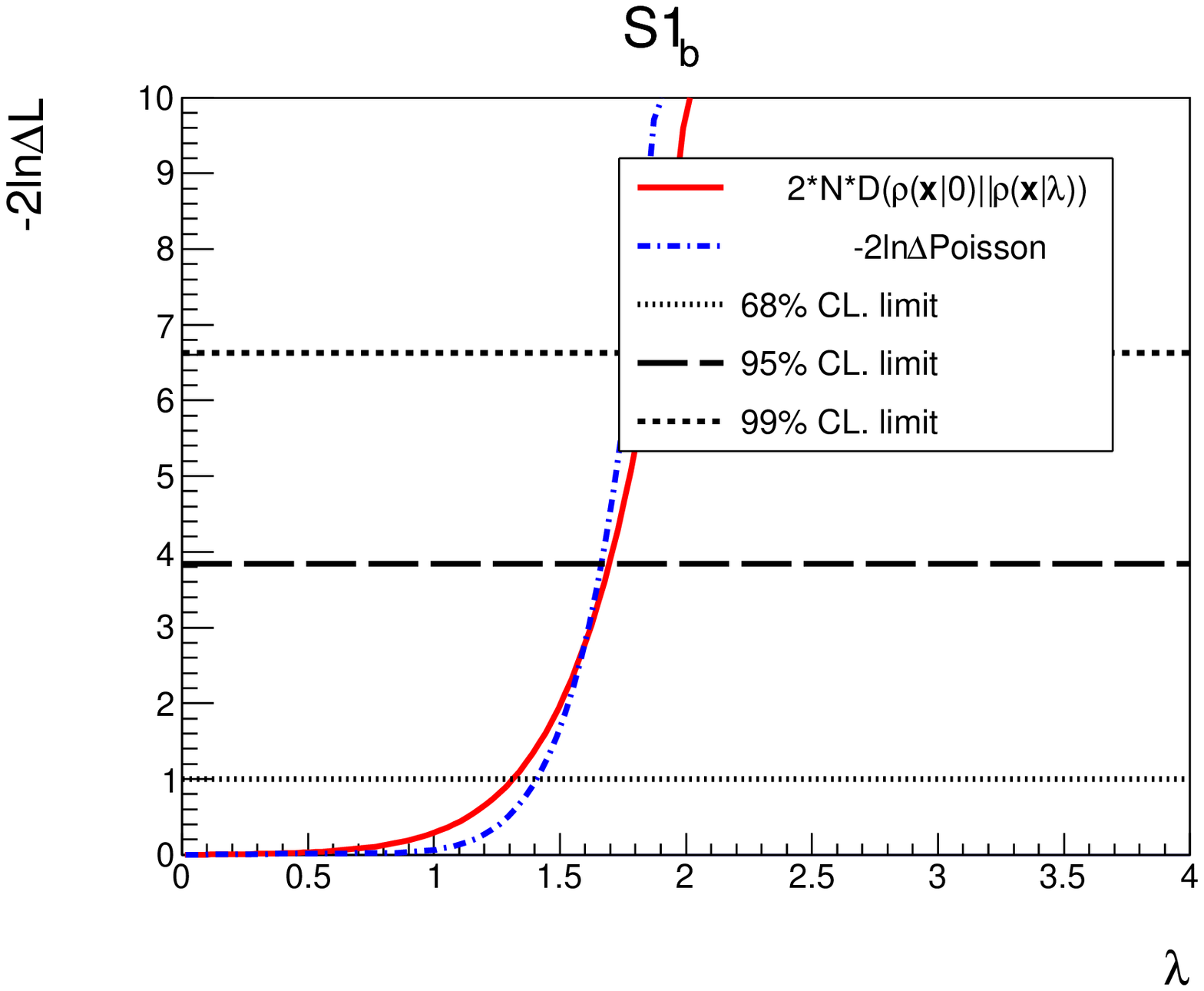}
  \includegraphics[width=4.5cm]{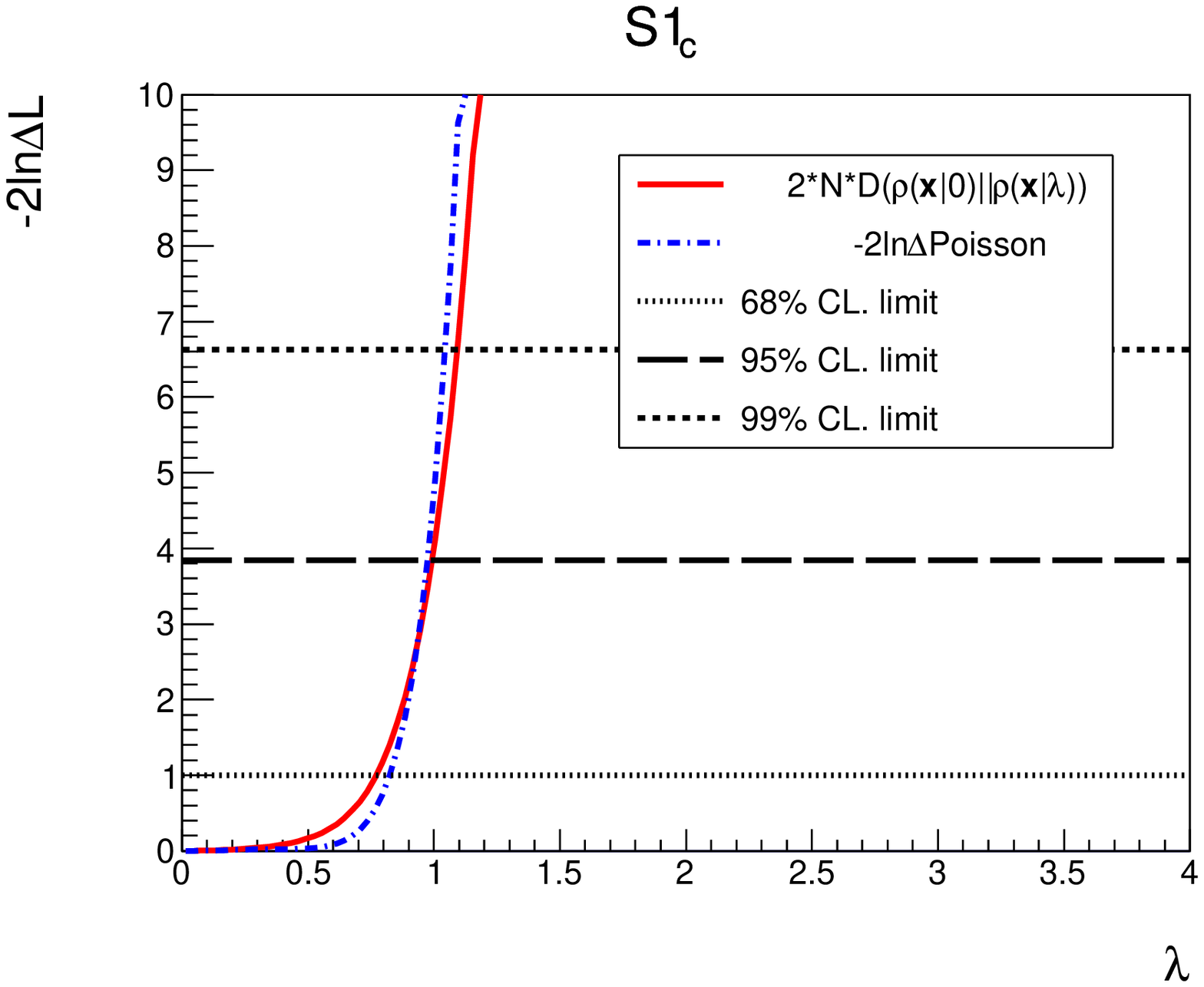}
  \caption{\label{fig:lims1} Upper limits on the coupling strength parameters of the S1 benchmark scenarios. }
\end{figure}

In our dark sector models, it is necessary to have two couplings: one for the interaction with SM particles, one for the DM decay.
For conciseness, we assume that both couplings in the benchmark model are scaled by a strength parameter $\lambda$.
This assumption makes the cross sections change with two orders severer in couplings than ones for limits of a single coupling.
We compare the upper limits set from the normalization term $-2\ln\text{Poisson}$ and
from the KL-divergence term $2N\cdot D(\rho(\mathbf{x}|0)||\rho(\mathbf{x}|\lambda))$
in Fig.~\ref{fig:lims0} for the S0 benchmark scenarios and
in Fig.~\ref{fig:lims1} for the S1 benchmark scenarios.
The shapes provide significant improvements in all cases.
The KL-divergence terms drive the final limits for the S0 benchmark scenarios
and are close to the normalization terms in the S1 benchmark scenarios.

We provide in Table~\ref{tab:limit} 95\% CL upper limits of the strength parameters.
In our evaluation, the numerical uncertainty of the normalization terms can be easily made negligible.
However, the evaluation of the KL-divergence terms can be computationally expensive.
It takes us roughly $700\times 6$ CPU hours, functioning at about 2.4 GHz, for us to obtain 30\%-50\% uncertainties on the KL-divergence terms around the limit values.
The signal cross sections at the limit values are also reported.
Since a counting experiment calculate limits based on signal background yields,
the results from the normalization term are almost the same.
The ones from the KL-divergence terms, however, depend on the shape difference between the signal and background.
As the KL-divergence is a measure the shape difference, a lower cross section means a larger the difference in shape.
These quantitative results are in agreement with qualitative features of the angular coefficients among models provided in Section~\ref{sec:angC}.

\begin{table}[htb]
\centering
\scalebox{1.0}{
  \begin{tabular}{r|ccc|ccc}
  \hline
  \hline
  Benchmark       ~~~~~~~~~~                     &  S0$_a$    &  S0$_b$    &   S0$_c$  &  S1$_a$       &   S1$_b$      &    S1$_c$   \\ \hline
Limit from the normalization term ($\lambda_1$)  &    4.4     &     4.6    &     103   &      1.1      &    1.7        &    0.97     \\
Signal cross section at $\lambda_1$ (fb)         &    1.86    &     1.87   &     1.86  &     1.87      &    1.87       &    1.87     \\
\hline                                                                                                
Limit from the KL-divergence term ($\lambda_2$)  &    3.5     &     3.6    &     81    &      1.1      &    1.7        &    0.99     \\
Signal cross section at $\lambda_2$ (fb)         &    0.75    &     0.70   &     0.72  &      1.9      &    2.0        &    2.0      \\
\hline                                                                                                
Combined limit ($\lambda_0$)                     &    3.5     &     3.5    &     79    &      1.0      &    1.5        &    0.89     \\
  \hline
  \hline
  \end{tabular}}
  \caption{ Upper limits on the coupling strength parameters of the dark sector models at 95\% CL, with signal cross sections at the limit values. }
  \label{tab:limit}
\end{table}

\subsection{Example application of MEKD}

Our computation considered only parton level matrix element at leading order (LO).
We comment that there are already efforts to extend the MEM to Next-to-Leading Order (NLO)~\cite{Campbell:2012cz}
and incorporates parton shower effects~\cite{Soper:2011cr}.
There is an easier approach to exploit the LO matrix elements, called the matrix element kinematic discriminator (MEKD)~\cite{Avery:2012um,Chatrchyan:2012sn,Chatrchyan:2012jja}. 
This method construct a variable named MEKD that can be calculated for events with required observables.
By construction, it utilizes the matrix element and can be used to distinguish the signal and background.
The advantage of this method is that detector effects and theoretical uncertainties in the
construction of likelihood function is independent of the application.

Based on the pdfs defined as in Eq.~\ref{eq:pdf} of the signal and combined background, we define the MEKD as:

\begin{eqnarray}
\text{MEKD} = \ln \dfrac{\rho_s(\mathbf{x},\lambda)}{\rho_b(\mathbf{x})}, 
\end{eqnarray}

where $\mathbf{x}=(\yz,\qt,\cos\theta_{CS},\phi_{CS})$ and the invisible part has been integrated out. 
Then we use the {\sc MG5} program to generate events for the applications. 
For the LO simulations, we consider the same setup as has been used in our program. 
For the NLO simulations, we consider NNPDF23\_nlo with default 
renormalization and factorization scales, defined as the sum of the transverse masses divided by two of all final state particles and partons.
Negatively weighted events in the NLO simulations have been incorporated consistently. 

\begin{figure}[!h]
  \centering
  \includegraphics[width=7.5cm]{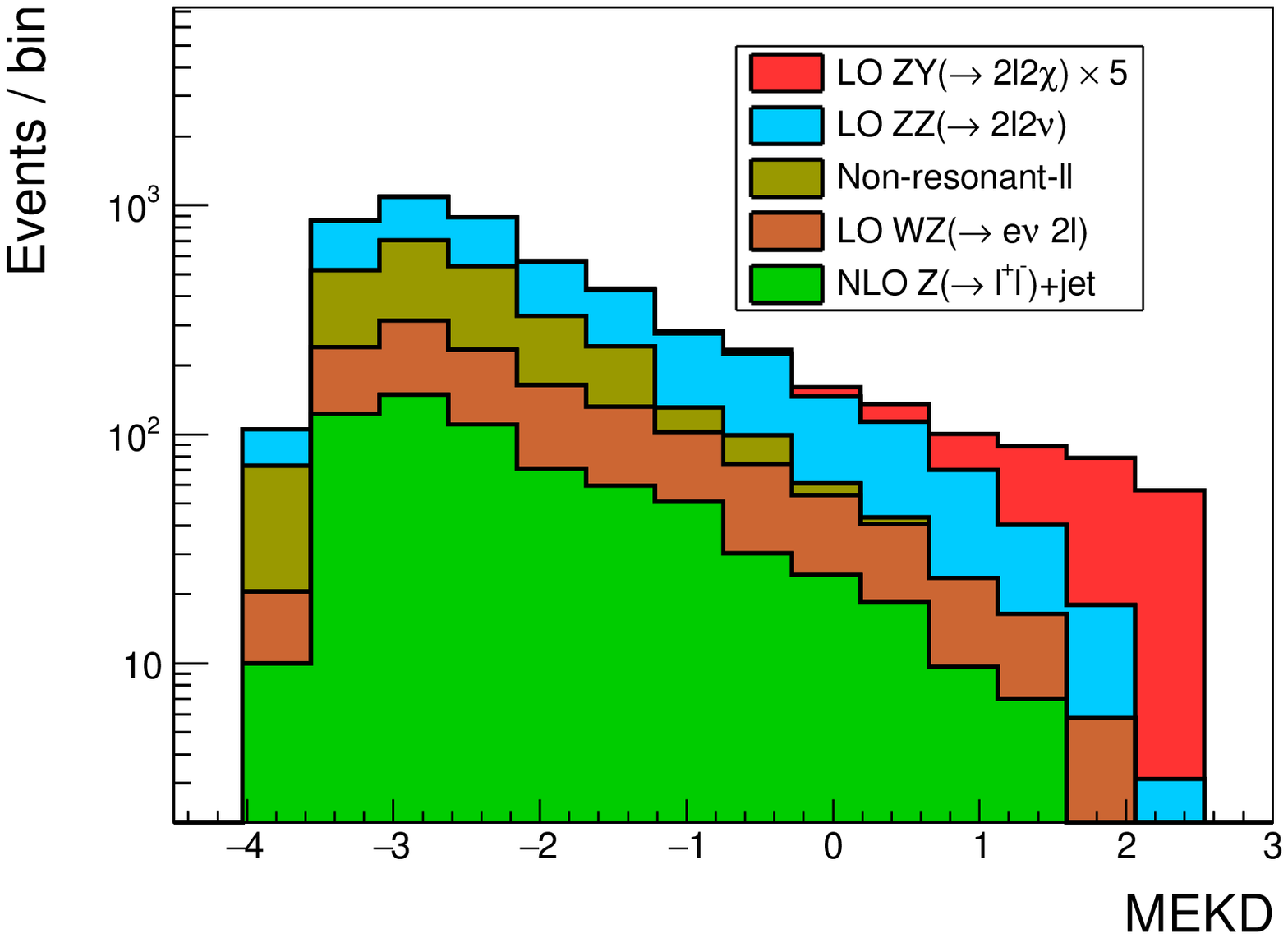}
  \includegraphics[width=7.5cm]{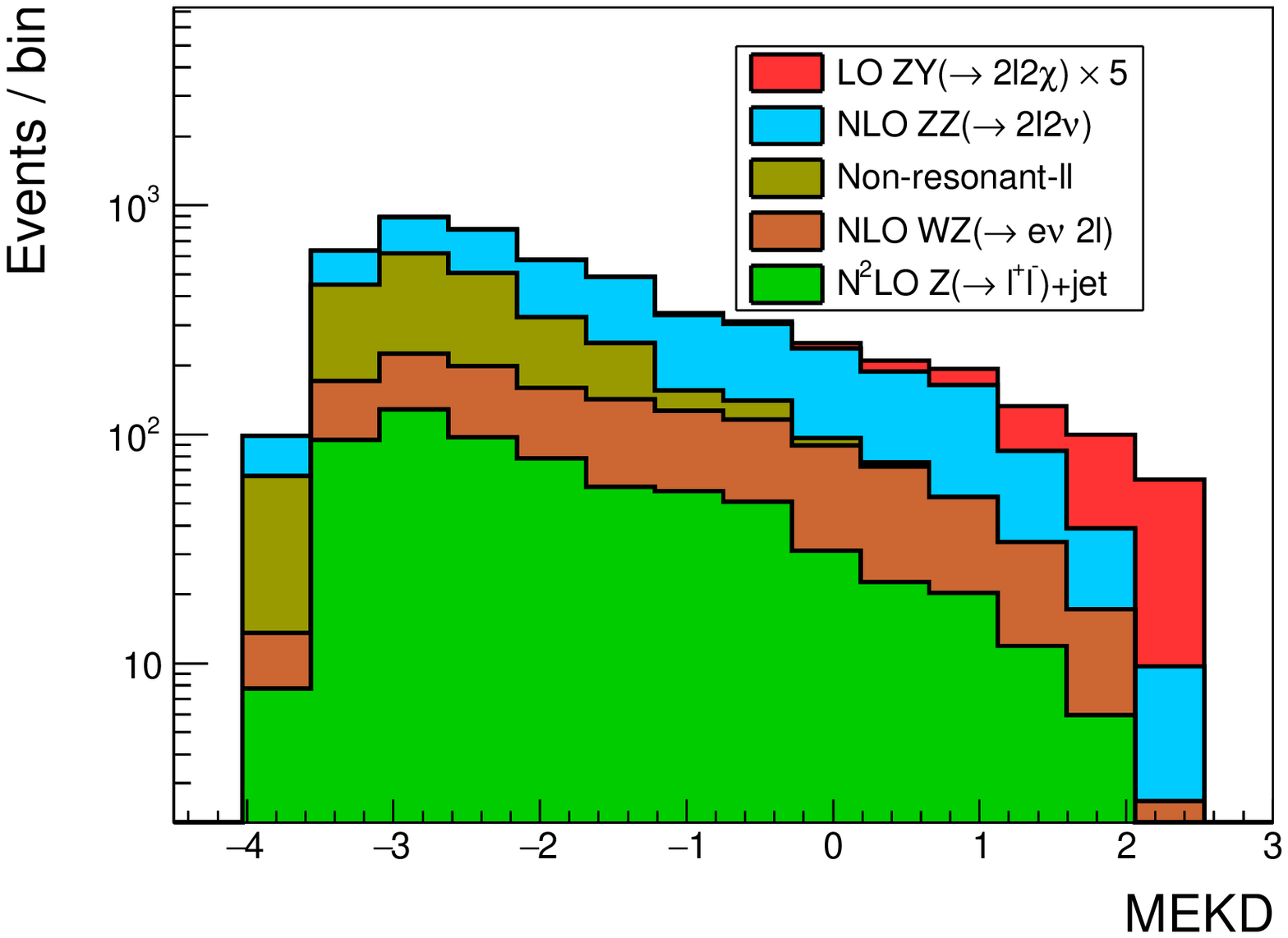}
  \caption{\label{fig:mekd} Example MEKD distributions with {\sc MG5} generated events.
          The left plot is obtained with simulated events at LO accuracy.
          The right plot considers events of Z($\to l^+l^-$)+jet processes at NLO accuracy.
          The signal considers S0$_a$ benchmark model with $\lambda=3.5$. We multiplied the signal yield by a factor of five for a better demonstration. }
\end{figure}

The Fig.~\ref{fig:mekd} stacks MEKD distributions of both signal and backgrounds.
On the left plot, all of the processes are generated with LO accuracy (NLO in QCD for Z($\to l^+l^-$)+jet).
The signal considers S0$_a$ benchmark model with $\lambda=3.5$. We multiplied the signal yield by a factor of five for a better demonstration.
The Non-resonant-$ll$ process is expected to be obtained from data-driven in the experiment.
We mimic its contribution by using a $\mathrm{t\bar{t}}(\to 2l2\nu2 \mathrm{b-jets})$ sample.
The right plot replaces the SM ZZ$\to 2l 2\nu$, WZ($\to e\nu 2l$) and Z($\to l^+l^-$)+jet with simulated events at NLO accuracy.
In both cases, the MEKD shows very nice discrimination power on the signal and background.
It is made clear that NLO simulated events are applicable, with a reasonable loss of sensitivity.

%Improved modeling of these effects and uncertainties only improve quality of the discriminator.
%This method is applicable in our case and we leave the work to additional feasibility studies.

%We remind the readers that NLO QCD corrections and parton shower effects of the models used in the paper
%can be included in an automated way in the framework of {\sc FeynRules-MadGraph5-aMC@NLO} with event generation.

\section{Summary}
\label{sec:summary}

In this paper, we have exploited the Z boson leptonic decay information to probe the dark sector with a scalar, vector, and tensor mediators.
We obtained angular coefficients of the SM $\mathrm{Z Z}\to 2l2\nu$ background and 
benchmark scenarios of the dark sector models in the $\yz-\qt$ plane.
Our results show that the angular coefficients $A_{0}-A_{4}$ behave very differently between the SM $\mathrm{Z Z}\to 2l2\nu$ process and the dark sector signal processes.
The angular coefficients among dark sector models of spin-0 and spin-1 mediators are also found to be different from scenario to scenario.
Specifically, the angular coefficients have sensitivities on the parity violation of the spin-1 model and the CP-violation of the spin-0 model.
The angular coefficients in the spin-2 model are found to be similar to the spin-independent scenario of the spin-1 model but still have minor differences.

To quantify the shape information that can be used for the search of dark sectors,
we consider unbinned fits to the four-dimensional $\yz-\qt-\cos\theta_{CS}-\phi_{CS}$ distributions
based on dynamically constructed matrix element likelihood functions
and set 95\% CL upper limits on the coupling strength parameters of the spin-0 and spin-1
benchmark scenarios.
To be realistic, we emulate the acceptance and efficiency effects referring to the 13 TeV LHC measurement~\cite{Sirunyan:2017onm,Aaboud:2017bja}.
To make our framework concise, we obtained all the results using asymptotic approximation without event generation.

Our evaluated KL-divergence term quantifies the shape effect in each case.
The obtained results demonstrate significant improvements in the limits, especially on the S0 benchmark models.                 
For easier usage of experimental data, we provide an example application of MEKD  with simulated events.
We show that our MEKD constructed with LO matrix elements are applicable for NLO events and preserves good discrimination power on
the signal and background.
We expect this kind of MEKDs to be useful for exploiting the lepton angular distributions in experimental analyses.

\section*{Acknowledgments}

This work would not be possible without what D. Yang have learned from Kaoru Hagiwara (KEK) back in PITT PACC (U.S.) and Xinjiang Univ. (China).
We have benefited from useful discussions with many people, to name a few, Kaoru Hagiwara, Tao Han, Junmo Chen, Xing Wang (PITT), Kai Ma (SUT)
and Yandong Liu, Jing Li (PKU).
D. Yang would also like to thank the PITT particle physics group and the Xinjiang Univ. theoretical physics group for warm hospitality during the stay.
We are also grateful to Junichi Kanzaki (KEK), Yajuan Zheng (NTU) for useful advises in using {\sc BASES}.
This work is supported in part by the National Natural Science Foundation of China, under Grants No. 11475190 and No. 1157500,
and by a short-term internship program from the graduate school of Peking University.

\appendix
\section{Cross checks with the {\sc MG5} program}
\label{app:valid}

To make the {\sc MG5} results comparable, we implemented similar setups as described in the paper.
These include coupling constants, the choice of PDF set, renormalization and factorization scales, Breit-Wigner cutoff, and BL selections as described in Table~\ref{tab:selections} of the paper.
The Table~\ref{tab:validation1} compares our results with the {\sc MG5} ones with one on-shell Z boson in the final states.
For all the cases, the differences lie within statistical uncertainty.
The Table~\ref{tab:validation2} compares our results with the {\sc MG5} with the Z boson leptonicalled decayed.
Our program considered all the BL-selections with NWA, while the {\sc MG5} ones replace the NWA with $|\mathrm{m}_{ll}-\mz|<15\times\Gamma_{\mathrm{Z}}$.
This replacement leads to slightly smaller {\sc MG5} cross sections comparing to ours,
but in general, the differences are not large.
Normalizations of the signal and all of the background pdfs are also checked to be consistent with one.
% We made a few coding errors in arXiv versoin 1 and 2 because only part of the pdfs and cross sections of subprocesses are checked. 

\begin{table}[htb]
\centering
\scalebox{0.8}{
    \begin{tabular}{c|c|c|c|c}
      \hline
      \hline
   Process/Benchmark   &  Cross section            &  Cross section             &  Relative                 &      Relative                       \\
                       &                   (fb)    &  from {\sc MG5} (fb)  &  Difference (\%)          & Statistical uncertainty (\%)        \\ \hline
   S0$_a$              &        0.1535             &         0.1536             &          0.052            &        0.34                         \\
   S0$_b$              &        0.1452             &         0.1454             &          0.14             &        0.29                         \\
   S0$_c$              &     4.436$\times 10^{-7}$ &     4.459$\times 10^{-7}$  &          0.52             &        0.14                         \\
   \hline
   S1$_a$              &        37.16              &         37.21              &          0.14             &        0.23                         \\
   S1$_b$              &        7.931              &         7.943              &          0.15             &        0.24                         \\
   S1$_c$              &        66.94              &         67.01              &          0.11             &        0.25                         \\
   \hline
   Z($\to 2\nu$)Z      &        3561               &         3564               &          0.081            &        0.16                         \\
   W($\to e\nu$)Z      &        2547               &         2556               &          0.39             &        0.26                         \\
   Z+jet               &     1.189$\times 10^{7}$  &     1.192$\times 10^{7}$   &          0.23             &        0.23                         \\
      \hline
      \hline
    \end{tabular}}
  \caption{ Comparison of cross sections obtained by our program and the {\sc MG5}, with one on-shell Z boson in the final states.
            Their differences and the statistical uncertainties taken from the {\sc MG5} are presented relative to the {\sc MG5} ones.   }
  \label{tab:validation1}
\end{table}

\begin{table}[htb]
\centering
\scalebox{0.8}{
    \begin{tabular}{c|c|c|c|c}
      \hline
      \hline
   Process/Benchmark   &  Cross section            &  Cross section             &  Relative                 &      Relative                       \\
                       &                   (fb)    &  from {\sc MG5} (fb)  &  Difference (\%)          & Statistical uncertainty (\%)        \\ \hline
   S0$_a$              &    4.748$\times 10^{-3}$  &    4.688$\times 10^{-3}$   &          1.3              &        0.31                         \\
   S0$_b$              &    4.333$\times 10^{-3}$  &    4.382$\times 10^{-3}$   &          1.1              &        0.33                         \\
   S0$_c$              &    1.667$\times 10^{-8}$  &    1.649$\times 10^{-8}$   &          1.1              &        0.27                         \\
   \hline
   S1$_a$              &         1.149             &         1.034              &          11               &        0.23                         \\
   S1$_b$              &         0.2431            &         0.2186             &          11               &        0.27                         \\
   S1$_c$              &         2.070             &         1.861              &          11               &        0.23                         \\
   \hline
   ZZ$\to 2l 2\nu$     &         27.71             &         26.50              &          4.6              &        0.13                         \\
%   Non-resonant-$ll$   &         1574              &         1574               &          0.017            &        0.27                         \\
   WZ($\to e\nu 2l$)   &         17.05             &         18.39              &          7.3              &        0.26                         \\
   Z($\to l^+l^-$)+jet &         36125             &         34440              &          4.9              &        0.30                         \\
      \hline
      \hline
    \end{tabular}}
  \caption{ Comparison of cross sections obtained by our program and the {\sc MG5}, with Z boson leptonically decayed.
            Our program considered all the BL-selections with NWA, while the {\sc MG5} ones replace the NWA with $|\mathrm{m}_{ll}-\mz|<15\times\Gamma_{\mathrm{Z}}$.
            Hence the {\sc MG5} results are in general slightly smaller than ours.
            Their differences and the statistical uncertainties taken from the {\sc MG5} are presented relative to the {\sc MG5} ones.   }
  \label{tab:validation2}
\end{table}

%%%%%%%%%%%%%% Begin References %%%%%%%%%%%%%%%%%%%%%%%%%%%%%%%%%%%%%%%%

\bibliographystyle{JHEP}
\bibliography{reference}

\providecommand{\href}[2]{#2}\begingroup\raggedright\begin{thebibliography}{10}

\bibitem{Patrignani:2016xqp}
{\scshape Particle Data Group} collaboration, C.~Patrignani et~al.,
  \emph{{Review of Particle Physics}},
  \href{https://doi.org/10.1088/1674-1137/40/10/100001}{\emph{Chin. Phys.}
  {\bfseries C40} (2016) 100001}.

\bibitem{Ade:2015xua}
{\scshape Planck} collaboration, P.~A.~R. Ade et~al., \emph{{Planck 2015
  results. XIII. Cosmological parameters}},
  \href{https://doi.org/10.1051/0004-6361/201525830}{\emph{Astron. Astrophys.}
  {\bfseries 594} (2016) A13},
  [\href{https://arxiv.org/abs/1502.01589}{{\ttfamily 1502.01589}}].

\bibitem{PhysRevLett.39.165}
B.~W. Lee and S.~Weinberg, \emph{Cosmological lower bound on heavy-neutrino
  masses}, \href{https://doi.org/10.1103/PhysRevLett.39.165}{\emph{Phys. Rev.
  Lett.} {\bfseries 39} (Jul, 1977) 165--168}.

\bibitem{Beltran:2010ww}
M.~Beltran, D.~Hooper, E.~W. Kolb, Z.~A.~C. Krusberg and T.~M.~P. Tait,
  \emph{{Maverick dark matter at colliders}},
  \href{https://doi.org/10.1007/JHEP09(2010)037}{\emph{JHEP} {\bfseries 09}
  (2010) 037}, [\href{https://arxiv.org/abs/1002.4137}{{\ttfamily 1002.4137}}].

\bibitem{Aaboud:2017buf}
{\scshape ATLAS} collaboration, M.~Aaboud et~al., \emph{{Measurement of
  detector-corrected observables sensitive to the anomalous production of
  events with jets and large missing transverse momentum in $pp$ collisions at
  $\sqrt{s} = 13$ TeV using the ATLAS detector}},
  \href{https://arxiv.org/abs/1707.03263}{{\ttfamily 1707.03263}}.

\bibitem{Sirunyan:2017hci}
{\scshape CMS} collaboration, A.~M. Sirunyan et~al., \emph{{Search for dark
  matter produced with an energetic jet or a hadronically decaying W or Z boson
  at $ \sqrt{s}=13 $ TeV}},
  \href{https://doi.org/10.1007/JHEP07(2017)014}{\emph{JHEP} {\bfseries 07}
  (2017) 014}, [\href{https://arxiv.org/abs/1703.01651}{{\ttfamily
  1703.01651}}].

\bibitem{Aaboud:2017rzf}
{\scshape ATLAS} collaboration, M.~Aaboud et~al., \emph{{Search for dark matter
  produced in association with bottom or top quarks in $\sqrt{s}$ = 13 TeV pp
  collisions with the ATLAS detector}},
  \href{https://arxiv.org/abs/1710.11412}{{\ttfamily 1710.11412}}.

\bibitem{Sirunyan:2017xgm}
{\scshape CMS} collaboration, A.~M. Sirunyan et~al., \emph{{Search for dark
  matter produced in association with heavy-flavor quarks in proton-proton
  collisions at sqrt(s)=13 TeV}},
  \href{https://arxiv.org/abs/1706.02581}{{\ttfamily 1706.02581}}.

\bibitem{Sirunyan:2017ewk}
{\scshape CMS} collaboration, A.~M. Sirunyan et~al., \emph{{Search for new
  physics in the monophoton final state in proton-proton collisions at $
  \sqrt{s}=13 $ TeV}},
  \href{https://doi.org/10.1007/JHEP10(2017)073}{\emph{JHEP} {\bfseries 10}
  (2017) 073}, [\href{https://arxiv.org/abs/1706.03794}{{\ttfamily
  1706.03794}}].

\bibitem{Carpenter:2012rg}
L.~M. Carpenter, A.~Nelson, C.~Shimmin, T.~M.~P. Tait and D.~Whiteson,
  \emph{{Collider searches for dark matter in events with a Z boson and missing
  energy}}, \href{https://doi.org/10.1103/PhysRevD.87.074005}{\emph{Phys. Rev.}
  {\bfseries D87} (2013) 074005},
  [\href{https://arxiv.org/abs/1212.3352}{{\ttfamily 1212.3352}}].

\bibitem{Sirunyan:2017onm}
{\scshape CMS} collaboration, A.~M. Sirunyan et~al., \emph{{Search for dark
  matter and unparticles in events with a Z boson and missing transverse
  momentum in proton-proton collisions at $ \sqrt{s}=13 $ TeV}},
  \href{https://doi.org/10.1007/JHEP09(2017)106,
  10.1007/JHEP03(2017)061}{\emph{JHEP} {\bfseries 03} (2017) 061},
  [\href{https://arxiv.org/abs/1701.02042}{{\ttfamily 1701.02042}}].

\bibitem{Aaboud:2017bja}
{\scshape ATLAS} collaboration, M.~Aaboud et~al., \emph{{Search for an
  invisibly decaying Higgs boson or dark matter candidates produced in
  association with a $Z$ boson in $pp$ collisions at $\sqrt{s} =$ 13 TeV with
  the ATLAS detector}},  \href{https://arxiv.org/abs/1708.09624}{{\ttfamily
  1708.09624}}.

\bibitem{Bai:2012xg}
Y.~Bai and T.~M.~P. Tait, \emph{{Searches with Mono-Leptons}},
  \href{https://doi.org/10.1016/j.physletb.2013.05.057}{\emph{Phys. Lett.}
  {\bfseries B723} (2013) 384--387},
  [\href{https://arxiv.org/abs/1208.4361}{{\ttfamily 1208.4361}}].

\bibitem{Aaboud:2017efa}
{\scshape ATLAS} collaboration, M.~Aaboud et~al., \emph{{Search for a new heavy
  gauge boson resonance decaying into a lepton and missing transverse momentum
  in 36 fb$^{-1}$ of $pp$ collisions at $\sqrt{s} =$ 13 TeV with the ATLAS
  experiment}},  \href{https://arxiv.org/abs/1706.04786}{{\ttfamily
  1706.04786}}.

\bibitem{Aaboud:2017yqz}
{\scshape ATLAS} collaboration, M.~Aaboud et~al., \emph{{Search for Dark Matter
  Produced in Association with a Higgs Boson Decaying to $b\bar b$ using 36
  fb$^{-1}$ of $pp$ collisions at $\sqrt s=13$ TeV with the ATLAS Detector}},
  \href{https://doi.org/10.1103/PhysRevLett.119.181804}{\emph{Phys. Rev. Lett.}
  {\bfseries 119} (2017) 181804},
  [\href{https://arxiv.org/abs/1707.01302}{{\ttfamily 1707.01302}}].

\bibitem{Sirunyan:2017hnk}
{\scshape CMS} collaboration, A.~M. Sirunyan et~al., \emph{{Search for
  associated production of dark matter with a Higgs boson decaying to
  $b\bar{b}$ or $\gamma\gamma$ at $\sqrt{s}$ = 13 TeV}},
  \href{https://arxiv.org/abs/1703.05236}{{\ttfamily 1703.05236}}.

\bibitem{Sirunyan:2016iap}
{\scshape CMS} collaboration, A.~M. Sirunyan et~al., \emph{{Search for dijet
  resonances in proton¿proton collisions at $\sqrt{s}$ = 13 TeV and constraints
  on dark matter and other models}},
  \href{https://doi.org/10.1016/j.physletb.2017.09.029,
  10.1016/j.physletb.2017.02.012}{\emph{Phys. Lett.} {\bfseries B769} (2017)
  520--542}, [\href{https://arxiv.org/abs/1611.03568}{{\ttfamily 1611.03568}}].

\bibitem{Khachatryan:2016jww}
{\scshape CMS} collaboration, V.~Khachatryan et~al., \emph{{Search for heavy
  gauge W' boson in events with an energetic lepton and large missing
  transverse momentum at $ \sqrt{s} = $ 13 TeV}},
  \href{https://doi.org/10.1016/j.physletb.2017.04.043}{\emph{Phys. Lett.}
  {\bfseries B770} (2017) 278--301},
  [\href{https://arxiv.org/abs/1612.09274}{{\ttfamily 1612.09274}}].

\bibitem{Sirunyan:2017nvi}
{\scshape CMS} collaboration, A.~M. Sirunyan et~al., \emph{{Search for low mass
  vector resonances decaying into quark-antiquark pairs in proton-proton
  collisions at $\sqrt{s} = $ 13 TeV}},
  \href{https://arxiv.org/abs/1710.00159}{{\ttfamily 1710.00159}}.

\bibitem{Aaboud:2017buh}
{\scshape ATLAS} collaboration, M.~Aaboud et~al., \emph{{Search for new
  high-mass phenomena in the dilepton final state using 36.1 fb$^{-1}$ of
  proton-proton collision data at $\sqrt{s}$ = 13 TeV with the ATLAS
  detector}},  \href{https://arxiv.org/abs/1707.02424}{{\ttfamily 1707.02424}}.

\bibitem{Abercrombie:2015wmb}
D.~Abercrombie et~al., \emph{{Dark Matter Benchmark Models for Early LHC Run-2
  Searches: Report of the ATLAS/CMS Dark Matter Forum}},
  \href{https://arxiv.org/abs/1507.00966}{{\ttfamily 1507.00966}}.

\bibitem{Neubert:2015fka}
M.~Neubert, J.~Wang and C.~Zhang, \emph{{Higher-Order QCD Predictions for Dark
  Matter Production in Mono-$Z$ Searches at the LHC}},
  \href{https://doi.org/10.1007/JHEP02(2016)082}{\emph{JHEP} {\bfseries 02}
  (2016) 082}, [\href{https://arxiv.org/abs/1509.05785}{{\ttfamily
  1509.05785}}].

\bibitem{Petriello:2008pu}
F.~J. Petriello, S.~Quackenbush and K.~M. Zurek, \emph{{The Invisible
  $Z^\prime$ at the CERN LHC}},
  \href{https://doi.org/10.1103/PhysRevD.77.115020}{\emph{Phys. Rev.}
  {\bfseries D77} (2008) 115020},
  [\href{https://arxiv.org/abs/0803.4005}{{\ttfamily 0803.4005}}].

\bibitem{Alves:2015dya}
A.~Alves and K.~Sinha, \emph{{Searches for Dark Matter at the LHC: A
  Multivariate Analysis in the Mono-$Z$ Channel}},
  \href{https://doi.org/10.1103/PhysRevD.92.115013}{\emph{Phys. Rev.}
  {\bfseries D92} (2015) 115013},
  [\href{https://arxiv.org/abs/1507.08294}{{\ttfamily 1507.08294}}].

\bibitem{Han:1999ne}
T.~Han, D.~L. Rainwater and D.~Zeppenfeld, \emph{{Drell-Yan plus missing energy
  as a signal for extra dimensions}},
  \href{https://doi.org/10.1016/S0370-2693(99)00950-8}{\emph{Phys. Lett.}
  {\bfseries B463} (1999) 93--98},
  [\href{https://arxiv.org/abs/hep-ph/9905423}{{\ttfamily hep-ph/9905423}}].

\bibitem{Yu:2014ula}
Z.-H. Yu, X.-J. Bi, Q.-S. Yan and P.-F. Yin, \emph{{Dark matter searches in the
  mono-$Z$ channel at high energy $e^+e^-$ colliders}},
  \href{https://doi.org/10.1103/PhysRevD.90.055010}{\emph{Phys. Rev.}
  {\bfseries D90} (2014) 055010},
  [\href{https://arxiv.org/abs/1404.6990}{{\ttfamily 1404.6990}}].

\bibitem{Aad:2014wca}
{\scshape ATLAS} collaboration, G.~Aad et~al., \emph{{Search for contact
  interactions and large extra dimensions in the dilepton channel using
  proton-proton collisions at $\sqrt{s}$ = 8 TeV with the ATLAS detector}},
  \href{https://doi.org/10.1140/epjc/s10052-014-3134-6}{\emph{Eur. Phys. J.}
  {\bfseries C74} (2014) 3134},
  [\href{https://arxiv.org/abs/1407.2410}{{\ttfamily 1407.2410}}].

\bibitem{Goodman:2010ku}
J.~Goodman, M.~Ibe, A.~Rajaraman, W.~Shepherd, T.~M.~P. Tait and H.-B. Yu,
  \emph{{Constraints on Dark Matter from Colliders}},
  \href{https://doi.org/10.1103/PhysRevD.82.116010}{\emph{Phys. Rev.}
  {\bfseries D82} (2010) 116010},
  [\href{https://arxiv.org/abs/1008.1783}{{\ttfamily 1008.1783}}].

\bibitem{Goodman:2010yf}
J.~Goodman, M.~Ibe, A.~Rajaraman, W.~Shepherd, T.~M.~P. Tait and H.-B. Yu,
  \emph{{Constraints on Light Majorana dark Matter from Colliders}},
  \href{https://doi.org/10.1016/j.physletb.2010.11.009}{\emph{Phys. Lett.}
  {\bfseries B695} (2011) 185--188},
  [\href{https://arxiv.org/abs/1005.1286}{{\ttfamily 1005.1286}}].

\bibitem{Cao:2009uw}
Q.-H. Cao, C.-R. Chen, C.~S. Li and H.~Zhang, \emph{{Effective Dark Matter
  Model: Relic density, CDMS II, Fermi LAT and LHC}},
  \href{https://doi.org/10.1007/JHEP08(2011)018}{\emph{JHEP} {\bfseries 08}
  (2011) 018}, [\href{https://arxiv.org/abs/0912.4511}{{\ttfamily 0912.4511}}].

\bibitem{Cotta:2012nj}
R.~C. Cotta, J.~L. Hewett, M.~P. Le and T.~G. Rizzo, \emph{{Bounds on Dark
  Matter Interactions with Electroweak Gauge Bosons}},
  \href{https://doi.org/10.1103/PhysRevD.88.116009}{\emph{Phys. Rev.}
  {\bfseries D88} (2013) 116009},
  [\href{https://arxiv.org/abs/1210.0525}{{\ttfamily 1210.0525}}].

\bibitem{Mattelaer:2015haa}
O.~Mattelaer and E.~Vryonidou, \emph{{Dark matter production through
  loop-induced processes at the LHC: the s-channel mediator case}},
  \href{https://doi.org/10.1140/epjc/s10052-015-3665-5}{\emph{Eur. Phys. J.}
  {\bfseries C75} (2015) 436},
  [\href{https://arxiv.org/abs/1508.00564}{{\ttfamily 1508.00564}}].

\bibitem{Backovic:2015soa}
M.~Backovic, M.~Krämer, F.~Maltoni, A.~Martini, K.~Mawatari and M.~Pellen,
  \emph{{Higher-order QCD predictions for dark matter production at the LHC in
  simplified models with s-channel mediators}},
  \href{https://doi.org/10.1140/epjc/s10052-015-3700-6}{\emph{Eur. Phys. J.}
  {\bfseries C75} (2015) 482},
  [\href{https://arxiv.org/abs/1508.05327}{{\ttfamily 1508.05327}}].

\bibitem{Das:2016pbk}
G.~Das, C.~Degrande, V.~Hirschi, F.~Maltoni and H.-S. Shao, \emph{{NLO
  predictions for the production of a spin-two particle at the LHC}},
  \href{https://doi.org/10.1016/j.physletb.2017.05.007}{\emph{Phys. Lett.}
  {\bfseries B770} (2017) 507--513},
  [\href{https://arxiv.org/abs/1605.09359}{{\ttfamily 1605.09359}}].

\bibitem{Kraml:2017atm}
S.~Kraml, U.~Laa, K.~Mawatari and K.~Yamashita, \emph{{Simplified dark matter
  models with a spin-2 mediator at the LHC}},
  \href{https://doi.org/10.1140/epjc/s10052-017-4871-0}{\emph{Eur. Phys. J.}
  {\bfseries C77} (2017) 326},
  [\href{https://arxiv.org/abs/1701.07008}{{\ttfamily 1701.07008}}].

\bibitem{Cheung:2010zf}
K.~Cheung, K.~Mawatari, E.~Senaha, P.-Y. Tseng and T.-C. Yuan, \emph{{The Top
  Window for dark matter}},
  \href{https://doi.org/10.1007/JHEP10(2010)081}{\emph{JHEP} {\bfseries 10}
  (2010) 081}, [\href{https://arxiv.org/abs/1009.0618}{{\ttfamily 1009.0618}}].

\bibitem{Lin:2013sca}
T.~Lin, E.~W. Kolb and L.-T. Wang, \emph{{Probing dark matter couplings to top
  and bottom quarks at the LHC}},
  \href{https://doi.org/10.1103/PhysRevD.88.063510}{\emph{Phys. Rev.}
  {\bfseries D88} (2013) 063510},
  [\href{https://arxiv.org/abs/1303.6638}{{\ttfamily 1303.6638}}].

\bibitem{doi:10.1143/JPSJ.57.4126}
K.~Kondo, \emph{Dynamical likelihood method for reconstruction of events with
  missing momentum. i. method and toy models},
  \href{https://doi.org/10.1143/JPSJ.57.4126}{\emph{Journal of the Physical
  Society of Japan} {\bfseries 57} (1988) 4126--4140},
  [\href{https://arxiv.org/abs/http://dx.doi.org/10.1143/JPSJ.57.4126}{{\ttfamily
  http://dx.doi.org/10.1143/JPSJ.57.4126}}].

\bibitem{doi:10.1143/JPSJ.60.836}
K.~Kondo, \emph{Dynamical likelihood method for reconstruction of events with
  missing momentum. ii. mass spectra for 2$\to$2 processes},
  \href{https://doi.org/10.1143/JPSJ.60.836}{\emph{Journal of the Physical
  Society of Japan} {\bfseries 60} (1991) 836--844},
  [\href{https://arxiv.org/abs/http://dx.doi.org/10.1143/JPSJ.60.836}{{\ttfamily
  http://dx.doi.org/10.1143/JPSJ.60.836}}].

\bibitem{Gao:2010qx}
Y.~Gao, A.~V. Gritsan, Z.~Guo, K.~Melnikov, M.~Schulze and N.~V. Tran,
  \emph{{Spin determination of single-produced resonances at hadron
  colliders}}, \href{https://doi.org/10.1103/PhysRevD.81.075022}{\emph{Phys.
  Rev.} {\bfseries D81} (2010) 075022},
  [\href{https://arxiv.org/abs/1001.3396}{{\ttfamily 1001.3396}}].

\bibitem{Chatrchyan:2012sn}
{\scshape CMS} collaboration, S.~Chatrchyan et~al., \emph{{Search for a Higgs
  boson in the decay channel $H$ to ZZ(*) to $q$ qbar $\ell^-$ l+ in $pp$
  collisions at $\sqrt{s}=7$ TeV}},
  \href{https://doi.org/10.1007/JHEP04(2012)036}{\emph{JHEP} {\bfseries 04}
  (2012) 036}, [\href{https://arxiv.org/abs/1202.1416}{{\ttfamily 1202.1416}}].

\bibitem{DeRujula:2010ys}
A.~De~Rujula, J.~Lykken, M.~Pierini, C.~Rogan and M.~Spiropulu, \emph{{Higgs
  look-alikes at the LHC}},
  \href{https://doi.org/10.1103/PhysRevD.82.013003}{\emph{Phys. Rev.}
  {\bfseries D82} (2010) 013003},
  [\href{https://arxiv.org/abs/1001.5300}{{\ttfamily 1001.5300}}].

\bibitem{kullback1951}
S.~Kullback and R.~A. Leibler, \emph{On information and sufficiency},
  \href{https://doi.org/10.1214/aoms/1177729694}{\emph{Ann. Math. Statist.}
  {\bfseries 22} (03, 1951) 79--86}.

\bibitem{Alwall:2010cq}
J.~Alwall, A.~Freitas and O.~Mattelaer, \emph{{The Matrix Element Method and
  QCD Radiation}},
  \href{https://doi.org/10.1103/PhysRevD.83.074010}{\emph{Phys. Rev.}
  {\bfseries D83} (2011) 074010},
  [\href{https://arxiv.org/abs/1010.2263}{{\ttfamily 1010.2263}}].

\bibitem{PhysRevD.16.2219}
J.~C. Collins and D.~E. Soper, \emph{Angular distribution of dileptons in
  high-energy hadron collisions},
  \href{https://doi.org/10.1103/PhysRevD.16.2219}{\emph{Phys. Rev. D}
  {\bfseries 16} (Oct, 1977) 2219--2225}.

\bibitem{Aad:2016izn}
{\scshape ATLAS} collaboration, G.~Aad et~al., \emph{{Measurement of the
  angular coefficients in $Z$-boson events using electron and muon pairs from
  data taken at $\sqrt{s}=8$ TeV with the ATLAS detector}},
  \href{https://doi.org/10.1007/JHEP08(2016)159}{\emph{JHEP} {\bfseries 08}
  (2016) 159}, [\href{https://arxiv.org/abs/1606.00689}{{\ttfamily
  1606.00689}}].

\bibitem{Khachatryan:2015paa}
{\scshape CMS} collaboration, V.~Khachatryan et~al., \emph{{Angular
  coefficients of Z bosons produced in pp collisions at $\sqrt{s}$ = 8 TeV and
  decaying to $\mu^+ \mu^-$ as a function of transverse momentum and
  rapidity}}, \href{https://doi.org/10.1016/j.physletb.2015.08.061}{\emph{Phys.
  Lett.} {\bfseries B750} (2015) 154--175},
  [\href{https://arxiv.org/abs/1504.03512}{{\ttfamily 1504.03512}}].

\bibitem{Beaujean:2015xea}
F.~Beaujean, M.~Chrzaszcz, N.~Serra and D.~van Dyk, \emph{{Extracting Angular
  Observables without a Likelihood and Applications to Rare Decays}},
  \href{https://doi.org/10.1103/PhysRevD.91.114012}{\emph{Phys. Rev.}
  {\bfseries D91} (2015) 114012},
  [\href{https://arxiv.org/abs/1503.04100}{{\ttfamily 1503.04100}}].

\bibitem{Dutta:2008bh}
S.~Dutta, K.~Hagiwara and Y.~Matsumoto, \emph{{Measuring the Higgs-Vector boson
  Couplings at Linear $e^{+} e^{-}$ Collider}},
  \href{https://doi.org/10.1103/PhysRevD.78.115016}{\emph{Phys. Rev.}
  {\bfseries D78} (2008) 115016},
  [\href{https://arxiv.org/abs/0808.0477}{{\ttfamily 0808.0477}}].

\bibitem{Alloul:2013bka}
A.~Alloul, N.~D. Christensen, C.~Degrande, C.~Duhr and B.~Fuks,
  \emph{{FeynRules 2.0 - A complete toolbox for tree-level phenomenology}},
  \href{https://doi.org/10.1016/j.cpc.2014.04.012}{\emph{Comput. Phys. Commun.}
  {\bfseries 185} (2014) 2250--2300},
  [\href{https://arxiv.org/abs/1310.1921}{{\ttfamily 1310.1921}}].

\bibitem{deAquino:2011ub}
P.~de~Aquino, W.~Link, F.~Maltoni, O.~Mattelaer and T.~Stelzer, \emph{{ALOHA:
  Automatic Libraries Of Helicity Amplitudes for Feynman Diagram
  Computations}},
  \href{https://doi.org/10.1016/j.cpc.2012.05.004}{\emph{Comput. Phys. Commun.}
  {\bfseries 183} (2012) 2254--2263},
  [\href{https://arxiv.org/abs/1108.2041}{{\ttfamily 1108.2041}}].

\bibitem{Alwall:2014hca}
J.~Alwall, R.~Frederix, S.~Frixione, V.~Hirschi, F.~Maltoni, O.~Mattelaer
  et~al., \emph{{The automated computation of tree-level and next-to-leading
  order differential cross sections, and their matching to parton shower
  simulations}}, \href{https://doi.org/10.1007/JHEP07(2014)079}{\emph{JHEP}
  {\bfseries 07} (2014) 079},
  [\href{https://arxiv.org/abs/1405.0301}{{\ttfamily 1405.0301}}].

\bibitem{HAGIWARA19861}
K.~Hagiwara and D.~Zeppenfeld, \emph{Helicity amplitudes for heavy lepton
  production in e+e annihilation},
  \href{https://doi.org/http://dx.doi.org/10.1016/0550-3213(86)90615-2}{\emph{Nuclear
  Physics B} {\bfseries 274} (1986) 1 -- 32}.

\bibitem{DREINER20101}
H.~K. Dreiner, H.~E. Haber and S.~P. Martin, \emph{Two-component spinor
  techniques and feynman rules for quantum field theory and supersymmetry},
  \href{https://doi.org/http://dx.doi.org/10.1016/j.physrep.2010.05.002}{\emph{Physics
  Reports} {\bfseries 494} (2010) 1 -- 196}.

\bibitem{Kawabata:1995th}
S.~Kawabata, \emph{{A New version of the multidimensional integration and event
  generation package BASES/SPRING}},
  \href{https://doi.org/10.1016/0010-4655(95)00028-E}{\emph{Comput. Phys.
  Commun.} {\bfseries 88} (1995) 309--326}.

\bibitem{Ball:2013hta}
{\scshape NNPDF} collaboration, R.~D. Ball, V.~Bertone, S.~Carrazza,
  L.~Del~Debbio, S.~Forte, A.~Guffanti et~al., \emph{{Parton distributions with
  QED corrections}},
  \href{https://doi.org/10.1016/j.nuclphysb.2013.10.010}{\emph{Nucl. Phys.}
  {\bfseries B877} (2013) 290--320},
  [\href{https://arxiv.org/abs/1308.0598}{{\ttfamily 1308.0598}}].

\bibitem{Barlow:1990vc}
R.~J. Barlow, \emph{{Extended maximum likelihood}},
  \href{https://doi.org/10.1016/0168-9002(90)91334-8}{\emph{Nucl. Instrum.
  Meth.} {\bfseries A297} (1990) 496--506}.

\bibitem{wilks1938}
S.~S. Wilks, \emph{The large-sample distribution of the likelihood ratio for
  testing composite hypotheses},
  \href{https://doi.org/10.1214/aoms/1177732360}{\emph{Ann. Math. Statist.}
  {\bfseries 9} (03, 1938) 60--62}.

\bibitem{Campbell:2012cz}
J.~M. Campbell, W.~T. Giele and C.~Williams, \emph{{The Matrix Element Method
  at Next-to-Leading Order}},
  \href{https://doi.org/10.1007/JHEP11(2012)043}{\emph{JHEP} {\bfseries 11}
  (2012) 043}, [\href{https://arxiv.org/abs/1204.4424}{{\ttfamily 1204.4424}}].

\bibitem{Soper:2011cr}
D.~E. Soper and M.~Spannowsky, \emph{{Finding physics signals with shower
  deconstruction}},
  \href{https://doi.org/10.1103/PhysRevD.84.074002}{\emph{Phys. Rev.}
  {\bfseries D84} (2011) 074002},
  [\href{https://arxiv.org/abs/1102.3480}{{\ttfamily 1102.3480}}].

\bibitem{Avery:2012um}
P.~Avery et~al., \emph{{Precision studies of the Higgs boson decay channel
  H$\to$ZZ$\to$4$l$ with MEKD}},
  \href{https://doi.org/10.1103/PhysRevD.87.055006}{\emph{Phys. Rev.}
  {\bfseries D87} (2013) 055006},
  [\href{https://arxiv.org/abs/1210.0896}{{\ttfamily 1210.0896}}].

\bibitem{Chatrchyan:2012jja}
{\scshape CMS} collaboration, S.~Chatrchyan et~al., \emph{{Study of the Mass
  and Spin-Parity of the Higgs Boson Candidate Via Its Decays to Z Boson
  Pairs}}, \href{https://doi.org/10.1103/PhysRevLett.110.081803}{\emph{Phys.
  Rev. Lett.} {\bfseries 110} (2013) 081803},
  [\href{https://arxiv.org/abs/1212.6639}{{\ttfamily 1212.6639}}].

\end{thebibliography}\endgroup

\end{document}